\documentclass[11pt,a4paper]{article}

\usepackage{jheppub}

\usepackage{epsfig}
\usepackage{amssymb}
\usepackage{amsfonts}
\usepackage{amsbsy}
\usepackage[all]{xy}
\usepackage{amsmath}
\usepackage{enumerate}

\usepackage{amssymb,amscd}
\usepackage{mathrsfs}
\usepackage{amsmath,amsthm}
\usepackage{xspace}
\usepackage{framed}

\definecolor{shadecolor}{rgb}{0.85,0.85,0.85}

\def\re{\mbox{Re }}

\def\twp{\tilde\wp}

\def\IC{\mathbb{C}}

\def\IH{\mathbb{H}}

\def\IP{\mathbb{P}}

\def\IR{{\mathbb{R}}}
\def\R{{\mathbb{R}}}

\def\IZ{{\mathbb{Z}}}
\def\Z{{\mathbb{Z}}}

\def\fg{\mathfrak{g}}

\def\CC {{\cal C}}
\def\CI {{\cal I}}
\def\CM {{\cal M}}

\def\CN{{\cal N}}
\def\CK {{\cal K}}
\def\CN {{\cal N}}
\def\CR {{\cal R}}
\def\CD {{\cal D}}
\def\CF {{\cal F}}
\def\CJ {{\cal J}}
\def\CP {{\cal P }}
\def\CL {{\cal L}}

\def\wnet {{\cal W}}
\def\knet {{\cal D}}

\def\CY {{\cal Y}}

\def\cY {{\cal Y}}

\def\CH {{\cal H}}
\def\CB {{\cal B}}
\def\CS {{\cal S}}
\def\cS {{\cal S}}
\def\CA{{\cal A}}
\def\CK{{\cal K}}
\def\cK{{\cal K}}

\def\CU{{\cal U}}

\def\CT{{\cal T}}

\def\half{\frac{1}{2}}

\renewcommand{\Im}{{\rm Im }}
\renewcommand{\Re}{{\rm Re }}

\def\one{{\hbox{ 1\kern-.8mm l}}}

\def\ab{{\mathrm{ab}}}
\def\tz{{\tilde z}}

\def\p{\partial}
\def\ba{\mathbf{a}}

\def\be{\bar{e}}

\def\bD{\mathbf{D}}

\def\bF{\mathbf{F}}
\def\bX{\mathbf{X}}

\def\half{\frac{1}{2}}

\newcommand{\abs}[1]{\lvert#1\rvert}
\newcommand{\norm}[1]{\lVert#1\rVert}
\newcommand{\tGamma}{\tilde\Gamma}

\newcommand{\ti}[1]{\textit{#1}}

\def\fb{\mathfrak{b}}
\def\fg{\mathfrak{g}}
\def\fm{\mathfrak{m}}
\def\fs{\mathfrak{s}}

\def\p{\partial}

\def\re{\mbox{Re }}

\def\IC{\mathbb{C}}

\def\IZ{{\mathbb{Z}}}
\def\IR{{\mathbb{R}}}
\def\IP{\mathbb{P}}

\def\CB {{\cal B}}
\def\CC {{\cal C}}
\def\CI {{\cal I}}
\def\CM {{\cal M}}

\def\CN{{\cal N}}
\def\CK {{\cal K}}
\def\CN {{\cal N}}
\def\CR {{\cal R}}
\def\CD {{\cal D}}
\def\CF {{\cal F}}
\def\CJ {{\cal J}}
\def\CP {{\cal P }}
\def\CL {{\cal L}}

\def\wnet {{\cal W}}

\def\CY {{\cal Y}}

\def\CH {{\cal H}}
\def\CB {{\cal B}}
\def\CS {{\cal S}}
\def\CA{{\cal A}}
\def\CK{{\cal K}}

\def\CU{{\cal U}}

\def\CT{{\cal T}}

\def\half{\frac{1}{2}}

\def\be{\begin{equation}
}

\def\ee{\end{equation}}

\newcommand\cD{{\mathcal D}}
\newcommand\I{{\mathrm i}}
\newcommand{\inprod}[1]{\langle#1\rangle}
\newcommand\N{{\cal N}}

\newcommand\eps{\epsilon}

\newcommand\fro{{\overline{\underline{\Omega}}}}

\newcommand\de{{\mathrm{d}}}
\newcommand\off{{\mathrm{off}}}

\newcommand\iin{{\mathrm{in}}}
\newcommand\out{{\mathrm{out}}}

\newcommand\formal{{\mathrm{fo}}}

\newcommand{\bS}{\mathbb S}

\DeclareMathOperator{\cl}{{cl}}
\DeclareMathOperator{\diag}{{diag}}

\DeclareMathOperator{\End}{{End}}

\DeclareMathOperator{\Hom}{{Hom}}

\newcommand{\insfig}[2]{\begin{figure}[htbp] \centering \includegraphics[scale=0.3]{figures/#1-crop.pdf} \caption{#2} \label{fig:#1} \end{figure}}
\newcommand{\insfigscaled}[3]{\begin{figure}[htbp] \centering \includegraphics[scale=#2]{figures/#1-crop.pdf} \caption{#3} \label{fig:#1} \end{figure}}

\title{Spectral networks}

\author[1]{Davide Gaiotto,}
\author[2]{Gregory W. Moore}
\author[3]{and Andrew Neitzke}

\affiliation[1]
{School of Natural Sciences, Institute for Advanced Study, \\
Princeton, NJ 08540, USA}
\affiliation[2]
{NHETC and Department of Physics and Astronomy, Rutgers University,\\
Piscataway, NJ 08855--0849, USA}
\affiliation[3]
{Department of Mathematics, University of Texas at Austin,\\
Austin, TX 78712, USA}

\emailAdd{dgaiotto@ias.edu}
\emailAdd{gmoore@physics.rutgers.edu}
\emailAdd{neitzke@math.utexas.edu}

\abstract{We introduce new geometric objects called \ti{spectral networks}.
Spectral networks are networks of trajectories on Riemann surfaces obeying
certain local rules.  Spectral networks arise naturally in
four-dimensional $\CN=2$ theories coupled to surface defects, particularly
the theories of class $S$.
In these theories spectral networks provide a useful tool for
the computation of BPS degeneracies:  the network directly determines
the degeneracies of solitons living on the surface
defect, which in turn   determine the degeneracies
for particles living in the 4d bulk.  Spectral
networks also lead to a new map between
flat $GL(K,\IC)$ connections on a two-dimensional surface $C$ and
flat abelian connections on an appropriate branched cover $\Sigma$ of $C$.
This construction produces natural coordinate systems on
moduli spaces of flat $GL(K,\IC)$ connections on $C$, which we conjecture
are cluster coordinate systems.
}

\begin{document}

\maketitle

\newpage

\section{Introduction and summary}\label{sec:Introduction}

In this paper we study objects which we call \ti{spectral networks}.
A spectral network $\wnet$
is made up of \ti{walls} drawn on  a punctured real surface $C$.
Here is a picture of one:
\insfigscaled{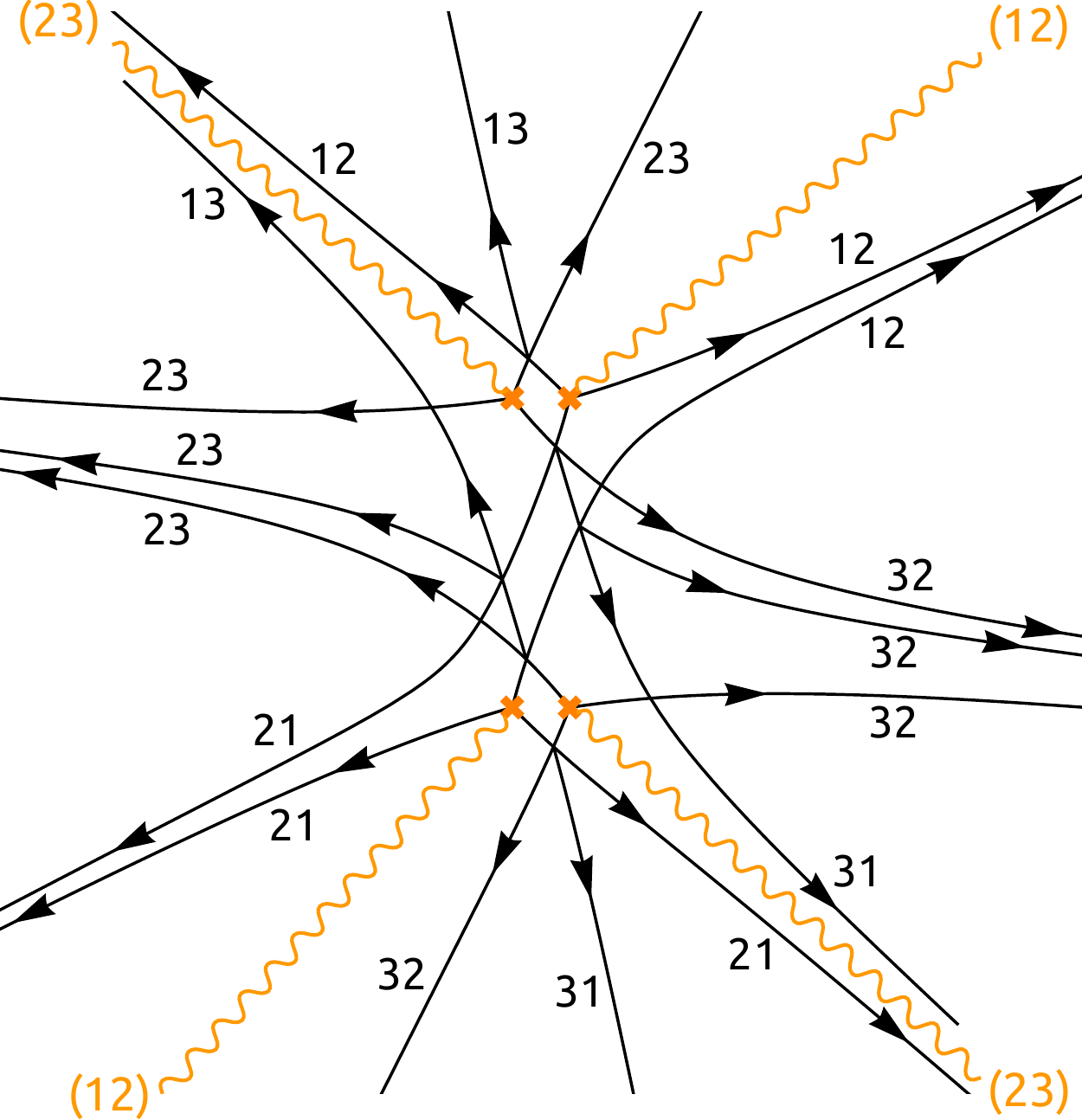}{0.6}{A spectral network, drawn on
the stereographic
projection of $C = S^2$, with a single puncture at infinity.
All of the walls eventually asymptote to this puncture.
The walls are labeled by pairs $ij$, where $i$, $j$ are sheets of a 3-fold covering $\Sigma \to C$.
The branch points of the covering are shown as orange crosses.
We have trivialized the covering over the complement
of some branch cuts, shown as wavy orange lines.}

Each wall carries some extra discrete data:  a pair $ij$ and (roughly) an integer $\mu$.  The labels $i$ and $j$ are drawn from the set of sheets
of a finite branched covering $\Sigma \to C$.  The walls and their
discrete data obey certain local constraints:  for example,
\begin{itemize}
 \item each simple branch point of the covering $\Sigma \to C$ gives birth to three walls;
 \item when an $ij$ wall and a $jk$ wall intersect, a new
$ik$ wall is born at their intersection point.
\end{itemize}

In \S\ref{sec:review-S}-\S\ref{sec:Examples} of this paper we study
some particular spectral networks which arise naturally in physics.
In \S\ref{sec:spectral-networks} we axiomatize the notion of spectral network.
In \S\ref{sec:moduli} we describe a more mathematical application, to coordinate systems
on moduli spaces of flat connections.  Those sections are mostly self-contained.

\medskip

Here is how spectral networks arise in physics.
Fix an $\N=2$ supersymmetric theory $T$ in $d=4$ and a point $u$ of the Coulomb branch.
Also fix a 1/2-BPS surface defect $\bS_z$ which has
a 1-complex-dimensional space $C$ of UV parameters, and is massive in the IR,
with finitely many vacua for any fixed $z$.
These data determine a 1-parameter family of spectral networks
$\wnet_\vartheta$ drawn on $C$, labeled by phases $\vartheta \in \R / 2 \pi \Z$.
Namely, $\wnet_\vartheta$ is the locus of $z \in C$ for which the surface defect $\bS_z$ carries a BPS soliton whose central charge $Z$ is aligned with $-e^{\I \vartheta}$.
The discrete data $\mu$ on the walls keep track of the degeneracies of these BPS solitons.
The points of the finite cover $\Sigma \to C$ over $z \in C$ are the vacua of $\bS_z$.
The data $ij$ on the walls keep track of which vacua of $\bS_z$ are being interpolated by the solitons.

\medskip

In this paper, for concreteness, we focus on a particular (large) class of $\N=2$ theories,
namely the theories of class $S$ \cite{Witten:1997sc,Gaiotto:2009we,Gaiotto:2009hg}.
These are the theories $S[\fg,C,D]$, associated to
a Lie algebra $\fg$ (which we take to be $\fg = A_{K-1}$),
punctured Riemann surface $C$, and a collection $D$
of defects placed at the punctures of $C$.
They are obtained by a partially topologically twisted compactification
of the six-dimensional $(2,0)$ theory $S[\fg]$.
A point $u$ of the Coulomb branch here means a
tuple $(\phi_2, \dots, \phi_K)$, where each $\phi_r$ is a
meromorphic $r$-differential on $C$, with poles at the defects.
In these theories there is
a canonical surface defect $\bS_z$ whose parameter space is $C$.
Moreover, in these theories the covering $\Sigma \to C$ can be identified with the
Seiberg-Witten curve \cite{Gaiotto:2009fs,Gaiotto:2011tf} and is given concretely by \eqref{eq:sw-curve} below.

The spectral networks $\wnet_\vartheta$ arising from theories of class $S$ can be described concretely.
The walls of the network are solutions of differential equations, given in \eqref{eq:trajectory-de} below.
The full network is built up by a continuous process:  three walls are born from each branch point of
the covering $\Sigma$ and flow according to \eqref{eq:trajectory-de}; whenever
two walls intersect, they can give birth to an additional wall.  This process is described in \S\ref{sec:constructing-w}.

\medskip

Here are some of the uses of the spectral networks $\wnet_\vartheta$:

\smallskip
{\bf Framed BPS states.} By letting the parameter $z$ of a surface defect vary along a path $\wp$ in $C$, one can define a supersymmetric interface $L_{\wp,\vartheta}$ \cite{Gaiotto:2011tf}.
Knowing $\wnet_\vartheta$ and $\mu$ allows one to determine the complete spectrum of BPS states of
these interfaces, called ``framed BPS states.''  In fact, the framed BPS spectrum turns out to be \ti{over}determined, so much so that one can use its consistency to compute $\mu$.

\smallskip
{\bf Jumps of $\wnet_\vartheta$ and the 4d BPS spectrum.} As the parameter $\vartheta$
is varied, $\wnet_\vartheta$ also varies.  There are some critical phases $\vartheta_c$
at which the topology of $\wnet_\vartheta$ suddenly changes (in a precise sense explained in
\S \ref{sec:varying-theta}).
These critical phases $\vartheta_c$
are the phases of the central charges $Z_\gamma$ of BPS states in the $d=4$ theory $S[ \fg,C,D]$.
Moreover, one can read off the degeneracies $\Omega(\gamma)$ of these BPS states from
the topology of $\wnet_{\vartheta = \vartheta_c}$.

We find in this way that BPS states in $S[ \fg,C,D]$ correspond to certain ``finite webs'' of
strings on $C$, which appear inside the spectral network $\wnet_\vartheta$ at the critical phase.
Some pictures of finite webs appear in Figure \ref{fig:finite-networks}.
The role of finite webs in the BPS spectrum was already expected
\cite{Klemm:1996bj,Brandhuber:1996xk,Aharony:1997bh,Gaiotto:2009hg}, but our analysis here
gives a much more precise understanding of how to compute the corresponding BPS degeneracies
$\Omega(\gamma)$ than was previously available.
It also provides a geometric argument (if not quite a proof) that the degeneracies so computed obey the
wall-crossing formula
of Kontsevich-Soibelman \cite{ks1}, as well as its extension to include the coupling between
2d and 4d BPS states, given in \cite{Gaiotto:2011tf}.

This story can be thought of as a broad generalization of
parts of \cite{Gaiotto:2009hg,Gaiotto:2011tf}, where we explained why the counts of finite webs
obey the wall-crossing formulas in the case $K=2$.
In fact, our approach here leads to a simpler and more conceptual understanding
even of that case.

The 4d BPS spectrum of $\N=2$ theories has been investigated by many authors recently; see in particular
\cite{Alim:2011ae,Alim2011a}.
One advantage of our approach here via spectral networks
is that, given a point $u$ of the Coulomb branch,
one can determine any particular BPS degeneracy $\Omega(\gamma)$ at $u$ just by drawing
the network $\wnet_{\vartheta = \arg Z_\gamma}$ corresponding to $u$.
The spectral network also gives more precise information ---
not only the 4d BPS degeneracies $\Omega(\gamma)$ but also some enhancements
$\omega(\gamma, \cdot)$,
which keep track of the interaction between the 4d BPS state and the surface defects
$\bS_z$.

\smallskip
{\bf Moduli of flat connections.}
Suppose we compactify on $S^1$ both the $d=4$ theory and the surface defects $\bS_z$.
The ground states of $\bS_z$ on $S^1$ form a $K$-dimensional vector space $E_z$;
letting $z$ vary we obtain
a rank $K$ vector bundle $E$ over $C$.
Now take a path $\wp$ in $C$ from $z_1$ to $z_2$.
The vacuum expectation value $\inprod{L_{\wp,\vartheta}}$ of the corresponding
interface wrapped on $S^1$ is an isomorphism from $E_{z_1}$ to $E_{z_2}$.
In other words, the bundle $E$ is equipped with a natural connection \cite{Gaiotto:2009fs},
which moreover is actually flat.\footnote{We are oversimplifying slightly here: the precise story,
laid out in the main text,
involves a slight twisting of the notion of flat connection.  This twisting
does not modify the basic picture and can safely be ignored at first.}

On the other hand, we could also take the perspective of the IR (abelian) theory.
From that point of view we would have surface defects
labeled by points on $\Sigma$ rather than on $C$, with $1$-dimensional spaces of vacua,
thus forming a line bundle $\CL$ over $\Sigma$.
Considering interfaces between these surface defects, we see that
$\CL$ too is naturally equipped with a flat connection.

So, to each vacuum of the compactification of the $d=4$ theory
on $S^1$, we have assigned on the one hand a flat rank $K$
connection over $C$, and on the other hand a flat rank $1$ connection over the covering $\Sigma$.
This induces a correspondence $\Psi_{\wnet_\vartheta}$
between the two types of connection, which turns out to be
(at least locally) an isomorphism of moduli spaces.  Moreover, we can compute
$\Psi_{\wnet_\vartheta}$ concretely:  the key is the framed BPS degeneracies mentioned above,
which (roughly) give the coefficients in the expansion of a UV interface $L_{\wp,\vartheta}$
in terms of IR interfaces.  Upon taking expectation values this becomes an expansion
of the vacuum expectation value $\inprod{L_{\wp,\vartheta}}$ in terms of ``Darboux coordinates''
as described e.g. in \S 5.8 of \cite{Gaiotto:2011tf}.
In particular, the correspondence $\Psi_{\wnet_\vartheta}$
is determined by the spectral network $\wnet_\vartheta$.

\smallskip
{\bf Cluster coordinate systems.}
Since the space of flat rank $1$ connections over $\Sigma$
is a product of copies of $\IC^\times$, another way to describe the last item
is to say that $\Psi_{\wnet_\vartheta}$ is a local \ti{coordinate system}
on the moduli space $\CM$ of flat rank $K$ connections over $C$.
The construction of $\Psi_{\wnet_\vartheta}$ from $\wnet_\vartheta$ can be
generalized:  it uses only some general properties of $\wnet_\vartheta$,
which we abstract into a definition of ``spectral network''.  Given any
spectral network $\wnet$ there is a corresponding $\Psi_\wnet$.

We conjecture that $\Psi_{\wnet}$ is a particularly nice coordinate system:
it is a \ti{cluster} coordinate system,
in the sense of \cite{MR2567745}.
Indeed, Fock and Goncharov proved \cite{MR2233852}
that the spaces $\CM$ we consider admit an atlas of cluster coordinate systems.\footnote{This
fact implies e.g. that these spaces admit a natural quantization \cite{MR2567745,qd-cluster}.}
For $K=2$ the cluster atlas
consists precisely of the coordinate systems $\Psi_\wnet$.
For $K>2$, Fock and Goncharov did not give a completely
explicit description of the atlas --- rather they described some particular cluster coordinate
systems.  The most general cluster coordinate system would be obtained by beginning with
one of these and performing some sequence of coordinate transformations
known as ``mutations''; each mutation generates a new coordinate system, and the
cluster atlas is the set of all coordinate systems obtained in this way.
We conjecture that our $\Psi_\wnet$ are coordinate systems in the
cluster atlas.\footnote{Again, here we are glossing over some slight differences between
our setup and that of Fock-Goncharov, and ignoring some extra discrete ``flag data'' attached to
the connections.}

How do the mutations arise in our story?
Small deformations of the spectral network $\wnet$ leave the coordinate
system $\Psi_{\wnet}$ invariant.  However,
there are some natural ``degenerations'' of spectral networks
which interpolate between ``inequivalent''  $\wnet$.  We have already seen these degenerations
in this introduction:  they occur in the families $\wnet_\vartheta$ when the parameter
$\vartheta$ is adjusted to a critical
phase $\vartheta_c$.  When such a degeneration occurs,
$\Psi_{\wnet}$ jumps by an automorphism $\CK$
of the torus of flat rank $1$ connections on $\Sigma$.  This $\CK$ is
simply determined by the degeneracies $\Omega(\gamma)$ read off from the
degenerate network (see \eqref{eq:holonomy-jump} for its explicit form.)
$\CK$ has a form very similar to that of a cluster mutation,
and part of our conjecture is that for the simplest types of degenerations
it is indeed a mutation.

Very recently Goncharov has also defined some new geometric objects which he calls \ti{spectral webs} \cite{goncharov},
and which correspond to coordinate systems in the cluster atlas.  Despite the similar nomenclature, this definition is
different from our definition of spectral networks.  It should be interesting to compare the two constructions.

\smallskip
{\bf WKB expansions and spectral networks as Stokes diagrams.} One of the key tools in \cite{Gaiotto:2009hg} was
the WKB approximation, applied to the $\zeta \to 0$ behavior of
families of flat $SL(2,\IC)$ connections
of the form
\begin{equation}
\nabla(\zeta) = R \frac{\varphi}{\zeta} + D +  R \zeta \bar \varphi.
\end{equation}
Such families are associated with solutions of Hitchin's equations \cite{MR89a:32021}
and naturally arise in theories of class $S$.

Some aspects of the WKB approximation are difficult to generalize
to rank $K>2$, and for a long time this proved a stumbling
block to giving nontrivial illustrations of the general statements
of \cite{Gaiotto:2008cd,Gaiotto:2011tf} in the higher
rank case.  As we briefly explain in \S \ref{subsec:WKB-Asymptotics}, the
spectral networks $\wnet_\vartheta$ in theories of class $S$ appear to be the key missing ingredient
to solve this problem:  they provide a way of constructing a basis of $\nabla(\zeta)$-flat sections in
each connected component of the complement of $\wnet_\vartheta$,
with ``good WKB asymptotics'' as $\zeta \to 0$ in a half-plane ${\mathbb H}_\vartheta$.
Thus, the walls of the spectral networks $\wnet_\vartheta$ also have an
interpretation as Stokes curves.  Similar statements have appeared before
in the mathematical literature, e.g. \cite{MR1209700,MR729194} and especially \cite{MR2132714}
which contains some examples of spectral networks with $K=3$.

\subsection*{Open problems}

Our work in this paper leaves many directions unexplored or incompletely explored.  Here are a few:

\begin{enumerate}

\item We discuss the theories $S[\fg, C, D]$ only for $\fg$ of type $A$.
There should be a closely
parallel story for $\fg$ of type $D$ or $E$.  Some parts of this story are easy to predict:
the walls of the spectral networks will be labeled (locally) by roots of $\fg$, and spectral networks will correspond
to coordinate systems on  moduli spaces of $\fg$-connections over $C$.  It would be desirable to work
out this picture in detail.

\item Even in the case of type $A$ our story is not quite complete.
We take the charges of BPS states to be valued in the lattice $\Gamma = H_1(\Sigma;\IZ)$,
but according to the more precise picture in \cite{Gaiotto:2009hg,Gaiotto:2010be},
it is better to identify the charge lattice as a certain subquotient of $H_1(\Sigma;\IZ)$.
This is related to the fact that our nonabelianization map $\Psi_{\wnet_\vartheta}$ most directly
produces connections with structure
group $GL(K,\IC)$, while the physically relevant objects should really be connections with
structure group $SL(K,\IC)$, considered modulo some further discrete equivalences \cite{Gaiotto:2010be}.
There are likely to be some topological subtleties here which deserve careful consideration.

\item Many of the results of \cite{Gaiotto:2009hg} for $K=2$ are subsumed in the present paper and extended to $K>2$.
However, there is one important piece of
\cite{Gaiotto:2009hg} which we have not extended to $K>2$.  Namely, we gave a recipe for computing an object called
the ``spectrum generator,'' which completely determines the whole 4d BPS spectrum of the theory at any point $u$.
This recipe uses only the combinatorics of a single spectral network $\wnet_\vartheta$, for a fixed generic $\vartheta$.
It would be desirable to extend it to the case $K>2$, and also to types $D$ and $E$.
One could further ask for an extension of the spectrum generator to include the 2d spectrum.\footnote{In
the case $K=2$ this has been worked out by Pietro Longhi.}

\item It would be desirable to generalize our discussion of BPS indices to include
information about the spins of the BPS states.  This means taking $y\not=\pm 1$
in the various indices defined below.  Some aspects of the ``motivic 2d-4d wall-crossing
formula'' which governs these degeneracies were spelled out in
\cite{Gaiotto:2011tf}, but important details remain to be filled in.

\item In \S \ref{subsec:PictureSN} we paint a heuristic picture of the ``space of all spectral networks,''
which would be nice to spell out more precisely.

\item We conjecture in \S \ref{subsec:Coord-System}
that the coordinate systems $\Psi_\wnet$ on the space $\CM$ of flat connections
are cluster coordinate systems.  It would be very interesting to prove this conjecture.
It would be even more interesting if it turned out that the $\Psi_\wnet$ actually exhaust the set of cluster coordinate systems.
This might be of some internal
use for the study of cluster varieties, e.g. for the positivity conjectures of \cite{MR2233852}.

\item In this paper we always require that the curve $C$ carries at least one defect $D_n$.
This restriction (which we also imposed in \cite{Gaiotto:2009hg})
leads to some simplifications.  One reason for these simplifications is
that the walls in $\wnet_\vartheta$ are solutions of differential equations, and it seems that
these equations generically imply that each wall is ``attracted'' asymptotically to some
defect.  In particular, any open path $\wp$ on $C$ meets $\wnet_\vartheta$ only finitely many times.

This contrasts sharply with
the case without defects:  in that case, a typical open path on $\wp$
may meet $\wnet_\vartheta$ infinitely many times.
We are hopeful that all of the constructions of this paper still make sense in that setting, but there will
be issues of convergence to consider.

Fortunately, some things can be said about the asymptotic behavior of the walls
even in the case without defects.
For example, in the case $K=2$ the walls in $\wnet_\vartheta$ are horizontal
trajectories of a holomorphic quadratic differential $e^{2 \I \vartheta} \phi_2$,
and the leading
asymptotic behavior of a generic such trajectory is classified by a virtual ``limit cycle'' in $H_1(C, \IR)$, with deviations
governed by certain Lyapunov exponents; see \cite{flat-surfaces} for a very useful review.  One can hope that using these kinds of results
(and their to-be-developed generalizations for $K>2$) it will be possible to extend everything in this paper to the
case without defects.

\item In this paper we mainly work at a generic point of the Coulomb branch, where the gauge
symmetry group in the IR is abelian.  If the parameters at the defects are adjusted carefully (or if there are no defects at all)
there may also be points where the unbroken gauge symmetry is nonabelian.
The physics of this situation is much less explored than the fully abelian case.
Nevertheless, much of what we have said should have an extension to this situation.

In particular, a spectral network $\wnet$ associated to a $K$-fold covering $\Sigma \to C$ should give not only
a map $\Psi_\wnet$ between moduli of flat $GL(1)$ connections on $\Sigma$ and moduli of
flat $GL(K)$ connections on $C$, but more generally a map $\Psi_{\wnet,N}$
between moduli of flat $GL(N)$ connections on $\Sigma$ and moduli of flat $GL(NK)$
connections on $C$.\footnote{One way of thinking about this is that the $K$-fold covering $\Sigma$
could arise as a non-reduced degeneration of an $NK$-fold covering, where the sheets coalesce in
groups of $N$.  Readers who prefer to think in terms of M-theory fivebranes might say that we
consider the theory of $NK$ fivebranes wrapped on $C \subset T^* C$,
and then move to a point of the Coulomb branch represented by $N$ fivebranes wrapped on $\Sigma \subset T^* C$.}
Our construction of $\Psi_\wnet$ in \S\ref{sec:moduli} is set up
in a way that should generalize directly to this setting.
We understand that similar constructions will appear in upcoming work of Goncharov and Kontsevich.

\item In this paper we encounter several tricky sign issues.  One of these
first pops up as an ambiguity in the notion of ``fermion number,'' which leads to an ambiguity in the sign of
the 2d BPS degeneracies, and recurs many times
thereafter.  We have found a scheme for fixing this sign ambiguity, which we use systematically
throughout the paper:  very roughly speaking, it amounts to
considering paths on $C$ and $\Sigma$ weighted by signs which keep track of the parity of
the number of times the tangent direction
to the path winds around the circle.  Our slavish implementation of this scheme
has various consequences,
leading us e.g. to consider \ti{twisted} flat connections on $C$ and $\Sigma$ in \S\ref{sec:moduli} rather than
ordinary flat connections.  (Twisted connections on $C$ also appeared in \cite{MR2233852}, and this was a useful clue which helped us to
find our sign prescription.)
While this scheme leads to a consistent picture both mathematically
and physically, we cannot say that we have really understood from physical first principles \ti{why} it works so well.  It would be very good to have a better understanding of how our sign rule arises from the physics of the six-dimensional theory $S[\fg]$.

\item In this paper we consider spectral networks $\wnet_\vartheta$ associated to a phase $\vartheta$ and a point $u$ of the Coulomb branch.  It is natural to ask whether every spectral network $\wnet$
(modulo the natural notion of equivalence described in \S\ref{subsubsec:EqivalentGSN}) arises as $\wnet_\vartheta$ for some $(u, \vartheta)$.  If the answer is ``no,'' can we classify those spectral networks which do occur?

\item It is natural to ask whether the framed BPS degeneracies can be categorified.
We have done some work along these lines with E. Witten, and we hope to return to it.

\item The networks $\wnet_\vartheta$
we study on $C$ are made up of ``$\CS$-walls,'' the loci where framed 2d-4d BPS state degeneracies jump.
They depend on a point $u$ of the Coulomb branch $\CB$, so really $\wnet_\vartheta = \wnet_\vartheta(u)$
although we usually do not write this dependence
explicitly.
On the other hand, very similar networks $\knet_\vartheta$ appear directly on $\CB$:
they are made up of ``$\CK$-walls,'' the loci where 4d framed BPS state degeneracies jump.
$\knet_\vartheta$ can be studied by methods very similar to the methods we employ here for $\wnet_\vartheta$.
In particular this gives another scheme for computing the pure 4d BPS state degeneracies $\Omega(\gamma)$,
parallel to what we do here for $\mu(a)$.  (This has also been pointed out by Kontsevich-Soibelman.)
Networks very similar to $\knet_\vartheta$ have appeared previously in the mirror symmetry literature,
e.g. \cite{MR2181810,MR2386535,Gross2007,Gross2009,Gross2011}.

The family of networks $\wnet_\vartheta(u)$ on $C$ and the single network $\knet_\vartheta$ on $\CB$
can be unified into a network $\knet\wnet_\vartheta$ on $\CB \times C$.  We believe that this
is really the most natural perspective, although we do not adopt it explicitly in this paper.

\item We give a recipe for determining the BPS degeneracies $\Omega(\gamma)$ from a degenerate
spectral network, and work out several examples, but not many.
It would be interesting to use spectral networks to study the spectrum of concrete theories, beyond
the few examples we consider in \S\ref{sec:Examples}.
One obvious possibility would be to consider the standard $SU(K)$ gauge theories coupled to
fundamental hypermultiplets.  Even the BPS spectrum of the pure $SU(3)$ theory has been the source of
some controversy; in this paper we study it at strong coupling but do not analyze the more intricate
weak-coupling spectrum.  It would be interesting to compare results obtained from
spectral networks with those in \cite{Fraser:1996pw,Taylor:2001hg,Taylor:2002sg,Chen:2011gk}.

\item In the degenerate spectral networks we do examine we do not
work out any example where the resulting $\abs{\Omega(\gamma)} > 2$, which would suggest
the possibility of higher spin states in the BPS spectrum.  It would be interesting to find
a degenerate spectral network which gives such higher spin states.

\item In this paper we have made some
significant progress in the determination of BPS spectra for a large class of $\N=2$
field theories. We hope that these techniques can shed light on some broader
questions of general interest, such as to what extent the BPS spectrum uniquely
determines a theory, and whether one can always engineer a theory to produce
a desired (consistent) BPS spectrum.

\end{enumerate}

Finally, we mention that in a companion paper \cite{gmn6-to-appear} we
apply the techniques of this paper to make a more direct connection to the
work of Fock and Goncharov on higher Teichm\"uller theory \cite{MR2233852}.
We also find a formula for the spectrum generator for a special class of spectral networks,
valid for all $K \geq 2$; this gives an implicit determination of the BPS spectrum
in the corresponding theories.

\section{A brief review of theories of class S} \label{sec:review-S}

In \cite{Witten:1997sc, Gaiotto:2009we, Gaiotto:2009hg} a large and interesting
class of $\CN=2$, $d=4$ theories was studied.
These theories, which we call ``theories of class S,''
can be constructed by compactification and twisting of
$\CN=(2,0)$, $d=6$ theories.
Here we quickly review the basic features of this construction.

Fix a compact Riemann surface $C$, with $s$ punctures at points $\fs_1, \dots, \fs_s$.  Also fix a Lie algebra
$\fg$ of $ADE$ type.  In this paper we will focus on the case $\fg = A_{K-1}$.
We consider the $\CN=(2,0)$ theory $S[\fg]$ compactified on $C$, and partially twisted as described in \cite{Gaiotto:2009hg}.  $S[\fg]$ admits half-BPS codimension-2 defects; we put one
such defect $D_n$ at each puncture $\fs_n$.

The main statements of this paper will be independent of the choice of which type of defect
we put at each puncture, but it is useful to have an example in mind to fix ideas.
There is a class of \ti{regular} defects labeled by Young diagrams with
$K$ boxes \cite{Gaiotto:2009we, Gaiotto:2009hg}.
In particular we can consider a \ti{full} regular defect,
corresponding to the Young diagram with a single row.
Each full defect admits a natural mass deformation depending
on $K$ complex parameters $m_n^{(i)}$, with
$\sum_{i=1}^K m_n^{(i)} = 0$, which we are free to fix arbitrarily.

In any case, whatever collection $D$ of defects we choose, the compactification procedure yields an $\CN=2$, $d=4$ theory $S[\fg, C, D]$.  This theory is our main object of study.

The Coulomb branch of $S[\fg, C, D]$ consists of tuples $(\phi_2, \dots, \phi_K)$, where $\phi_r$ is an $r$-differential
on $C$ (i.e. a section of $K_C^{\otimes r}$), which is holomorphic away from the punctures $\fs_n$, and has some prescribed singular behavior
at the $\fs_n$.  For example, if we choose $D_n$ to be a full defect,
then the prescription is that
for each $r$, $\phi_r$ has a pole of order $r$ at $\fs_n$, with residue determined by a combination of the parameters $m_n^{(i)}$ \cite{Gaiotto:2009we, Gaiotto:2009hg}.
Having fixed $(\phi_2, \dots, \phi_K)$, the Seiberg-Witten curve is given by
\begin{equation} \label{eq:sw-curve}
 \Sigma = \{ \lambda: \lambda^K + \sum_{r=2}^K \phi_r \lambda^{K-r} = 0 \} \subset T^* C.
\end{equation}
$\Sigma$ is a $K$-fold branched cover of $C$.
Since $\Sigma$ sits inside $T^* C$ it carries a canonical 1-form,
the restriction of the Liouville 1-form, which by slight abuse of notation
we will also call $\lambda$.

In this paper it will be crucial to introduce a half-BPS \ti{surface defect} into
$S[\fg, C, D]$.  There is a canonical such defect $\bS_z$ \cite{Alday:2009fs,Gaiotto:2009fs,Gaiotto:2011tf}, depending only on a point $z \in C$
(and on a representation of $\fg$, but since we choose $\fg = A_{K-1}$ we can just
take the fundamental representation.)  When $z$ is generic, the defect $\bS_z$
has $K$ distinct massive vacua, which correspond to the $K$ solutions of \eqref{eq:sw-curve} at $z$;
locally we may denote these by $z^{(1)}, \cdots, z^{(K)}$.
From the point of view of the
$d=6$ theory $S[\fg]$, $\bS_z$ is obtained by inserting a surface defect
at the point $z$.

\section{Geometric description of BPS states}

\subsection{4d BPS states}\label{subsec:4d-BPS-States}

Now we recall the geometric description of BPS states in $S[\fg, C, D]$.

All such BPS states arise from BPS strings of the six-dimensional $S[\fg]$, which are extended
along $C$ and hence look like point particles in the remaining 3+1 dimensions.
Roughly there are $\binom{K}{2}$ distinct
kinds of BPS string, but the topological twisting and compactification makes the story slightly
trickier, as we now recall.

An oriented
segment of string passing through a point $z \in C$
is labeled by the choice of a pair of distinct sheets of the $K$-fold covering
$\Sigma \to C$, i.e. two solutions
$\lambda^{(i)}$, $\lambda^{(j)}$ ($i \neq j$) of the degree-$K$ polynomial equation \eqref{eq:sw-curve},
in a neighborhood of $z$.  To keep track of this discrete label, we use the term
``$ij$-string'' rather than just ``string.''
Reversing orientation exchanges $ij$-strings and $ji$-strings.

The $\N=2$ central charge $Z$ of an $ij$-string is obtained by integrating the complex 1-form $\frac{1}{\pi}(\lambda^{(i)} - \lambda^{(j)})$ along the string.
The mass $M$ of an $ij$-string is obtained by integrating the real density $\frac{1}{\pi} \abs{\lambda^{(i)} - \lambda^{(j)}}$.

When is such a string BPS?
Introduce a local coordinate on $C$ by $w^{(ij)} = \int \lambda^{(i)} - \lambda^{(j)}$, and
then define an \ti{$ij$-trajectory with phase $\vartheta$} to be a straight line in the
$w^{(ij)}$-coordinate, with inclination $\vartheta$, i.e. a line along which
\begin{equation} \label{eq:trajectory-de}
\Im(e^{- \I \vartheta} \dot{w}^{(ij)}) = 0.
\end{equation}
An $ij$-trajectory is naturally oriented:
the positive direction is the
direction in which $\Re(e^{-\I \vartheta} w^{(ij)})$ increases.
Reversing orientation of an $ij$-trajectory gives a $ji$-trajectory.
An $ij$-string is BPS if and only if it is stretched along an $ij$-trajectory
with some phase $\vartheta$; we call such a string a ``BPS string of phase $\vartheta$.''
We see at once that for BPS strings $\abs{Z} = M$ as expected.
The phase $\vartheta$ determines which supercharges the BPS string preserves.

An $ij$-string can end in two ways.  First, it can end on an \ti{$(ij)$-branch point}, i.e.
a point where $\lambda^{(i)} - \lambda^{(j)} = 0$.\footnote{We use the notation $(ij)$ for the branch
points but $ij$ for the trajectories; the $(ij)$ is meant to denote the \ti{transposition} associated
to the branch point, which exchanges sheet $i$ and sheet $j$ of the covering $\Sigma \to C$.
In particular $(ij) = (ji)$.}
Second, it can end on a \ti{junction}
where an $ij$-string, $jk$-string and $ki$-string meet (all oriented into the junction.)
In the latter case, for the combined web of strings to be BPS, all three strings must be BPS
strings with the \ti{same} phase $\vartheta$.

BPS states in $S[\fg, C, D]$ arise from webs of BPS strings, such that
all strings in the web have finite total central charge.
This condition means that the
strings are either closed loops or have both ends on branch points or junctions.
We call these \ti{finite webs}.
Some possible topologies for finite webs
are shown in Figure \ref{fig:finite-networks}.
\insfigscaled{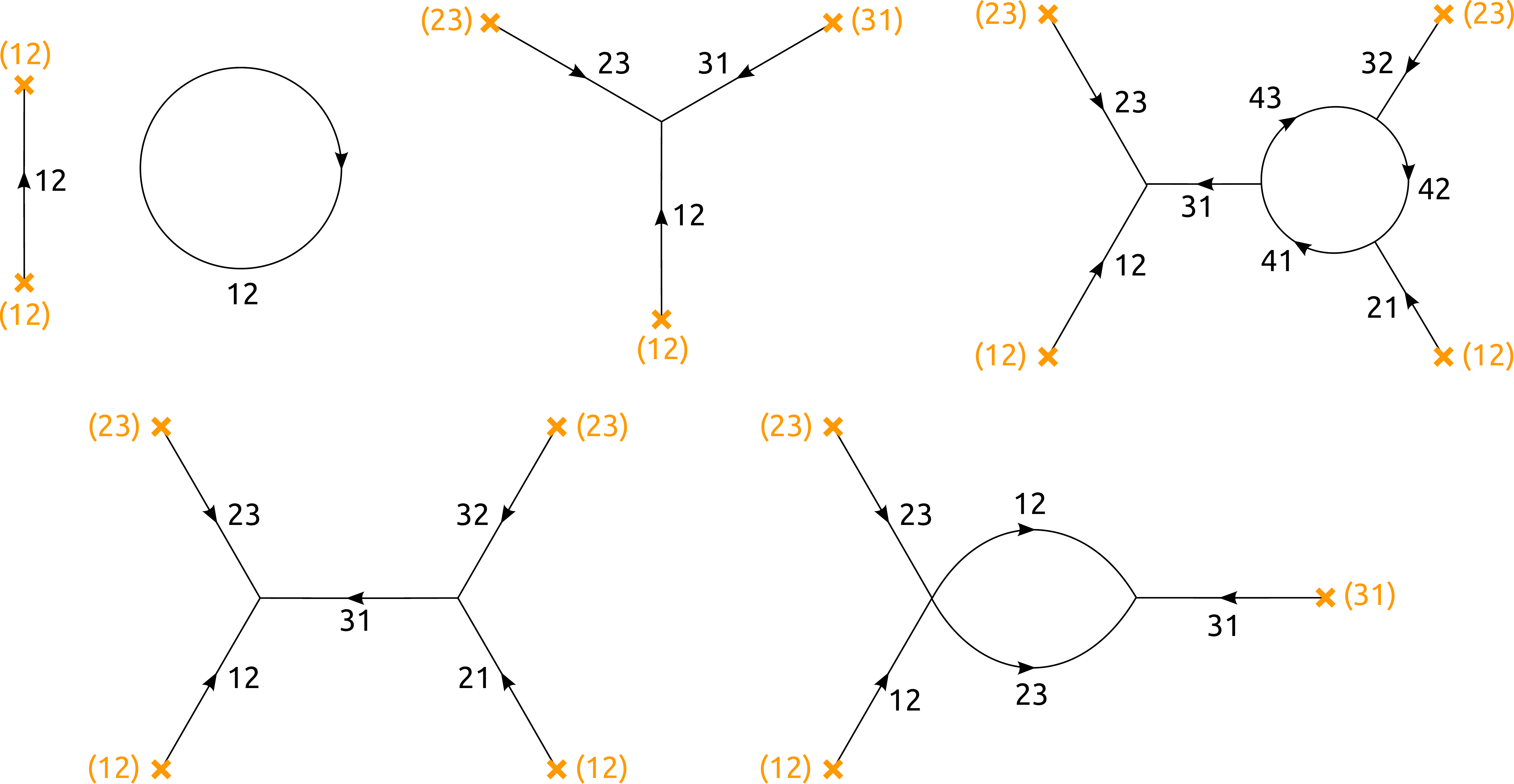}{0.24}{Some possible topologies for finite webs of BPS strings.
An orange cross with label $(ij)$ denotes an $(ij)$-branch point.  Wherever
a string with label $ij$ appears, it could equally well have been represented by
a string with label $ji$ and the opposite orientation.}

The charges of the BPS states are determined by the topology of the webs, in the following way.
An $ij$-string stretched along an oriented path $p$ on $C$
can be lifted in a canonical way to $p_\Sigma$, a union of oriented curves on $\Sigma$:
namely, $p_\Sigma$ is the union of the lift $p^{(i)}$ of
$p$ to the $i$-th sheet and the lift $-p^{(j)}$ of $-p$ ($p$ with reversed orientation)
to the $j$-th sheet.  Letting $p$ run over the strings in a finite web $N$, the union of the $p_\Sigma$ is a closed 1-cycle $N_\Sigma$ on $\Sigma$.
$N_\Sigma$ has a homology class $[N_\Sigma] \in \Gamma := H_1(\Sigma; \Z)$.  This $[N_\Sigma]$
is the charge of the BPS state.\footnote{The precise charge lattice of
the theory $S[\fg, C, D]$ is actually a subquotient of
$H_1(\Sigma; \Z)$, as explained in \cite{Gaiotto:2009hg}. Nevertheless, in this paper,
we consider $\Gamma = H_1(\Sigma; \Z)$ for simplicity.}

The central charge of a finite web $N$ is the sum of the central charges of the
strings in $N$; this gives a simple result which depends only on $\gamma = [N_\Sigma]$,
\begin{equation} \label{eq:Zgamma}
 Z_\gamma = \frac{1}{\pi} \oint_\gamma \lambda.
\end{equation}

To determine the BPS spectrum of the theory, in particular the second helicity supertrace
$\Omega(\gamma)$, one should in principle proceed by quantizing
the zero modes of each finite web.  This would give some definite formula
for the contribution of each finite web to $\Omega(\gamma)$.
In practice, such a quantization has not been completely carried out, and we will not do it here
either.\footnote{In \cite{Gaiotto:2009hg} we used wall-crossing to determine $\Omega(\gamma)$ in the case $K=2$,
reproducing earlier results of \cite{Klemm:1996bj}.
That case is particularly simple since there are only two possible topologies for finite
webs:  one can either have a single string connecting two branch points or a single closed loop.
These topologies contribute $\Omega(\gamma) = +1$ and $\Omega(\gamma) = -2$ respectively.
We will reproduce this result yet again in \S\ref{sec:k2-examples} below.}
Rather, we will explain a more indirect route, which determines $\Omega(\gamma)$ using the interactions between 2d and 4d BPS states.

One could also ask about extending $\Omega(\gamma)$ to an object which keeps track of the spins
of the BPS multiplets appearing.  Such ``refined BPS degeneracies'' and ``protected
spin characters'' have been considered in the $\N=2$ context
e.g. in \cite{ks1,Dimofte:2009tm,Cecotti:2009uf,Cecotti:2010qn,Gaiotto:2010be}.  We will not consider
that extension in this paper.

\subsection{Solitons} \label{sec:2d4d-bps-states}

In the presence of the canonical surface defect $\bS_z$, there is a second kind of BPS state in the story:
we can consider BPS particles which are bound to the defect and interpolate between distinct vacua.
We call these particles \ti{solitons}.

As described in \cite{Gaiotto:2011tf}, BPS solitons
are also realized geometrically in terms of finite webs of strings.  The main difference is that now
we consider webs in which one of the strings ends on the point $z$, which we call
\ti{finite open webs}.  See Figure \ref{fig:finite-networks-2d}.
\insfigscaled{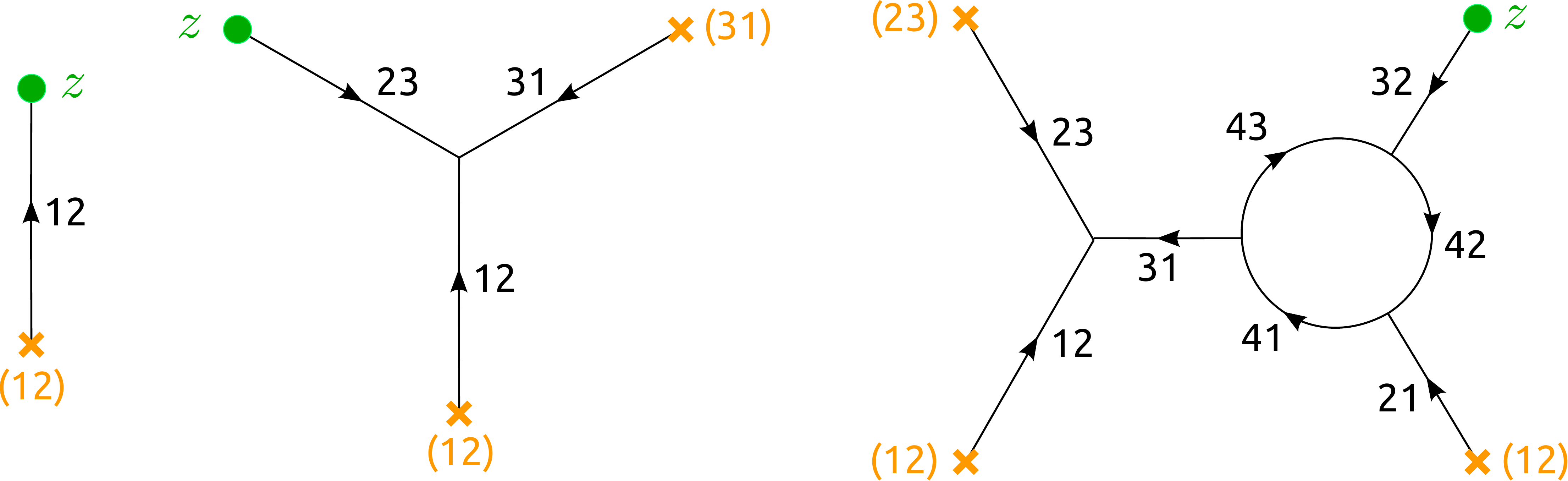}{0.24}{Some possible topologies for finite open webs of BPS strings, representing BPS solitons
on the surface defect $\bS_z$.
Each finite open web includes one string that ends on the point $z \in C$.
An orange cross with label $(ij)$ denotes an $(ij)$-branch point.}

Let $N$ denote a finite open web.
As for the pure 4d case, the charge of the corresponding BPS soliton is
determined by the topology of the web, as follows.
Let $N_\Sigma$ denote the union of the lifts of all strings in $N$ to $\Sigma$.
Suppose the string of $N$ ending on $z$ is an $ij$-string oriented out of $z$.
Let $z^{(i)}$ and $z^{(j)}$ be the preimages of $z$
on the $i$-th and $j$-th sheets of $\Sigma$.
$N_\Sigma$ is a 1-chain with boundary:
\begin{equation} \label{eq:chain-boundary}
\partial N_\Sigma = z^{(j)} - z^{(i)}.
\end{equation}
We let $\Gamma_{ij}(z,z)$ denote the set of relative homology
classes on $\Sigma$ obeying \eqref{eq:chain-boundary}, and
\begin{equation}
\Gamma(z,z) = \cup_{i,j} \Gamma_{ij}(z,z).
\end{equation}
Thus the charges $[N_\Sigma]$ of BPS solitons on $\bS_z$ are elements of $\Gamma(z,z)$,
and in fact of the smaller set $\cup_{i,j,i \neq j} \Gamma_{ij}(z,z)$.

The central charges of BPS solitons are given by a formula analogous to \eqref{eq:Zgamma}:
for any $\bar a \in \Gamma(z,z)$ we have\footnote{In \cite{Gaiotto:2011tf} we denoted
elements of $\Gamma_{ij}(z,z)$ by $\gamma_{ij}$, but we are now deprecating that notation in favor of $\bar a$.
See Appendix \ref{app:Conv-conv} for a summary of conventions.}
\begin{equation}
 Z_{\bar a} = \frac{1}{\pi} \int_{\bar a} \lambda.
\end{equation}
There is a BPS index $\mu(\bar a) \in \Z$ which counts 2d-4d BPS states
of charge $\bar a$, defined as a trace over the Hilbert space of 1-particle
BPS states \cite{Cecotti:1992qh,Cecotti:1993rm},
\begin{equation} \label{eq:def-mu}
 \mu(\bar a) = {\rm Tr}_{\CH^{1,\rm BPS}_{\bS_z,\bar a}}\,F e^{\I \pi F},
\end{equation}
where $F$ is a fermion number operator.
In principle, $\mu(\bar a)$ could be determined by quantizing the zero modes
of the BPS strings.  In practice, as with $\Omega(\gamma)$, we will take a more
indirect route in this paper.

We hasten to warn the reader that there are two important subtleties
in the definition \eqref{eq:def-mu}.  The first subtlety is
addressed in \S\ref{sec:tricky-sign}:
to define $F$ properly we will need to keep track of slightly more information about
the 2d-4d BPS states, by extending the charge $\bar{a}$ to a new ``charge'' denoted $a$.
It will turn out that $\mu$ really depends on $a$, not $\bar a$.  The second subtlety is that the
space of 1-particle BPS states is only well defined when the parameters are not on
a wall of marginal stability.  At these walls, $\mu(a)$ can jump. In particular,
if $\bar a \in \Gamma(z,z)$, then generically, if we move $z$ holding all other parameters
fixed, and apply the natural parallel transport to the
relative homology cycle $\bar a$, $\mu(a)$ remains constant.  However, there are
walls on $C$ across which $\mu(a)$ jumps, and in this case we will need to
define limits $\mu^\pm(a)$ as $z$ approaches
the wall from either side. This will be important in \S \ref{sec:varying-theta}.

\subsection{Framed 2d-4d BPS states} \label{sec:framed-bps}

Our approach to the BPS spectrum will involve an auxiliary device, the
\ti{framed} BPS states.  These were introduced in the pure 4d context in
\cite{Gaiotto:2010be}, and in the 2d-4d context in \cite{Gaiotto:2011tf}.
Our interest in this paper is in the 2d-4d version.

Consider a pair of points $z_1$, $z_2$ in $C$, and a path
$\wp$ in $C$ from $z_1$ to $z_2$.  Also fix a parameter $\vartheta \in \IR$.
These data determine a pair of surface defects
$\bS_{z_1}$ and $\bS_{z_2}$ in $S[\fg, C, D]$ along with a supersymmetric
interface $L_{\wp,\vartheta}$ between the two surface defects.
The interface $L_{\wp,\vartheta}$ preserves 2 out of the 4 supercharges preserved by
the surface defects; which 2 supercharges are preserved is determined by the parameter
$e^{\I \vartheta}$, as explained
in \cite{Gaiotto:2011tf}.\footnote{The parameter which was called $\zeta$ in \cite{Gaiotto:2011tf}
is here given by $\zeta = e^{\I \vartheta}$.}
It is generally believed that the defect
$L_{\wp,\vartheta}$ does not depend on the precise path $\wp$ but only on its homotopy class.
In the
present paper we will take this as an assumption, and will find a very consistent picture
(although of course the quantities we study are somewhat protected by supersymmetry);
see e.g. \cite{Gaiotto2011} for some related discussion.
Indeed, we will find that the constraint of homotopy invariance is very strong.

Now we study the 4d theory $S[\fg, C, D]$ with the defects $\bS_{z_1}$ and $\bS_{z_2}$
inserted on two half-lines $\{ x^1 = 0, x^2 = 0, x^3 > 0 \}$
and $\{ x^1 = 0, x^2 = 0, x^3 < 0 \}$ respectively,
separated by the interface $L_{\wp,\vartheta}$.  See Figure \ref{fig:two-surface-operators}.
\insfig{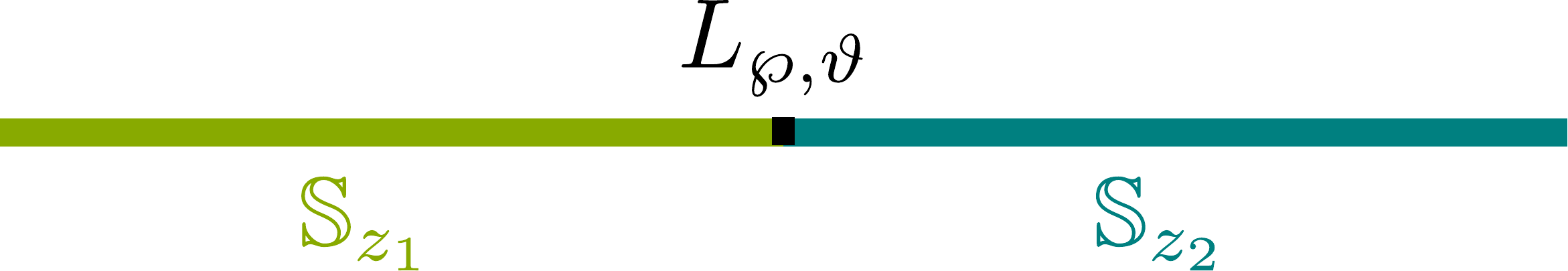}{Two surface defects connected by an interface.  (This picture
lives in the three-dimensional space where the field theory $S[\fg,C,D]$ is defined; we have factored
out the time direction.)}
We then define
(2d-4d) \ti{framed BPS states} to be states of the 1-particle Hilbert space $\CH^1_{L_{\wp,\vartheta}}$
of this combined system
which are fully supersymmetric, i.e. preserve the 2 supercharges present in the system.

The 2d-4d framed BPS states should be thought of as different ``vacuum states'' of the interface
$L_{\wp, \vartheta}$.  To make this statement sharp, however, we need to impose a constraint.
We say $\vartheta$ is \ti{generic} if there is no charge $\gamma \in \Gamma$ with
$e^{-\I \vartheta} Z_\gamma \in \R_-$.
We say the pair $(\wp, \vartheta)$ is \ti{generic} if
$\vartheta$ is generic and also there is no charge
$\bar a \in \Gamma(z_1,z_1)$ or $\bar a \in \Gamma(z_2,z_2)$ with
$e^{-\I \vartheta} Z_{\bar a} \in \R_-$, where $z_1,z_2$ are the
initial and final points of $\wp$.
When the pair $(\wp, \vartheta)$ is generic, the 2d-4d framed BPS states
are indeed localized near the interface, in the sense that they do not mix with the continua
of unframed 2d-4d or 4d BPS states.

The classification of 2d-4d framed BPS states by charges is similar to that
for the 2d-4d unframed BPS states in \S\ref{sec:2d4d-bps-states}.
Instead of
paths on $\Sigma$ between two lifts $z^{(i)}$, $z^{(j)}$ of a single point $z$, now we consider
paths on $\Sigma$ from a lift $z^{(i)}_1$ of $z_1$ to a lift $z^{(j)}_2$ of $z_2$.
We let $\Gamma_{ij}(z_1, z_2)$ denote the set of relative homology classes represented by such paths,
and $\Gamma(z_1, z_2) = \cup_{i,j} \Gamma_{ij}(z_1,z_2)$.\footnote{In \cite{Gaiotto:2011tf} we denoted
elements of $\Gamma(z_1,z_2)$ by $\gamma_{ij'}$, but we are now deprecating that
notation in favor of $\bar a$. See Appendix \ref{app:Conv-conv} for a summary of conventions.}

There is a ``framed BPS index'' $\fro(L_{\wp,\vartheta}, \bar a;y)$
counting 2d-4d framed BPS states with charge $\bar a \in \Gamma_{ij}(z_1, z_2)$.
It was defined in
\S 4.4 of \cite{Gaiotto:2011tf}:\footnote{As in \cite{Gaiotto:2011tf}, $F$ denotes a generator of $u(1)_V$ of the 2d $(2,2)$ supersymmetry algebra
preserved by $L_{\wp,\vartheta}$, and $\CJ = 2 J_{12} + 2 I_{12}$
is a linear combination of rotation and $su(2)_R$ generators of the 4d $\N=2$
supersymmetry.  There is a subtlety here
which we address in \S\ref{sec:tricky-sign}:
to define $F$ properly we will need to keep track of slightly more information about the 2d-4d framed
BPS states, by extending the charge $\bar{a}$
to a new ``charge'' denoted $a$.  It will turn out that $\fro$ really depends on $a$, not $\bar a$.}
\begin{equation}\label{eq:2d-4d-motindx}
\fro(L_{\wp,\vartheta}, \bar a;y) = {\rm Tr}_{\CH^{1,\rm BPS}_{L_{\wp,\vartheta}},\bar a} \, e^{\I \pi F} (-y)^{\CJ}.
\end{equation}
In this paper we will concentrate on the indices at $y=1$, which will be denoted
by $\fro'(L_{\wp,\vartheta}, \bar a)$.  We leave the generalization of our story to arbitrary $y$
as an important open problem.

Framed 2d-4d BPS states admit a geometric description somewhat similar to those given above
for unframed 2d-4d BPS states.  We will not describe it
explicitly here.\footnote{In the case $K=2$ (i.e. $\fg = A_1$) we did explain the relevant objects
in \cite{Gaiotto:2010be}, where we called them ``millipedes.''}

\subsection{Enhanced degeneracies} \label{sec:enh-4d}

In the presence of a surface defect $\bS_z$ there is an important enhancement to the 4d BPS degeneracies $\Omega(\gamma)$ \cite{Gaiotto:2011tf}:  they are replaced by numbers
$\omega(\gamma, \bar a) \in \Z$ for any $\gamma \in \Gamma$
and $\bar a \in \Gamma(z_1,z_2)$.  $\omega$ is ``linear'' in its second argument, i.e. it obeys
\begin{equation}
 \omega(\gamma, \bar a + \bar b) = \omega(\gamma, \bar a) + \omega(\gamma, \bar b)
\end{equation}
when $\bar a+\bar b$ is defined (that is, when the end of $\bar a$ coincides with the
start of $\bar b$).  Moreover, $\omega$ obeys
\begin{equation} \label{eq:Omega-def}
\omega(\gamma, \bar a + \gamma') - \omega(\gamma, \bar a) = \Omega(\gamma) \inprod{\gamma, \gamma'}
\end{equation}
for $\gamma' \in \Gamma$.
In particular this equation is sufficient to determine $\Omega$ if we know $\omega$.
(More precisely, it determines $\Omega(\gamma)$ for all $\gamma$ not in the kernel
of $\inprod{\cdot, \cdot}$, i.e. all $\gamma$ which are not pure flavor charges.)

The $\omega(\gamma, \bar a)$ are a bit subtle to interpret directly in terms of
traces over Hilbert spaces.  In \cite{Gaiotto:2011tf} the most general interpretation we found
was in terms of a Hilbert space of ``halo states'' (analogues of the ones studied by Frederik
Denef\footnote{See \cite{Denef:2007vg,Gaiotto:2010be,Andriyash:2010yf,Andriyash:2010qv}
for a description of the halo approach to wall-crossing.}), which induce 2d-4d wallcrossing.
The individual states in this Hilbert space may be interpreted either as 4d particles carrying
charge $\gamma$ or as 2d particles living on the surface defect and
carrying the same charge.
This fact will become relevant in one of our concrete examples,
in \S\ref{sec:nine-hypers}.

\subsection{A problem of signs} \label{sec:tricky-sign}

We must now confront a pesky but important detail.

There is an ambiguity in \eqref{eq:2d-4d-motindx}:  the generators $F$ and $\CJ$
are well defined only up to $c$-number shifts.  As a result, the index $\fro'(L_{\wp,\vartheta}, \bar a)$
suffers from some potential ambiguity (which even depends on $\wp$).  We can partially fix this ambiguity by requiring that
$\fro'(L_{\wp,\vartheta}, \bar a)$ is real, but this still leaves the possibility of an integer shift
of $F$ or $\CJ$.  Such a shift would reverse the sign of $\fro'(L_{\wp,\vartheta}, \bar a)$.
So, \ti{a priori}, we would expect that we need some additional data in order
to fix $\fro'(L_{\wp,\vartheta}, \bar a)$ uniquely.
A similar ambiguity afflicts the 2d-4d BPS degeneracies $\mu(\bar a)$:  the definition \eqref{eq:def-mu}
depends on the choice of generator $F$, and changing this choice can change $\mu(\bar a)$ by a sign.

There is no difficulty in fixing these ambiguities locally in any particular corner of parameter space.
In this paper, though, we will mainly be concerned with phenomena which occur when parameters are
varied, sometimes over long distances in parameter space.  It is therefore
desirable to have a global way of fixing these sign ambiguities, which is consistent with all of the
physical constraints, and ideally one that does not depend on any arbitrary choices.
We have found such a rule, which we now describe.  It would be desirable to give a first-principles
\ti{derivation} of this rule from the physics of the $(2,0)$ theory $S[\fg]$ in six dimensions.

\insfig{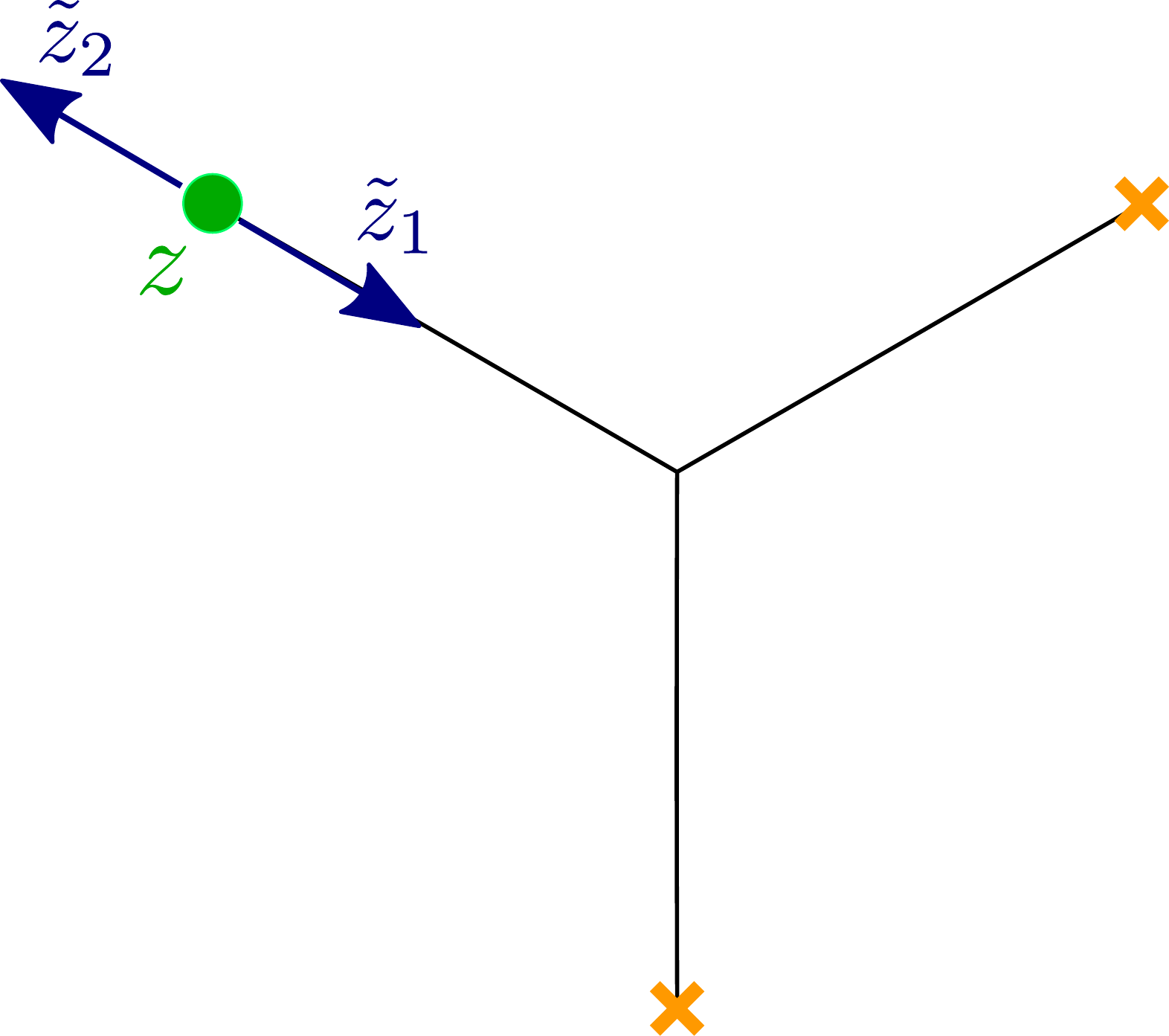}{Two tangent directions at $z$ determined by a finite web $N$:  $\tz_1$
points ``into'' the finite web while $\tz_2$ points ``away.''}

It will be useful to keep track of
a bit more information about the finite webs $N$ representing 2d-4d BPS states.
For any real surface $S$, let $\tilde S$ denote the circle
bundle of tangent directions to $S$.
For $z\in S$ we will let $\tilde z$ denote a lift
of $z$ to $\tilde S$, that is, a choice of tangent direction at $z$.  Any smooth path $\wp$
on $S$ carries a natural tangent direction field and hence has a canonical lift to a path
$\tilde\wp$ on $\tilde S$; we will use this lift often.
There is a distinguished class $H \in H_1(\tilde\Sigma; \Z)$, represented by a path which winds
once around a fiber of $\tilde\Sigma$ (the choice of fiber does not matter).

By smoothing out the junctions slightly,
we can deform $N_\Sigma$ into a finite union of \ti{smooth} paths on $\Sigma$,
which thus has a canonical lift to a path $N_{\tilde\Sigma}$ on $\tilde\Sigma$.
Let $\tz_1$ and $\tz_2$ denote the tangent directions to $C$ at $z$ shown in
Figure \ref{fig:tangent-vectors}.  Let $\tz_1^{(i)}$ be the lift of $\tz_1$ to the $i$-th sheet of $\tilde\Sigma$,
and similarly define $\tz_2^{(j)}$.  The path $N_{\tilde\Sigma}$ then has
\begin{equation} \label{eq:lifted-boundary}
 \partial N_{\tilde\Sigma} = \tz_2^{(j)} - \tz_1^{(i)}.
\end{equation}
Let $\tilde\Gamma(\tz_1,\tz_2)$ denote the set of relative homology classes on $\tilde\Sigma$ obeying \eqref{eq:lifted-boundary}, \ti{modulo} shifts by the class $2H$.
$\tilde\Gamma(\tz_1, \tz_2)$ is a principal $\Z/2\Z$-bundle over $\Gamma(z_1, z_2)$,
with the $\Z/2\Z$ action given by adding $H$.
The relative homology class $[N_{\tilde\Sigma}] \in \tilde\Gamma(\tz_1,\tz_2)$ thus keeps track of
one extra $\Z / 2\Z$ worth of information beyond that in $[N_\Sigma] \in \Gamma(z,z)$.
There can be different open finite BPS webs $N$ carrying the same charge $[N_\Sigma]$ but
with different lifts $[N_{\tilde\Sigma}]$; see Figure \ref{fig:distinct-lifts} for an example.
\insfig{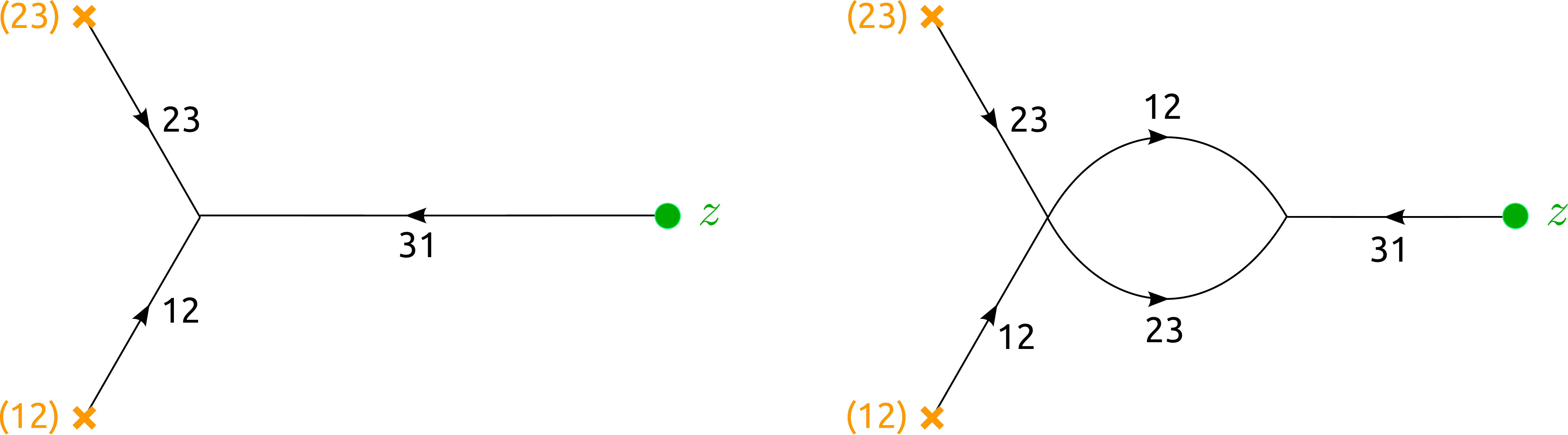}{Two open finite BPS webs $N$ on the same patch of $C$, drawn displaced
from one another for clarity.  These two webs
carry the same charge $[N_\Sigma]$ but
have different lifts $[N_{\tilde\Sigma}]$. Informally, one can say that the lifts of these two
webs ``differ by one unit of winding.''}

Now we can explain our proposal for how the sign of $\mu$ behaves.  Given a charge $\bar a \in \Gamma(z,z)$,
$\mu(\bar{a})$ is not well defined (although it is well defined up to sign).
In order to make it well defined, we propose that we must choose a class
$a \in \tGamma(\tz_1,\tz_2)$ which projects to $\bar a$.  Having done so,
there should be a way of fixing the ambiguity of $F$ to obtain a well defined
BPS degeneracy, which we call $\mu(a)$.
Two choices $a$, $a'$ differ by a winding number $w(a,a')$;
the corresponding $\mu$ should obey
\begin{equation} \label{eq:mu-sign-rule}
 \mu(a) / \mu(a')  = (-1)^{w(a,a')}.
\end{equation}
(Of course, knowing $\abs{\mu(a)}$ for all $a$ and knowing \eqref{eq:mu-sign-rule} is still not enough by itself to determine $\mu$.
Later in this paper we will fix one more convention, in \eqref{eq:S-factor},  after which we
will be able to calculate $\mu(a)$ for all $a$.)

A similar discussion applies to $\fro'(L_{\wp,\vartheta}, \bar a)$, but with the extra complication
that now we have to discuss both the dependence on the path $\wp$ and on the charge $\bar a$.
First suppose we \ti{fix} the path $\wp$ from $z_1$ to $z_2$, with initial tangent vector $\tz_1$
and final tangent vector $\tz_2$.  Then we propose that the situation is strictly parallel
to our discussion of $\mu$ above:
the degeneracies $\fro'(L_{\wp,\vartheta}, \bar a)$ are not well defined (although they are well
defined up to sign), and
what is really defined is an integer $\fro'(L_{\wp,\vartheta}, a)$, where $a \in \tGamma(\tz_1,\tz_2)$
is a lift of $\bar a \in \Gamma(z_1,z_2)$.  The dependence on the choice of lift is given by
\begin{equation} \label{eq:winding-rule-1}
\fro'(L_{\wp,\vartheta}, a) / \fro'(L_{\wp,\vartheta}, a') = (-1)^{w(a,a')}.
\end{equation}

Now let us also consider the dependence on $\wp$.
Recall from \S \ref{sec:framed-bps} that the line defect $L_{\wp, \vartheta}$ depends only on the
homotopy class of the path $\wp$. Nevertheless,
 the prescription for fixing the sign of $\fro'(L_{\wp,\vartheta},a)$ can depend on
more information than just the homotopy class.  We propose that the sign actually depends on
the homotopy class of the lift of $\wp$ from $C$ to the bundle of tangent directions $\tilde{C}$.
For two paths $\wp$, $\wp'$ whose lifts to $\tilde{C}$
have the same initial and final endpoints, there is a mod-2
winding number $w(\wp,\wp')$ (defined similarly to the winding number we considered above for paths on $\Sigma$).
We propose that for such paths we have
\begin{equation}
\fro'(L_{\wp,\vartheta},a) / \fro'(L_{\wp',\vartheta},a) = (-1)^{w(\wp,\wp')}.
\end{equation}
We summarize this proposal by saying that the framed 2d-4d BPS degeneracies of the
interface $L_{\wp,\vartheta}$ are not quite homotopy invariants of $\wp$, but rather are
``twisted homotopy invariants'' of $\wp$.

\section{Basics of framed 2d-4d indices} \label{sec:F-basics}

In this section we study the basic properties of the indices
$\fro'(L_{\wp,\vartheta}, a)$ counting framed 2d-4d BPS states.  We will organize
these indices into a natural generating function $F(\wp, \vartheta)$ which is
the main player in this paper.

Throughout this section we assume $(\wp,\vartheta)$ is generic, in the sense explained in
\S\ref{sec:framed-bps}.

\subsection{Generating functions of framed 2d-4d indices} \label{sec:gen-func}

The fundamental properties of $\fro'(L_{\wp,\vartheta}, a)$
are most elegantly expressed in terms of a formal generating function
$F(\wp,\vartheta)$.

The idea is to
introduce a formal variable $X_{\bar a}$ for each charge $\bar a \in \Gamma(z_1, z_2)$,
and then define
$ F(\wp,\vartheta) = \sum_{\bar a \in \Gamma(z_1,z_2)}  \fro'( L_{\wp,\vartheta},\bar a) X_{\bar a}$.
As we have noted in \S\ref{sec:tricky-sign}, though, we need to take some care here to deal with
sign ambiguities.  So more precisely, we introduce formal variables $X_a$ for each $a \in \tGamma(\tz_1, \tz_2)$,
subject to the relation that if $a$ and $a'$ project to the same class $\bar{a} \in \Gamma(z_1,z_2)$
then we have
\begin{equation} \label{eq:winding-rule-2}
 X_a / X_{a'} = (-1)^{w(a,a')}.
\end{equation}
We then choose one representative $a \in \tGamma(\tz_1,\tz_2)$ for each $\bar a \in \Gamma(z_1,z_2)$,
and define
\begin{equation}\label{eq:FLP-gen-prelim}
F(\wp,\vartheta) := \sum_{\bar a \in \Gamma(z_1,z_2)} \fro'( L_{\wp,\vartheta},a) X_{a}.
\end{equation}
$F(\wp,\vartheta)$ is independent of our choices of
representatives $a$, thanks to \eqref{eq:winding-rule-1}, \eqref{eq:winding-rule-2}.

\subsection{Formal products and composition}

Suppose $a$ and $b$ are the relative homology classes of
two open paths on $\tilde\Sigma$.  If the end of $a$ coincides with the start of $b$,
we let $a+b$ denote the relative homology class of
the concatenation of $a$ and $b$.
Then we introduce a product law on our formal variables:
\begin{equation} \label{eq:X-product-law}
 X_a X_b = \begin{cases} X_{a+b} & \text{if the end of $a$ is the start of $b$}, \\
            0 & \text{otherwise}.
           \end{cases}
\end{equation}

\insfig{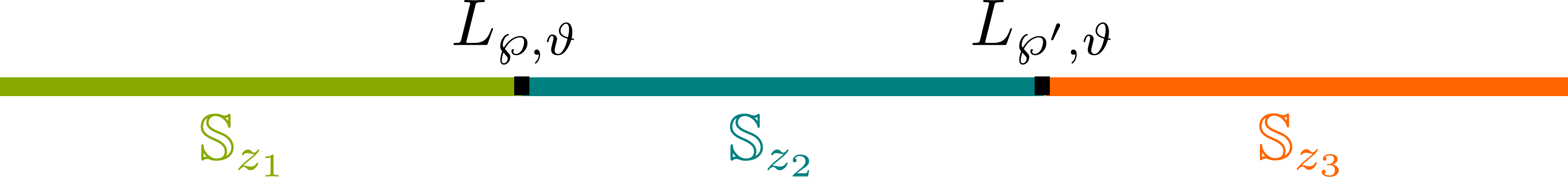}{Three surface defects connected by two interfaces.  (This picture
lives in the three-dimensional space where the field theory $S[\fg,C,D]$ is defined; we have factored
out the time direction.)}
We can now state one of the key properties of the generating functions $F(\wp, \vartheta)$:
if $\wp$ and $\wp'$ are paths on $C$
which can be concatenated to make a smooth path $\wp \wp'$
(with the end of $\wp$ attached to the start of $\wp'$),
and if both $(\wp, \vartheta)$ and $(\wp', \vartheta)$ are generic,
then
\begin{equation} \label{eq:composition}
 F(\wp, \vartheta) F(\wp', \vartheta) = F(\wp \wp', \vartheta).
\end{equation}
Here is the physical reason for \eqref{eq:composition}.  We consider three surface defects
$\bS_{z_1}$, $\bS_{z_2}$, $\bS_{z_3}$, connected by two interfaces $L_{\wp,\vartheta}$
and $L_{\wp',\vartheta}$, as shown in Figure \ref{fig:three-surface-operators}.  The Hilbert space $\CH$ of framed BPS states
in this situation should be independent of the separation between the interfaces.
For large separations, since the framed BPS states
are localized near the interfaces,
$\CH$ is a tensor product between a space of framed BPS states for the interface $\wp$ and
one for the interface $\wp'$.  On the other hand, by considering the limit of zero separation,
we see that $\CH$ is the space of framed BPS states for the interface $\wp \wp'$.
Equating these two descriptions of $\CH$ gives \eqref{eq:composition}.

(One could wonder whether there should be a $\pm$ sign on the right side of \eqref{eq:composition}; this amounts to asking whether
our rules for fixing the signs of $F(\wp, \vartheta)$, $F(\wp', \vartheta)$ and $F(\wp\wp', \vartheta)$ are compatible with one another.
Fortunately, it will follow from our explicit rules below that there is no sign needed.)

\subsection{The spectral network $\wnet_\vartheta$} \label{sec:spectral-network}

Our next aim is to explain how $F(\wp,\vartheta)$ can actually be computed.
As it turns out, the answer depends crucially
on how $\wp$ meets a certain codimension-1 locus
$\wnet_\vartheta \subset C$, which we call a \ti{spectral network}.

We say that a point $z \in C$
\ti{supports} those charges $\bar a \in \Gamma(z,z)$ for which $Z_{\bar a}(z) / e^{\I \vartheta} \in \R_-$.
Define $\wnet_\vartheta$ to be the set of $z \in C$ such
that $z$ supports some $\bar a \in \Gamma(z,z)$ with $\mu(a) \neq 0$ (for either lift $a$ of $\bar a$).

$\wnet_\vartheta$ is a codimension-$1$ network on $C$, the union of
segments which we call \ti{$\cS$-walls}.
The $\cS$-walls can end at branch points or at special points which we call \emph{joints},
and can also asymptote to the punctures of $C$.  We will assume that along any $\cS$-wall, a generic point
supports exactly one charge $\bar a$.

There are two ways in which this
genericity could be violated.  One is for $\vartheta$ to be non-generic:  in that case some $\cS$-walls will support
both a charge $\bar a \in \Gamma_{ij}(z,z)$ and a charge $\bar b \in \Gamma_{ji}(z,z)$.  This phenomenon is crucial for our
story and will be analyzed in \S\ref{sec:varying-theta}.  For now, however, we are assuming $\vartheta$ generic, so we explicitly exclude this
possibility.  The second way in which genericity could be violated is less common:  it might happen accidentally that
e.g. $\lambda_i - \lambda_j = \lambda_k - \lambda_l$ on a whole patch of $C$.
In this case a single wall could support both a charge
$\bar a \in \Gamma_{ij}(z,z)$ and $\bar b \in \Gamma_{kl}(z,z)$.  If $i,j,k,l$ are all distinct, all
our discussion in this section has a straightforward
extension to that case; if they are not all distinct the situation is more subtle.  In any case,
from now on we assume that a
generic point along an $\cS$-wall supports exactly one charge $\bar a$.

In this case, all points of any single $\cS$-wall support ``the same'' charge $\bar a$, in the sense that if
$z$, $z'$ are generic points on a common wall, the natural parallel transport along the wall
takes the charge $\bar a \in \Gamma(z,z)$ supported at $z$
into the charge $\bar a' \in \Gamma(z',z')$ supported at $z'$.
If we choose a charge $a \in \tGamma(\tz,\tz)$ lifting $\bar a$ as discussed in \S\ref{sec:tricky-sign},
then the parallel transport also takes $a$ to an $a' \in \tGamma(\tz',\tz')$ lifting $\bar a'$.

Now recall the soliton degeneracies $\mu(a) \in \Z$, which are
defined for any $a \in \tGamma(\tz,\tz)$, so long as
$z \in C$ does not lie on a wall of marginal stability.\footnote{To reduce potential confusion we
emphasize that the walls of marginal stability are \ti{not} the same thing as the $\cS$-walls.}
$\mu(a)$ does not depend on the parameter $\vartheta$, and its definition does not involve the spectral network $\wnet_{\vartheta}$.
In our computation of $F(\wp, \vartheta)$ below, though, we will find that the $\mu(a)$
which are really important are the ones where $\bar a \in \Gamma(z,z)$ is supported at $z$ ---
or said otherwise, we will be mainly interested in evaluating $\mu(a)$ \ti{along the $\cS$-walls supporting $\bar a$.}
We will also find below that $\mu(a)$ is \ti{constant} along each $\cS$-wall supporting $\bar a$;
with this in mind we immediately simplify our notation by letting $\mu(a,p)$ denote the constant value of $\mu(a)$ along
a wall $p$ supporting $\bar a$.

\subsection{Computing $F(\wp,\vartheta)$}\label{subsec:Compute-FP}

Now we can describe how $F(\wp,\vartheta)$ is computed.
The recipe we will summarize here follows from the 2d-4d wall-crossing formula
of \cite{Gaiotto:2011tf}.

The simplest situation occurs when $\wp$ does not cross $\wnet_\vartheta$ anywhere.  Define
\begin{equation}
 \cD(\wp) = \sum_{i=1}^K X_{\wp^{(i)}},
\end{equation}
where the $\wp^{(i)}$ are the canonical lifts of $\wp$ to the $K$ sheets of $\tilde\Sigma$.
Then if $\wp \cap \wnet_\vartheta = \emptyset$ we have simply
\begin{equation}\label{eq:TrivPar}
F(\wp,\vartheta) = \CD(\wp).
\end{equation}
In other words, interfaces corresponding to short enough paths $\wp$ on $C$
just support $K$ framed BPS states, one for each vacuum of the surface operator.

The more interesting question is how to compute $F(\wp)$ if $\wp$ does cross $\wnet_\vartheta$.
Because of the composition property \eqref{eq:composition}, it is enough to
answer this question in the case where $\wp$ crosses $\wnet_\vartheta$ exactly once.

So suppose $\wp$ crosses $\wnet_\vartheta$ at a point $z$ supporting
a charge $\bar a \in \Gamma_{ij}(z,z)$.  In this case $\wp$ is divided into two subpaths
$\wp_+$ and $\wp_-$, which we deform slightly to obtain $\wp'_+$ and $\wp'_-$
as shown in Figure \ref{fig:p-crossing-s-wall}.
\insfigscaled{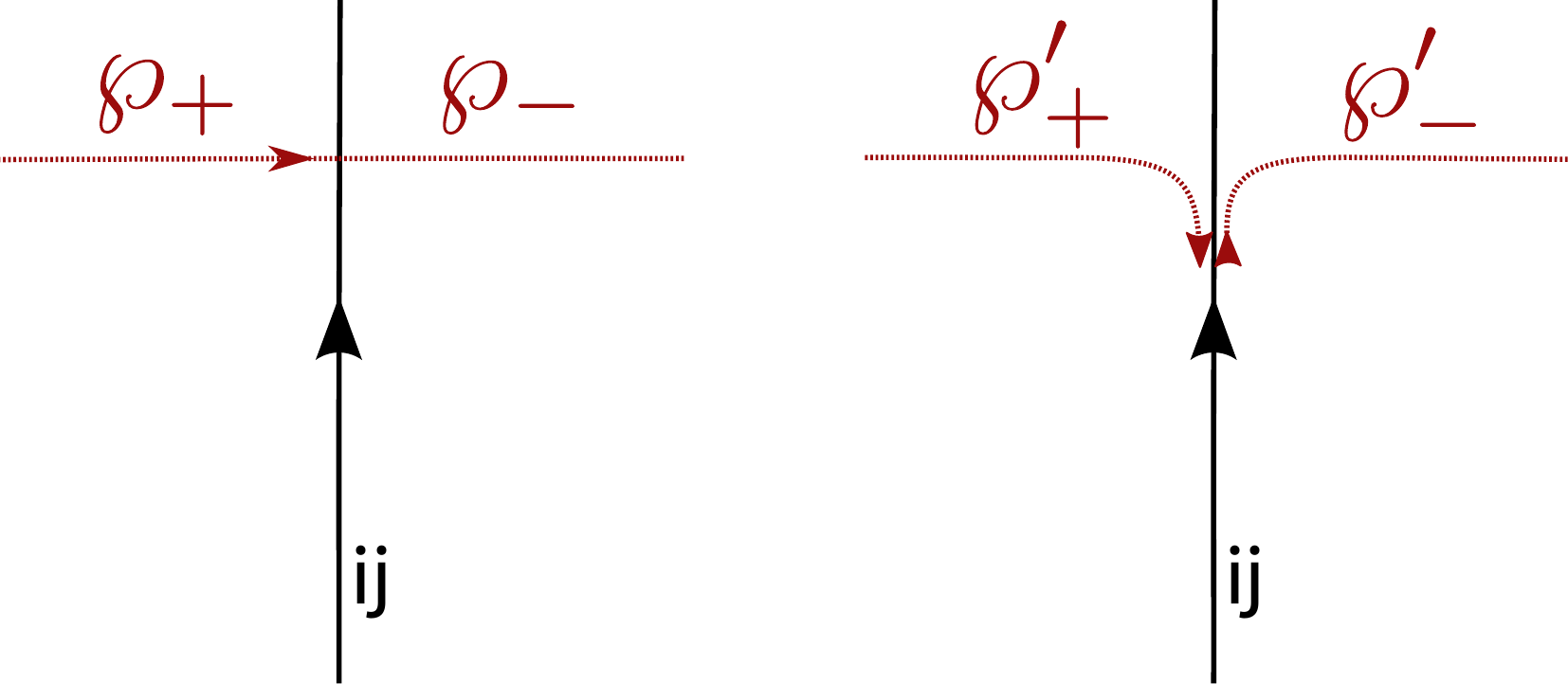}{0.4}{A path $\wp$ crossing an $\cS$-wall is locally divided into two pieces
$\wp_\pm$.  We deform $\wp_\pm$ slightly to paths $\wp'_\pm$, so that the final tangent vector of $\wp'_+$ and the initial tangent vector
of $\wp'_-$ point along the $\cS$-wall, in opposite directions.  The paths $\wp'_\pm$ are shown
slightly displaced from the wall for clarity.}
Then we have\footnote{The factor $\mu(a) X_a$
appearing in \eqref{eq:S-factor}
is independent of the choice of lift $a$ of $\bar{a}$, because of \eqref{eq:mu-sign-rule}, \eqref{eq:winding-rule-2}.}
\begin{equation} \label{eq:S-factor}
 F(\wp, \vartheta) = \cD(\wp) + \cD(\wp'_+) (\mu(a) X_a) \cD(\wp'_-).
\end{equation}
The second term is the interesting one:  it says that when $\wp$ crosses $\wnet_\vartheta$, $F(\wp, \vartheta)$ includes paths which
are segments of lifts of $\wp$ combined with ``detours'' along the lifts of
BPS solitons.  (The point of our deformation from $\wp_\pm$ to $\wp'_\pm$ here was to arrange that the second term makes sense, i.e. that $\cD(\wp'_+)$ and $\cD(\wp'_-)$ can be
concatenated with the soliton charge $a$.)

For later convenience, we introduce a second notation for \eqref{eq:S-factor}.  Let $\tz$
denote the tangent vector to $\wp$ at $z$.  Let $t_1$ be the shortest arc running from
$\tz$ to the initial point of $a$, in the fiber of $\tilde\Sigma$ over $z$. Similarly,  let $t_2$
be the shortest arc running from the final point of $a$ to $\tz$, in the fiber of
$\tilde\Sigma$ over $z$; and define
\begin{equation}\label{eq:aztil-def}
a_{\tilde z} = t_1 + a + t_2.
\end{equation}
Note that $t_1,t_2$ are well-defined because the intersection of $\wp$ and $\wnet_{\vartheta}$
is assumed to be transverse.
With this definition  \eqref{eq:S-factor} can be rewritten as
\begin{equation} \label{eq:S-factor-simple}
 F(\wp, \vartheta) = \cD(\wp_+) (1 + \mu(a) X_{a_{\tilde z}}) \cD(\wp_-).
\end{equation}

The formula \eqref{eq:S-factor-simple}
can be interpreted as a kind of wall-crossing formula for the framed 2d-4d BPS spectrum:
it implies that when an endpoint of the path $\wp$ is moved across an $\cS$-wall,
$F(\wp, \vartheta)$ jumps by multiplication with a factor $(1 + \mu(a) X_{a_{\tilde z}})$.
This is the way that this formula appeared in \cite{Gaiotto:2011tf}.

\section{The spectral network $\wnet_\vartheta$ at fixed $\vartheta$} \label{sec:wnet}

In \S\ref{sec:F-basics}
we have given a recipe for computing the generating functions $F(\wp, \vartheta)$.
The recipe depends on the data of the spectral network $\wnet_\vartheta$ and the framed
2d-4d BPS degeneracies $\mu(\cdot, p)$ along each wall $p$ of $\wnet_\vartheta$.
To make this recipe explicit, then, we need to be able to determine
$\wnet_\vartheta$ and $\mu$.  In this section we explain how this can be done.

Throughout this section we continue to assume that $\vartheta$ is generic,
in the sense explained in \S\ref{sec:framed-bps}.

\subsection{Walls as trajectories}

Suppose the $\cS$-wall $p$
supports the charge $\bar a \in \Gamma_{ij}$ (by which we mean more precisely that
every point $z \in p$ supports a charge $\bar a(z) \in \Gamma_{ij}(z,z)$).
Then $e^{- \I \vartheta} Z_{\bar a}$ is real everywhere along $p$.
In particular, since $\de Z_{\bar a} = \frac{1}{\pi}(\lambda^{(j)} - \lambda^{(i)})$,
this means that $p$ is an $ij$-trajectory with phase $\vartheta$, in the sense of
\eqref{eq:trajectory-de}.

So the walls in $\wnet_\vartheta$ obey differential equations, and in fact
exactly the \ti{same} equations which are obeyed
by the BPS strings which make up both the 2d and the 4d BPS states.
This is not a coincidence, as we will see below.

By virtue of being an $ij$-trajectory, $p$ is naturally oriented.
As we move along $p$ in the positive direction, the BPS mass $\abs{Z_{\bar a}} = - e^{- \I \vartheta} Z_{\bar a}$ increases.

\subsection{Joints and wall-crossing for $\mu$} \label{sec:wnet-basics}

Define a \ti{joint} to be a point $z \in C$ where at least two $\cS$-walls intersect.
A joint thus supports at least two charges, say $\bar a \in \Gamma_{ij}(z,z)$ and $\bar b \in \Gamma_{kl}(z,z)$.

\insfig{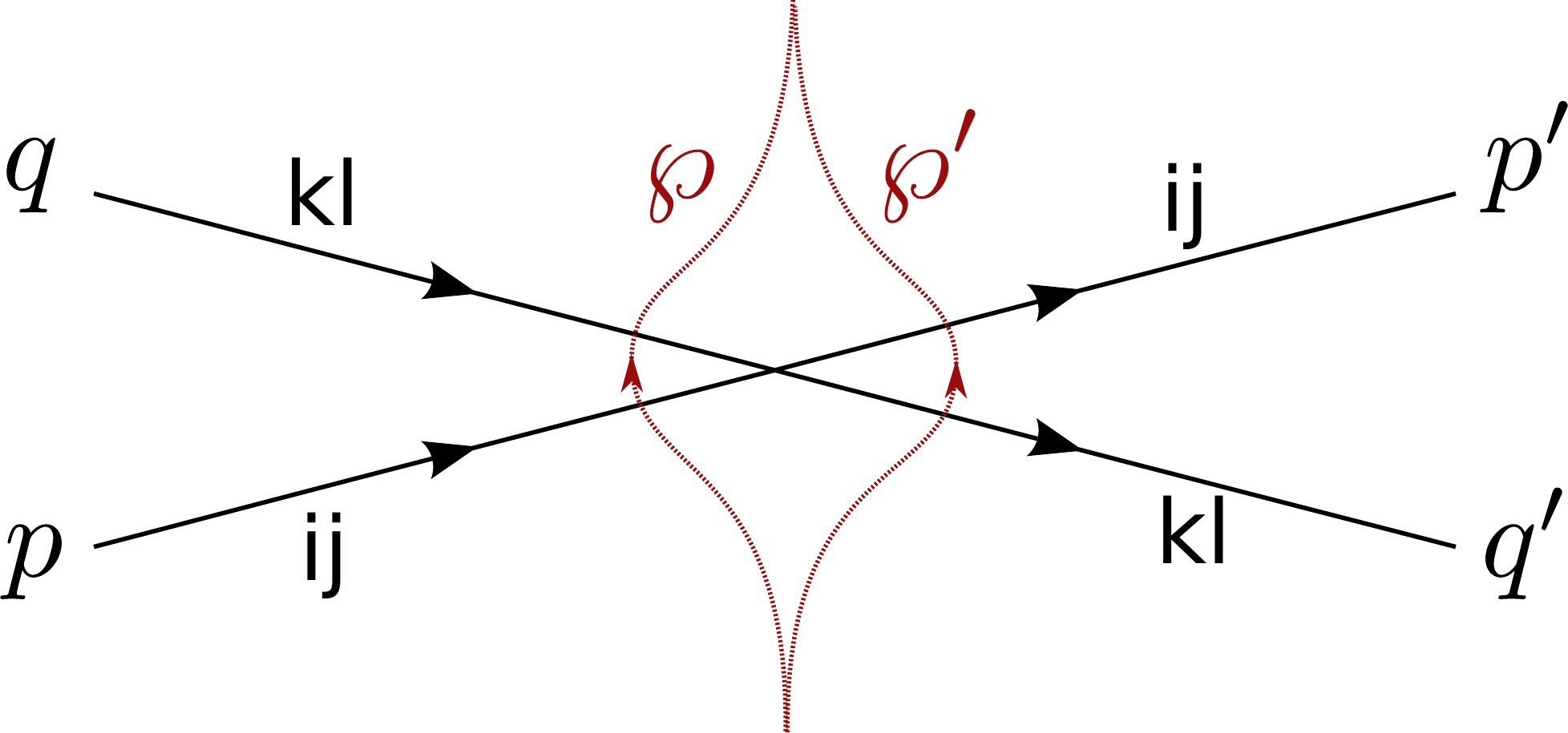}{The local picture around a joint where exactly
two $\cS$-walls meet.  We also show two paths $\wp$, $\wp'$ on $C$
which are related to one another by regular homotopy across the joint.}
The simplest situation arises if the joint $z$ supports \ti{only} these two charges, so exactly two $\cS$-walls meet there.  This can only occur if
$i \neq l$ and $j \neq k$.  In that case the local picture around $z$ is as shown in Figure \ref{fig:trajectories-meeting-commuting}.

Now let us consider what happens to the soliton degeneracies $\mu$ as we
move across a joint.  As it turns out, we can answer this question completely
by considering the defect operators attached to the paths $\wp$, $\wp'$ on $C$ shown
in Figure \ref{fig:trajectories-meeting-commuting}.
The requirement of homotopy invariance says that
\begin{equation} \label{eq:flat-simple}
F(\wp, \vartheta) = F(\wp', \vartheta).
\end{equation}
To understand what this really means,
let us evaluate both sides using the rules of \S\ref{subsec:Compute-FP},
and then deform all the resulting paths on $\tilde\Sigma$ so that they run directly into
the joint $z$, with a common tangent vector $\tilde z$ at $z$.
Then we obtain
\begin{multline} \label{eq:x-simp}
\cD(\wp_+) (1 + \mu(a,p) X_{a_{\tilde z}}) (1 + \mu(b,q) X_{b_{\tilde z}}) \cD(\wp_-) = \\
\cD(\wp_+) (1 + \mu(b,q') X_{b_{\tilde z}}) (1 + \mu(a,p') X_{a_{\tilde z}}) \cD(\wp_-).
\end{multline}
The paths $a_{\tilde z}$ and $b_{\tilde z}$ are not composable in either direction
(since $a_{\tilde z}$ runs from $\tz^{(i)}$ to $\tz^{(j)}$ while $b_{\tilde z}$ runs from $\tz^{(k)}$ to $\tz^{(l)}$),
so we have $X_{a_{\tilde z}} X_{b_{\tilde z}} = X_{b_{\tilde z}} X_{a_{\tilde z}} = 0$.  Considering the terms linear in $X_{a_{\tilde z}}$ and $X_{b_{\tilde z}}$ in \eqref{eq:x-simp} then
gives
\begin{align}
 \mu(a,p) &= \mu(a,p'), \\ \mu(b,q) &= \mu(b,q').
\end{align}
In short:  when an $\cS$-wall of type $ij$ and an $\cS$-wall of type $kl$ meet transversely, they
cross without any changes in $\mu$.

\insfig{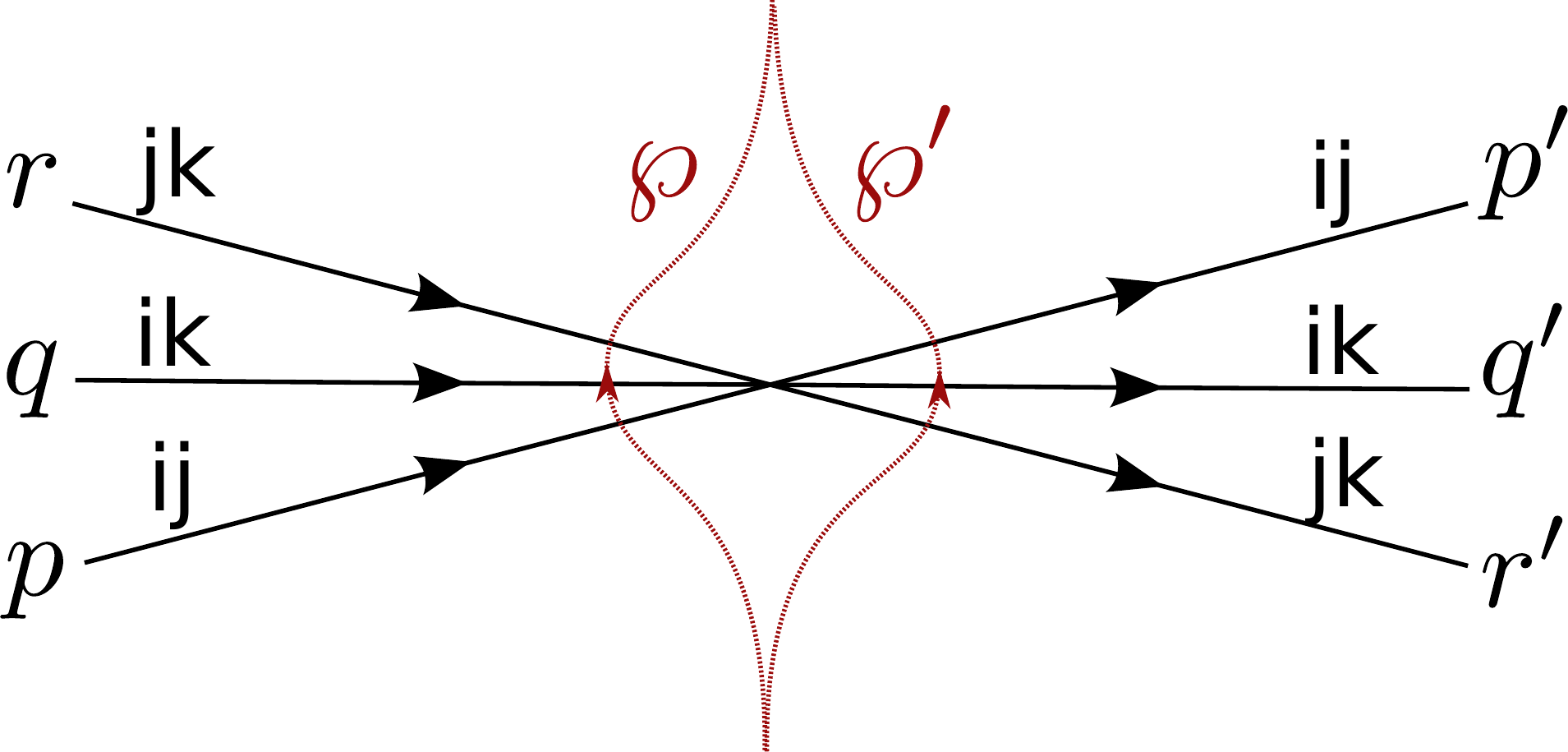}{The local picture around a
generic collision between three $\cS$-walls.}
A more interesting situation arises when an $\cS$-wall of type $ij$ and an $\cS$-wall of type
$jk$ meet transversely.  Suppose the two walls support charges
$\bar a \in \Gamma_{ij}(z,z)$ and $\bar b \in \Gamma_{jk}(z,z)$.  (Note that in this case
we must have $i \neq k$; otherwise the two walls would obey the \ti{same} differential equation,
and it would be impossible for them to intersect transversely.)
In this case the joint $z$ supports both $\bar a$ and $\bar b$, which are \ti{composable} since $z^{(j)}$ is both
the end of $\bar a$ and the start of $\bar{b}$.  Hence the joint
also supports a third charge $\bar c = \bar a + \bar b \in \Gamma_{ik}(z,z)$, and
so there could be a third $\cS$-wall meeting $z$.  When $\mu$ is generic enough, the
picture is as shown
in Figure \ref{fig:ijkl-collisions}.\footnote{One might think that there is a second, inequivalent
possibility:  one could have
exchanged the $ij$ and $jk$ labels in Figure \ref{fig:ijkl-collisions}.  This gives a new picture, which is not
related to Figure \ref{fig:ijkl-collisions} by a rotation, but \ti{is}
related to Figure \ref{fig:ijkl-collisions} by an orientation-reversing map.  Fortunately, our rules
for computing $F(\wp, \vartheta)$ do not use the orientation of $C$.  So our analysis is fully general.}

The condition
\begin{equation} \label{eq:flat-simple-2}
F(\wp, \vartheta) = F(\wp', \vartheta)
\end{equation}
becomes
\begin{align} \label{eq:x-simp-2}
&\cD(\wp_+) (1 + \mu(a,p) X_{a_{\tilde z}}) (1 + \mu(c,q) X_{c_{\tilde z}}) (1 + \mu(b,r) X_{b_{\tilde z}}) \cD(\wp_-) = \\ \notag
&\cD(\wp_+) (1 + \mu(b,r') X_{b_{\tilde z}}) (1 + \mu(c,q') X_{c_{\tilde z}}) (1 + \mu(a,p') X_{a_{\tilde z}}) \cD(\wp_-).
\end{align}
Using $X_{a_{\tilde z}} X_{b_{\tilde z}} = (-1)^{w(a_{\tilde z}+b_{\tilde z},c_{\tilde z})} X_{c_{\tilde z}}$ and $X_{b_{\tilde z}} X_{a_{\tilde z}} = 0$,
\eqref{eq:x-simp-2} implies
\begin{align}
 \mu(a,p') &= \mu(a,p), \\
 \mu(b,r') &= \mu(b,r), \\
 \mu(c,q') &= \mu(c,q) + (-1)^{w(a_{\tilde z}+b_{\tilde z},c_{\tilde z})} \mu(a,p) \mu(b,r). \label{eq:cv-wcf}
\end{align}

The result \eqref{eq:cv-wcf}
says that the number of solitons carrying charge $\bar{c}$ changes as we move
the modulus $z$ of the surface defect $\bS_z$ across the joint.
This reflects the phenomenon of decay/formation of bound states between solitons of charges
$\bar{a}$ and $\bar{b}$.  Indeed \eqref{eq:cv-wcf} is the same wall-crossing formula which was
discovered in \cite{Cecotti:1993rm} in the context of pure 2d theories, and was reinterpreted in
the 2d-4d
context in \cite{Gaiotto:2011tf}.  Our derivation of it here, using consistency of the framed 2d-4d
BPS spectrum, is essentially the same as the one given in \cite{Gaiotto:2011tf}.

There are a few important special cases which deserve separate discussion.
One arises when $\mu(c,q)=0$, so that we actually have only two $\cS$-walls intersecting, of
types $ij$ and $jk$.
In this case the situation is as shown in Figure \ref{fig:ijkl-collision-birth}:
the wall $q'$ is born from the joint, and \eqref{eq:cv-wcf} reduces to
$\mu(c,q') = (-1)^{w(a_{\tilde z}+b_{\tilde z},c_{\tilde z})} \mu(a,p) \mu(b,r)$.

\insfig{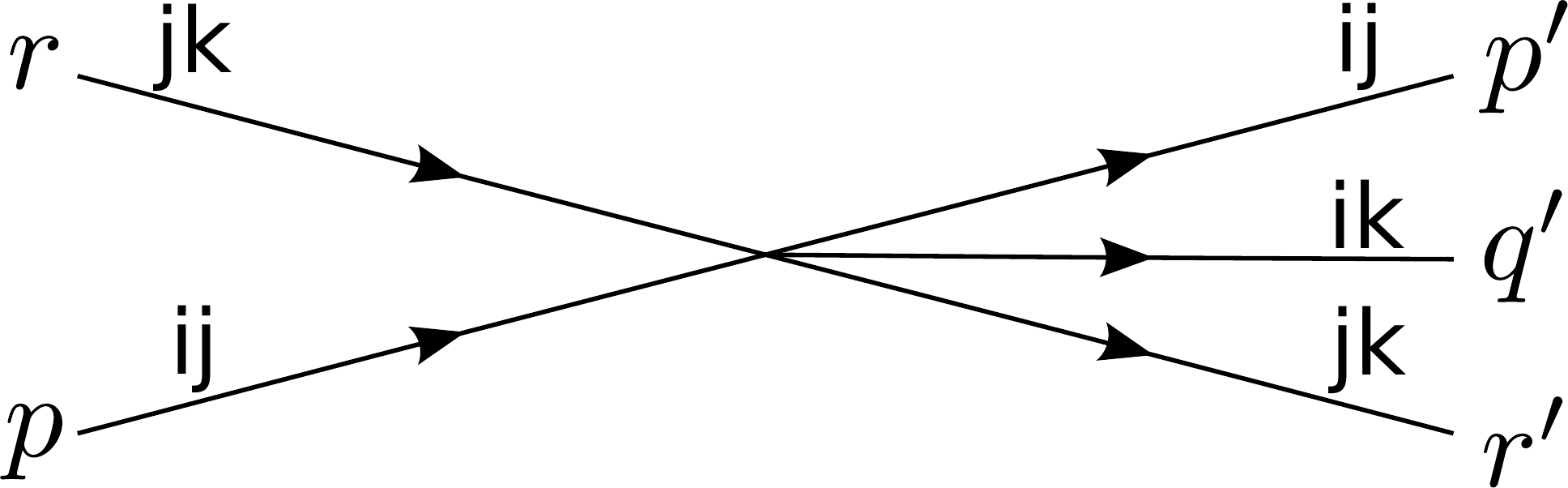}{A collision between $\cS$-walls, at which a new $\cS$-wall
is born.}
The reverse situation is also allowed:  if $\mu(c,q) = - (-1)^{w(a_{\tilde z}+b_{\tilde z},c_{\tilde z})} \mu(a,p) \mu(b,r)$, then \eqref{eq:cv-wcf} gives
$\mu(c,q') = 0$, so the wall $q$ ``dies'' at the joint,
as shown in Figure \ref{fig:ijkl-collision-death}.
\insfig{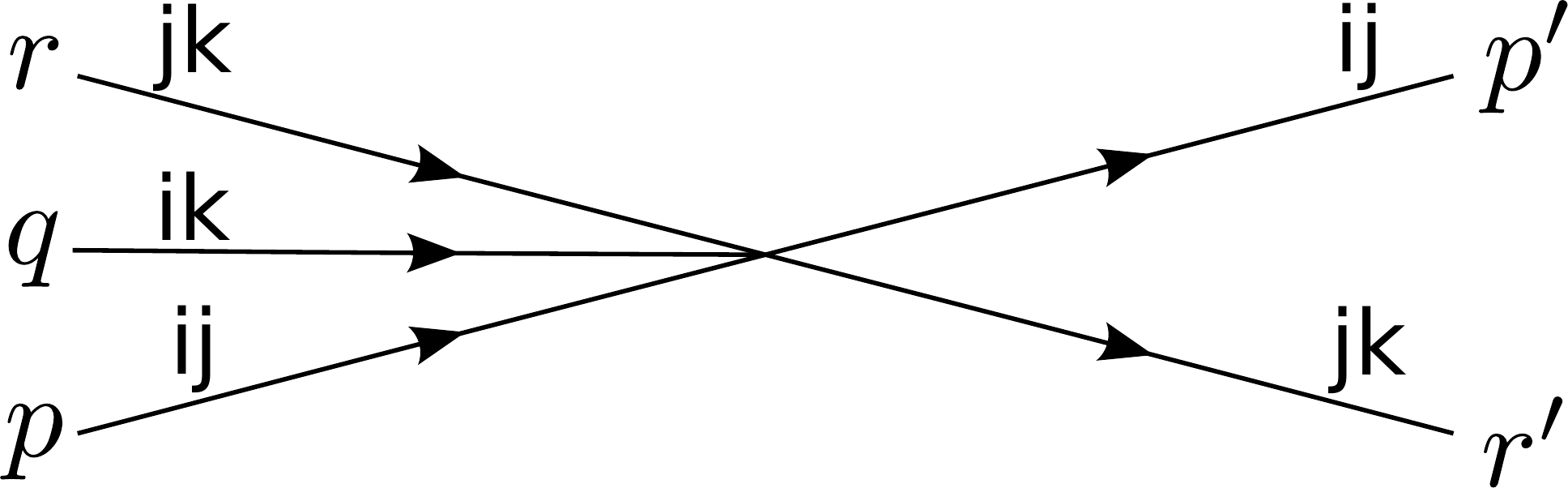}{A collision between $\cS$-walls, at which an $\cS$-wall
dies.}

So far we have described some local properties of $\wnet_\vartheta$ and $\mu$.  In the
next few sections we describe a recipe for explicitly \ti{constructing} them.

\subsection{The mass filtration; $\wnet_\vartheta[\Lambda]$ and $\mu[\Lambda]$ for small $\Lambda$} \label{sec:determining-wnet}

We begin with the observation that
$\wnet_\vartheta$ carries a useful filtration.  Namely, for any $\Lambda > 0$ (with dimensions
of mass), we can define a new object $\wnet_\vartheta[\Lambda]$ by truncating all the $\cS$-walls:
for a wall $w$ supporting a charge $\bar a$,
$\wnet_\vartheta[\Lambda]$ includes only the portion of $w$ with $\abs{Z_{\bar a}} < \Lambda$.
So
\begin{equation}
\wnet_\vartheta[\Lambda] \subset \wnet_\vartheta[\Lambda'] \text{ for } \Lambda \le \Lambda'
\end{equation}
and
\begin{equation}
 \wnet_\vartheta = \lim_{\Lambda \to \infty} \wnet_\vartheta[\Lambda].
\end{equation}
We also define a truncated version $\mu[\Lambda]$ of the soliton degeneracies
$\mu$:  $\mu[\Lambda](a)$ is defined only for charges $a$ with $\abs{Z_{\bar a}} < \Lambda$,
and for such charges it agrees with $\mu(a)$.

For small enough $\Lambda$, we can
describe $\wnet_\vartheta[\Lambda]$ and $\mu[\Lambda]$ simply and explicitly.
The reason is that charges $\bar a \in \Gamma(z,z)$ with $\abs{Z_{\bar a}} < \Lambda$
are represented by very short paths on $\Sigma$
between distinct lifts $z^{(i)}$ and $z^{(j)}$ of $z$.  The only way to get such a
short path is for $z$ to be close to an $(ij)$ branch point $\fb$:  then if we take $\bar a \in \Gamma(z,z)$ to be a path running
from $z^{(i)}$ to the ramification point over $\fb$
and returning to $z^{(j)}$, we indeed have $\abs{Z_{\bar a}} \to 0$
as $z \to \fb$.
We could also have considered the charge $-\bar a \in \Gamma(z,z)$, corresponding to a path
running in the opposite direction.
These two charges are exchanged by the monodromy when $z$ goes once around $\fb$.
They are the only charges which become massless at $\fb$.

So, to determine $\wnet_\vartheta[\Lambda]$ for $\Lambda$ small,
we just have to describe the $\cS$-walls supporting these light charges.
Letting $z$ be a local coordinate with $z=0$ at $\fb$, we have $Z_{\bar a} \sim z^{3/2}$
and $Z_{-\bar a} \sim -z^{3/2}$ (the
square-root branch cut arises from the monodromy $\bar a \to -\bar a$ noted above.)
It follows that there are $3$ $\cS$-walls
emerging from the branch point, as shown in Figure \ref{fig:branch-point-trajectories}.
\insfig{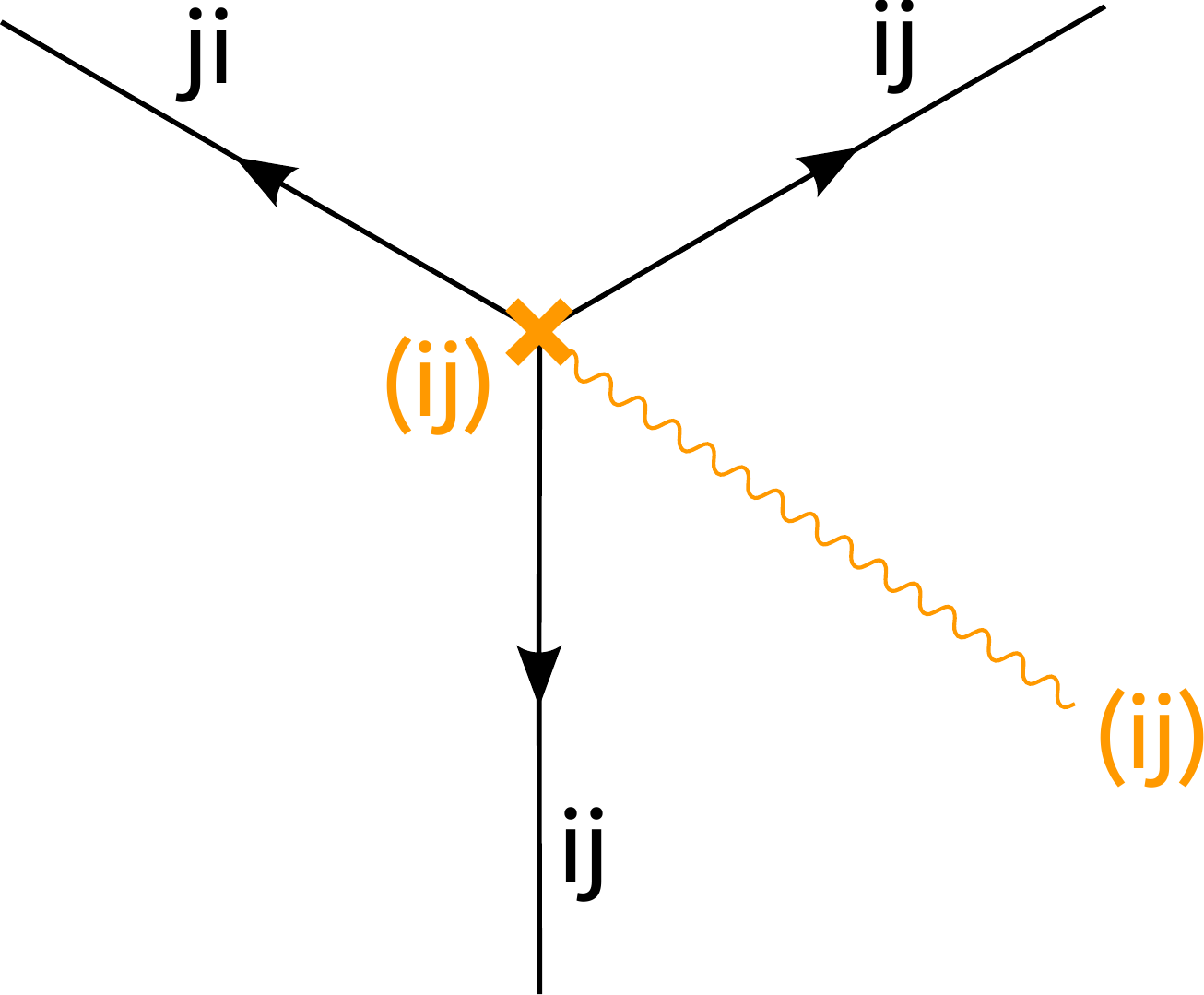}{The $3$ $\cS$-walls emerging from a branch point.
Because the two relevant sheets of the covering $\Sigma \to C$
are exchanged by monodromy around the branch point,
we cannot assign labels to the trajectories globally; instead we have chosen a branch cut and
a labeling of the sheets $i$ and $j$ on the complement of the cut.  We also label the cut with
the transposition $(ij)$ which relates the sheets on the two sides of the cut.}

If $z$ lies on one of these three $\cS$-walls near $\fb$, then
there is a light 2d-4d BPS soliton on the surface defect $\bS_z$, with central charge
in $\R_- e^{\I \vartheta}$.\footnote{In a sense
it would be better to refer to these light states simply as ``2d'' solitons rather than ``2d-4d,'' since their existence
does not depend much on the coupling to the 4d theory.  Indeed, the
existence of these light states can be deduced from a universal computation
involving the simplest nontrivial Landau-Ginzburg
model.  See Section 8.1 of \cite{Gaiotto:2011tf} for further discussion.}
This BPS soliton is represented by a short finite open web $N(z)$, consisting of a single
BPS string connecting $z$ to $\fb$, as illustrated in
Figure \ref{fig:short-trajectory}.
\insfig{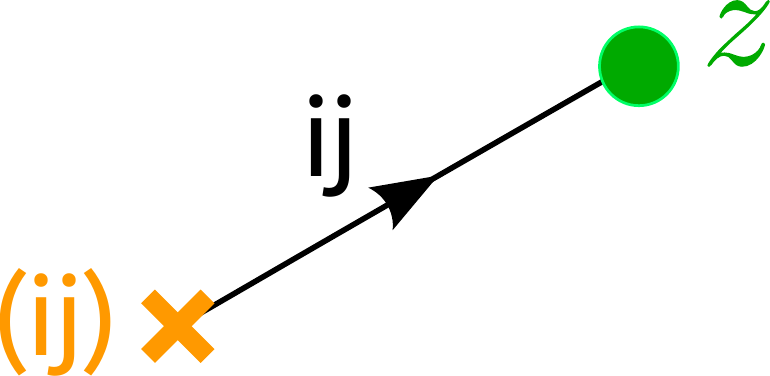}{A finite open web $N(z)$ consisting of a single short $ij$-trajectory, with one end on the $(ij)$-branch point $\fb$
and one end on $z$.  This finite open web represents a light BPS soliton
on the surface defect $\bS_z$.}
In accordance with the general rules of \S\ref{sec:2d4d-bps-states}, the charge $\bar a$ of this
BPS state is $\bar a = [N(z)_\Sigma]$, where $N(z)_\Sigma$ is the lift of $N(z)$ to $\Sigma$, i.e. a short path on $\Sigma$
running from $z^{(i)}$ to the ramification point over $\fb$ and then back to $z^{(j)}$.

As we have mentioned in \S\ref{sec:tricky-sign}, however,
to fix the sign of the BPS degeneracy for this soliton,
we need to choose a lift from the charge $\bar a$ to a class $a$.
Suppose we make the most obvious choice, namely the one determined by the web $N(z)$:  $a = [N(z)_{\tilde\Sigma}]$.  Let $p$ denote the $\cS$-wall on which $z$ sits.
Then finally $\mu(a, p)$ is a well defined integer, and it is meaningful to ask what it is.
We have a single isolated 2d-4d BPS web here, so it should contribute just
a single state; the only question is whether we will have $\mu(a, p) = +1$ or $\mu(a, p) = -1$.
We claim that the correct answer is
\begin{equation} \label{eq:mu-near-bp}
 \mu(a, p) = +1.
\end{equation}
Indeed, this is forced on us by the requirement of homotopy
invariance, which we discuss in \S\ref{sec:homotopy-invariance} below.

\subsection{$\wnet_\vartheta[\Lambda]$ and $\mu[\Lambda]$ for general $\Lambda$} \label{sec:constructing-w}

We have determined $\wnet_\vartheta[\Lambda]$ and $\mu[\Lambda]$ for small
$\Lambda$.
Now we can ask how $\wnet_\vartheta[\Lambda]$ and $\mu[\Lambda]$
evolve as $\Lambda$ increases.
Using the properties of $\wnet_\vartheta$ and $\mu$ we have already determined, it is
straightforward to deduce the answer.

The three $\cS$-walls
emerging from each branch point $\fb$ flow according to the
differential equation \eqref{eq:trajectory-de}, for a distance
determined by the cutoff $\Lambda$.
When we increase $\Lambda$ enough, it might happen that
an $\cS$-wall of type $ij$ intersects
another $\cS$-wall of type $jk$.  At this point
a new $\cS$-wall of type $ik$ is born, as shown in
Figure \ref{fig:ijkl-collision-birth}.  The soliton degeneracy $\mu$
on this new wall is determined by the soliton degeneracies on
its parents, as in \eqref{eq:cv-wcf}.\footnote{Note that if two $\cS$-walls
carrying charges $\bar a \in \Gamma_{ij}$, $\bar b \in \Gamma_{jk}$ meet at a joint $z \in C$,
the joint is visible in $\wnet_\vartheta[\Lambda]$ beginning
at $\Lambda = \textrm{Max}(\abs{Z_{\bar a}(z)},
\abs{Z_{\bar b}(z)})$, but the $\cS$-wall born from the joint
does not appear until
$\Lambda = \abs{Z_{\bar a}(z)}+\abs{Z_{\bar b}(z)}$.}
This new ``secondary'' $\cS$-wall, in turn, also evolves according to \eqref{eq:trajectory-de}.
As $\Lambda$ increases the secondary walls can intersect with other $\cS$-walls and give birth to yet
more progeny, and so on.
In Figure \ref{fig:network-growth} we give an illustration
of the growth of $\wnet_\vartheta[\Lambda]$ with $\Lambda$ in one particular theory $S[\fg,C,D]$.
\insfigscaled{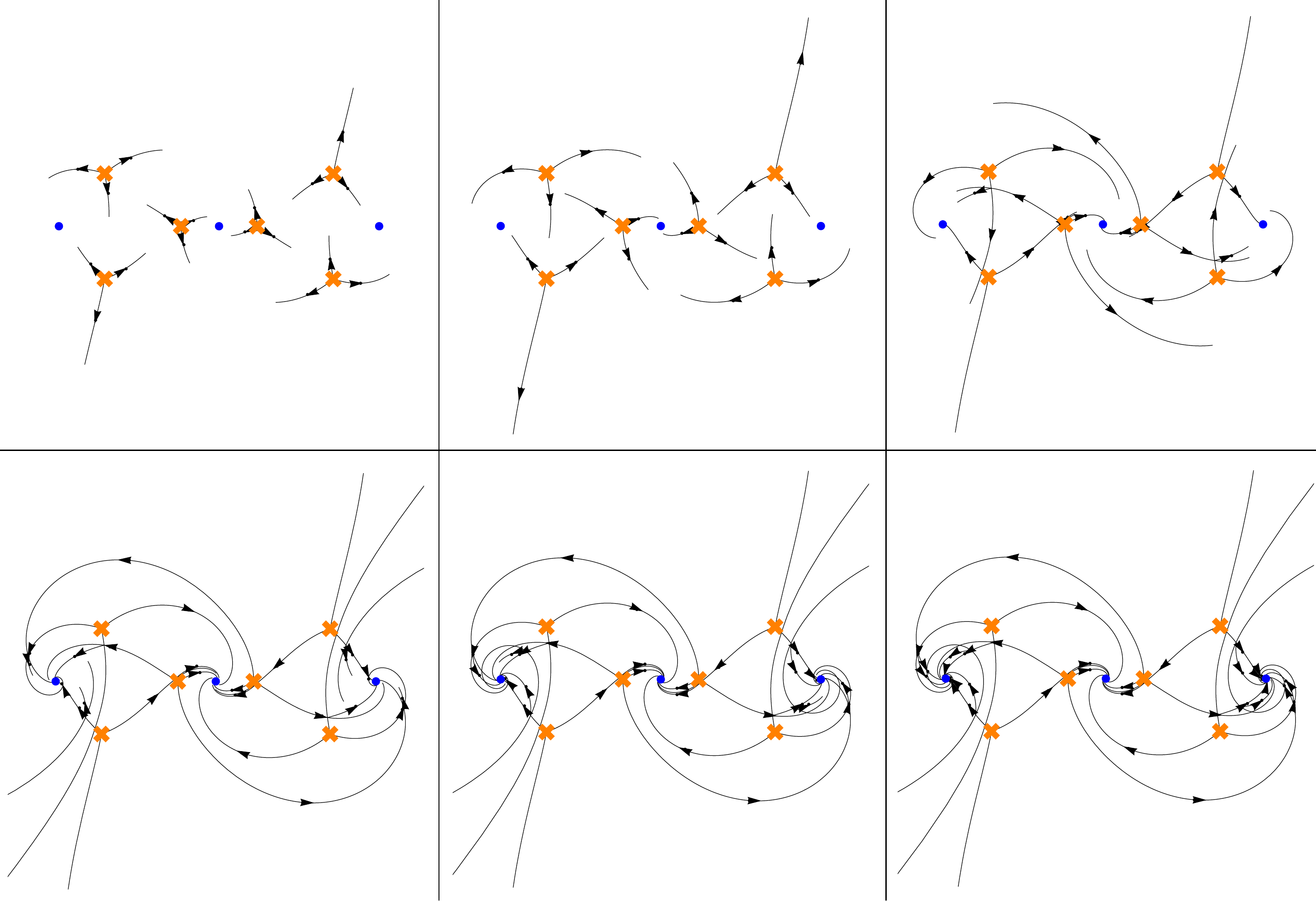}{0.52}
{Sample $\wnet_\vartheta[\Lambda]$ in the theory $S[A_2,\IC\IP^1,D]$
where the $D_n$ are 3 full defects (blue dots at $z = -1, 0, 1$).
We choose $\vartheta = \pi/4$, $\phi_2 = \frac{2}{(z - 1)^2} + \frac{2}{(z + 1)^2} - \frac{4}{z^2}$, $\phi_3 = \frac{1 - 3 (z - 1) + 6 (z - 1)^2}{(z - 1)^3} + \frac{1 + 3 (z + 1) + 6 (z + 1)^2}{(z + 1)^3} - \frac{2 + 12  z^2}{z^3}$.
The plots shown are at $\Lambda = 1, 2, 5, 12, 20, \infty$.
At $\Lambda = 0$ three $\cS$-walls are born at each of the 6 branch points
(orange crosses); at larger values of $\Lambda$, additional $\cS$-walls are born at intersections.
As $\Lambda \to \infty$ all $\cS$-walls asymptotically approach the defects.
Some $\cS$-walls are cut off
by the boundary of the plot; the visible range is a square with side length $2.6$.
See \cite{spectral-network-movies}
for an animated version of this figure.}

We may also have intersections where three $\cS$-walls meet at a point, as
in Figure \ref{fig:ijkl-collisions}.
(The reader might feel that this phenomenon should not occur generically; indeed, three arbitrary
trajectories of types $ij$, $jk$, $ik$ would be unlikely to intersect at a single point, but the $\cS$-walls
are not arbitrary trajectories.  We include an example in Figure \ref{fig:threefold-intersection}.)
\insfigscaled{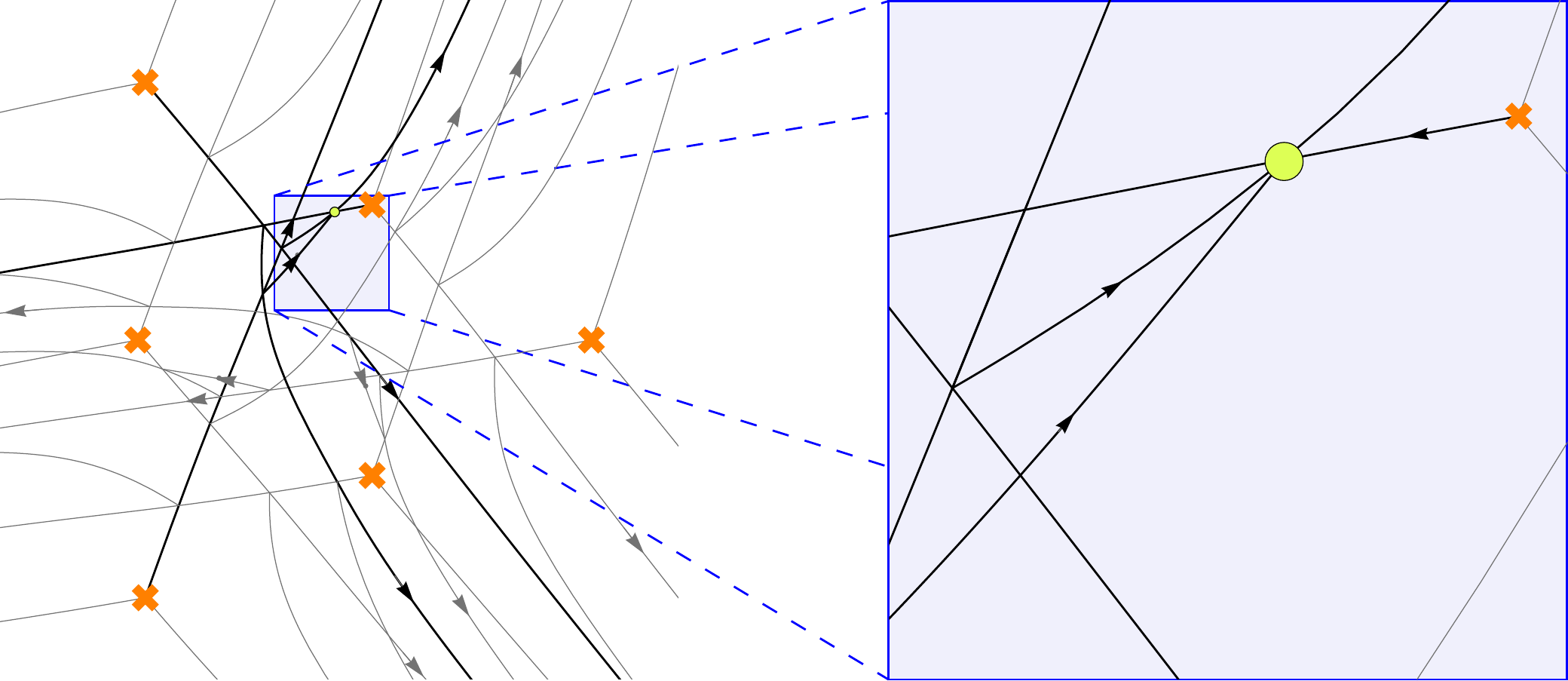}{0.69}{A portion of a sample $\wnet_\vartheta$,
exhibiting an intersection between three $\cS$-walls, marked by a yellow dot.  The region in the blue box is
blown up at right.
The $\cS$-walls which appear somewhere
in the genealogy of the three intersecting $\cS$-walls are shown in black; other $\cS$-walls
are shown in gray.  There are six branch points (orange crosses).
This example arises in the theory $S[A_3, C = \IC\IP^1, D]$ with
$\phi_2 = -10z \, \de z^2$, $\phi_3 = 4 \, \de z^3$, $\phi_4 = 9 z^2 \, \de z^4$.
We took $\vartheta = 1.841$, but we emphasize that this triple intersection persists
for nearby $\vartheta$ as well.}

There is one more complicated phenomenon which one can imagine:
what would happen if an $\cS$-wall of type $ij$ ran directly into an $(ij)$ branch point?
In this case it would not be immediately clear how to continue the network
$\wnet_\vartheta[\Lambda]$ and the BPS degeneracies $\mu[\Lambda]$.
This puzzling-looking situation cannot occur when $\vartheta$ is generic.
It does occur for non-generic $\vartheta$,
and this fact plays a crucial role in the considerations of \S\ref{sec:varying-theta} below.

We have now
given a recipe for constructing $\wnet_\vartheta[\Lambda]$ and $\mu[\Lambda]$
for any $\Lambda > 0$ and generic $\vartheta$.  $\wnet_\vartheta[\Lambda]$
consists of a finite number of $\cS$-walls for any $\Lambda$.
Taking the limit $\Lambda \to \infty$ we obtain the full $\wnet_\vartheta$
and $\mu$.  So we have now managed to determine \ti{all} of the soliton degeneracies of the
theory (including their tricky signs) using only the constraint of homotopy invariance!

More precisely, so far we have fixed some $\vartheta$ and determined all $\mu(a,p)$ --- i.e. we determined $\mu(a)$
when the parameter $z$ of the surface defect lies on an $\CS$-wall supporting the charge $a$.  This is not what one
would usually mean by ``determining the soliton degeneracies'':  what one would usually mean is that we fix some $z$
once and for all, and then compute all $\mu(a)$ for $a \in \tGamma(\tz,\tz)$.  The point is that, for any $a \in \tGamma(\tz,\tz)$,
there is some $\vartheta$ for which $z$ \ti{does} lie on an $\CS$-wall supporting $a$:  namely,
$\vartheta = \arg -Z_{\bar a}$.  So for each $a$ we can draw the corresponding network $\wnet_{\vartheta= \arg -Z_{\bar a}}$
and use it to compute $\mu(a)$.

\subsection{$\wnet_\vartheta$ near full defects} \label{sec:falling-down-holes}

On a close look at Figure \ref{fig:network-growth}, one notices that
as $\Lambda \to \infty$, all of the $\cS$-walls in $\wnet[\Lambda]$
asymptotically approach the full defects $D_n$.

To understand this concretely, let us consider the general behavior of $ij$-trajectories around a full defect.
In a local coordinate $z$ where the defect is at $z=0$,
we have
\begin{equation}
 \lambda^{(i)} = m^{(i)} \frac{\de z}{z} + \cdots
\end{equation}
where $\cdots$ denotes regular terms.  $ij$-trajectories near $z=0$ thus behave asymptotically like
\begin{equation}
 z(t) = z_0 \exp (\xi^{(ij)} t),
\end{equation}
where we defined
\begin{equation}
\xi^{(ij)} = \frac{e^{\I \vartheta}}{m^{(i)} - m^{(j)}}.
\end{equation}
As $t \to \infty$, we have $z(t) \to 0$ if and only if $\Re \, \xi^{(ij)} < 0$.
This suggests a natural ordering on the sheets in a neighborhood of $z=0$:  we say that $i < j$
if $\Re \, e^{- \I \vartheta}m^{(i)} < \Re \, e^{- \I \vartheta}m^{(j)}$.
$ij$-trajectories near $z=0$ asymptote to $z=0$ if and only
if $i < j$ in this ordering.

In the coordinate $z$,
these infalling trajectories asymptotically approach logarithmic spirals.  Passing to the covering coordinate $w = \log z$, the $ij$-trajectories are straight lines,
\begin{equation}
 w(t) = w_0 + \xi^{(ij)} t.
\end{equation}
Now suppose we have an infalling asymptotic $\cS$-wall of type $ij$ and another of type $jk$
(so $i < j < k$.)  Assuming $\arg \xi^{(ij)} \neq \arg \xi^{(jk)}$, these
two walls intersect at infinitely many
points as they spiral into $z = 0$.  Each intersection gives birth to a new $\cS$-wall of type $ik$,
which also spirals into $z = 0$.  This new $\cS$-wall can in turn intersect other
inspiraling $\cS$-walls
and give birth to yet more progeny.  The strict ordering of the sheets ensures
that the progeny cannot commit incest with one another.
Nevertheless, for $K>2$ there are an infinite number of joints accumulating at the
full defect.  See Figure \ref{fig:infalling-trajectories} for an illustration.
\insfigscaled{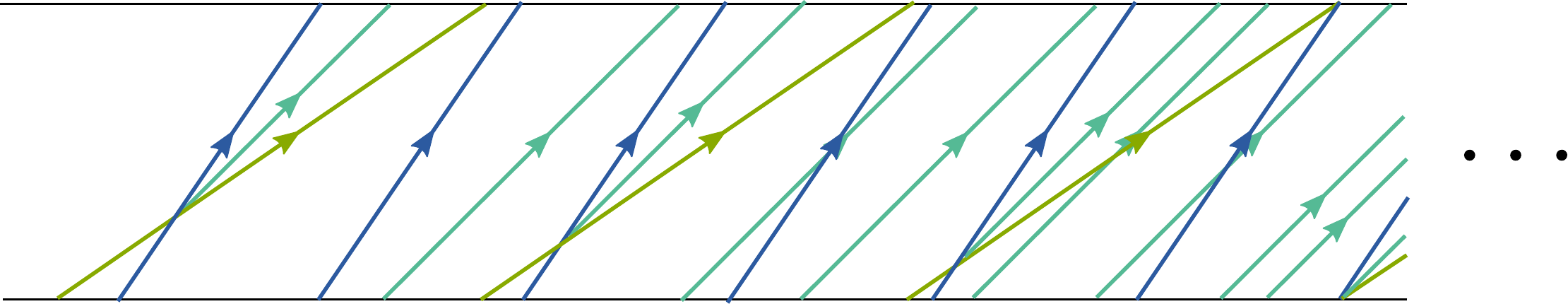}{0.45}{$\cS$-walls falling into a full defect, shown in the covering coordinate
$w = \log z$.  There is a single $ij$ wall and a single $jk$ wall.  We have $i < j < k$
in the ordering described in the text.
Each intersection between these two walls generates a new $ik$ wall.  There are infinitely many
such intersections, accumulating at the defect, which generate infinitely many $ik$ walls.}
Similar (but more involved) remarks should apply for more general types of defect.

Based on computer experimentation, we expect that the behavior observed in Figure \ref{fig:network-growth}
is indeed generic:  whenever there is a defect
with sufficiently generic mass parameters, and $\vartheta$ is generic, all $\cS$-walls in $\wnet_\vartheta$
should asymptotically
approach punctures.  In the case $K=2$ this follows directly from known mathematical results on the trajectories of
quadratic differentials \cite{MR743423} (see \cite{Gaiotto:2009hg} for an account of this.)
For $K>2$ the analogous foundational results
are not yet available as far as we know.  It would be very desirable to work this out.

\subsection{Homotopy invariance} \label{sec:homotopy-invariance}

Now let us consider a key consistency check of our story so far.
We have claimed on general physical grounds that $F(\wp, \vartheta)$ should be
a homotopy invariant of $\wp$, or more precisely a twisted homotopy invariant as
described in \S\ref{sec:tricky-sign}.
On the other hand, we have also given a recipe which completely determines $F(\wp, \vartheta)$.
So we can ask whether this recipe indeed obeys the necessary twisted homotopy invariance.

To check this twisted homotopy invariance it is enough to check the invariance under a few
elementary moves, which we now consider in turn.
This is the first place where the tricky
minus signs mentioned in \S\ref{sec:tricky-sign} play a decisive role.

First, consider a pair of paths $\wp$ and $\wp'$, neither of which meets any $\cS$-walls, and
which are not only homotopic but related by a \ti{regular}
homotopy, i.e. a homotopy through immersions.
In this case the constraint of twisted homotopy invariance requires that
\begin{equation}
 F(\wp, \vartheta) = F(\wp', \vartheta).
\end{equation}
But since neither path meets any $\cS$-walls, this reduces to
\begin{equation}
 \cD(\wp) = \cD(\wp'),
\end{equation}
which is indeed true:  the regular homotopy of $\wp$ to $\wp'$ lifts to
a regular homotopy of each $\wp^{(i)}$ to $\wp'^{(i)}$.

\insfig{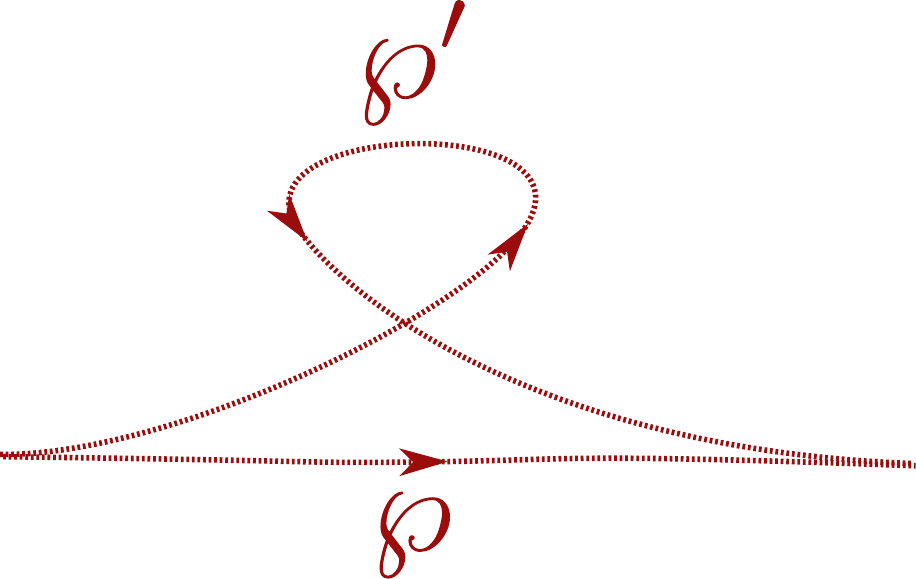}{A pair of paths that are homotopic but differ by a unit of winding.}
Next, consider the pair of paths in Figure \ref{fig:collapsing-loop}.
In this case the constraint of twisted homotopy invariance requires that
\begin{equation}
 F(\wp, \vartheta) = - F(\wp', \vartheta).
\end{equation}
But neither of these paths meets any $\cS$-walls, so this reduces to
\begin{equation}
 \cD(\wp) = - \cD(\wp'),
\end{equation}
which indeed follows by lifting the homotopy of $\wp$ to $\wp'$ to
homotopies of $\wp^{(i)}$ to $\wp'^{(i)}$ and using \eqref{eq:winding-rule-2}.
We stress that this is a non-regular homotopy, which accounts for the sign change.

Next, consider two paths $\wp$, $\wp'$ which differ by homotopy across an $\cS$-wall as
shown in Figure \ref{fig:s-wall-homotopy}.
\insfigscaled{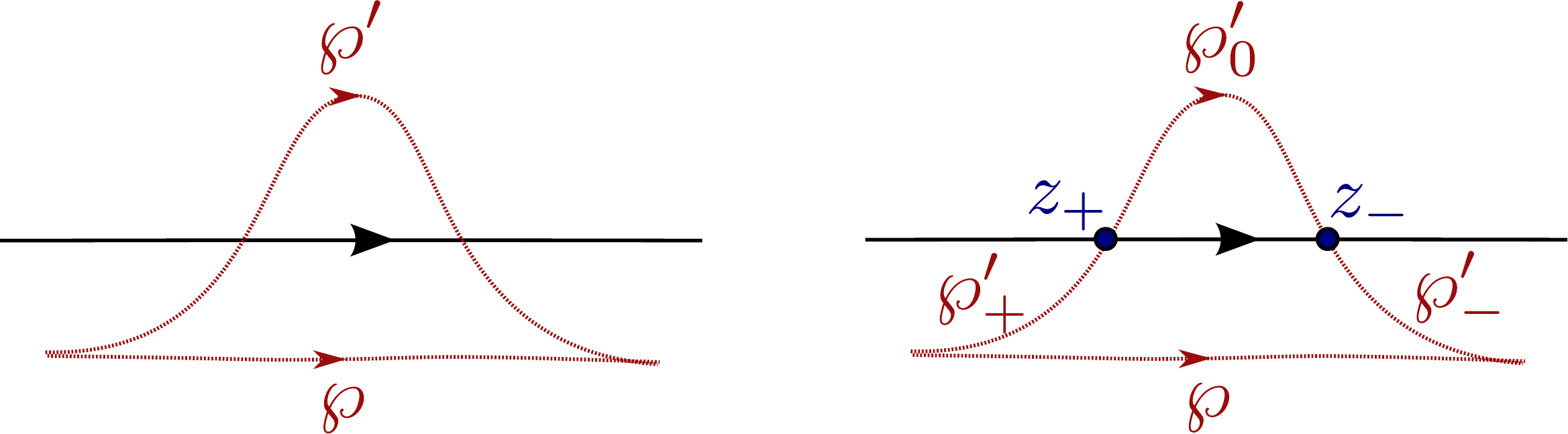}{0.36}{Two open paths $\wp$, $\wp'$ which differ by a homotopy across an $\cS$-wall.}
The constraint of twisted homotopy invariance says
\begin{equation} \label{eq:hom-inv-s}
F(\wp, \vartheta) = F(\wp', \vartheta).
\end{equation}
Let $\bar{a}$ be the charge supported on the $\cS$-wall in the figure.
The two intersections $z_\pm$ between $\wp'$ and the wall divide $\wp'$ into $\wp'_+ \wp'_0 \wp'_-$;
let $\tz_\pm$ be the tangent vectors to $\wp'$ at the intersection points.
Then evaluating both sides directly, letting $a$ be the charge supported along the wall,
and using the fact that $\mu(a)$ is constant along the wall,
\eqref{eq:hom-inv-s} becomes
\begin{align}\label{eq:homotop-check}
 \cD(\wp) &= \cD(\wp'_+) (1 + \mu(a) X_{a_{\tz_+}}) \cD(\wp'_0) (1 + \mu(a) X_{a_{\tz_-}}) \cD(\wp'_-) \nonumber \\
&= \cD(\wp') + \mu(a) \cdot \cD(\wp'_+) \left(X_{a_{\tz_+}} \cD(\wp'_0) + \cD(\wp'_0) X_{a_{\tz_-}}\right) D(\wp'_-) \\ & \qquad + \mu(a)^2 \cD(\wp'_+) X_{a_{\tz_+}}\cD(\wp'_0)X_{a_{\tz_-}} \cD(\wp'_-). \nonumber
\end{align}
The two soliton paths $a_{\tz_+}$ and $a_{\tz_-}$
are both of type $ij$, so $X_{a_{\tz_+}}\cD(\wp'_0)X_{a_{\tz_-}} = 0$.
Using the definition \eqref{eq:aztil-def} one can check that
the two terms in the parentheses correspond to paths which differ by one
unit of winding; according to the rule \eqref{eq:winding-rule-2},
these two terms thus differ by a minus sign,
and so cancel one another.
Thus \eqref{eq:homotop-check} reduces to
\begin{equation}
 \cD(\wp) = \cD(\wp')
\end{equation}
which is indeed true since the obvious homotopy of $\wp$ into $\wp'$ also gives homotopies between each
$\wp^{(i)}$ and $\wp'^{(i)}$.  Note that this would not have worked out without using the fact that $\mu(a)$ is constant
along the wall:  indeed this shows that homotopy invariance requires $\mu(a)$ to be constant along walls, thus making good
on a promise we made in \S\ref{sec:spectral-network}.

Next, consider two paths $\wp$, $\wp'$ which differ by homotopy across a joint,
as in Figure \ref{fig:trajectories-meeting-commuting} or Figure \ref{fig:ijkl-collisions}.
We have already verified in \S\ref{sec:wnet-basics} that in this case $F(\wp, \vartheta) = F(\wp', \vartheta)$:  indeed we used this constraint as part of our system for determining
$\mu$.

\insfig{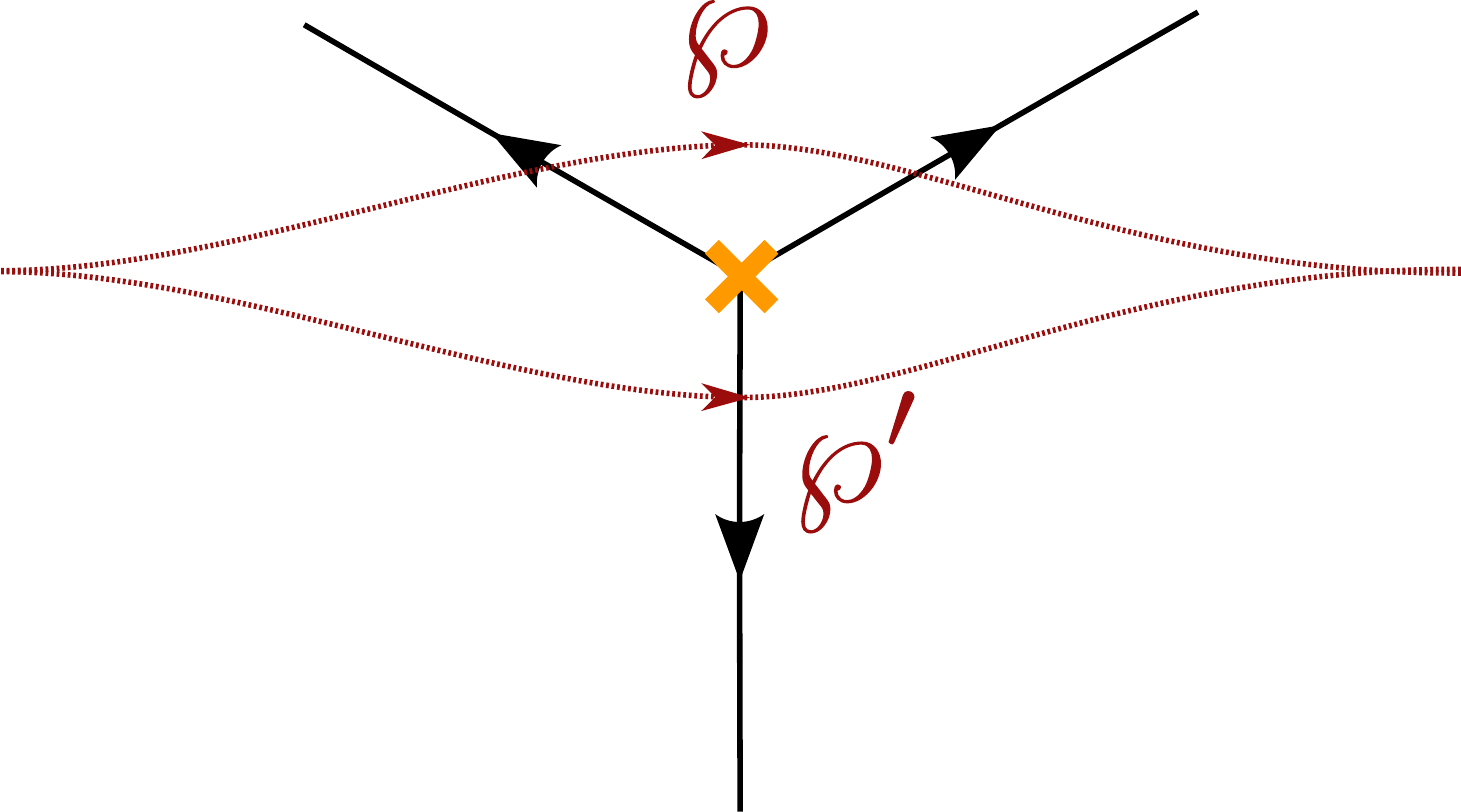}{Two paths which differ by homotopy across a branch point.}
Finally we reach the most interesting case.  Consider two paths $\wp$, $\wp'$ which
differ by homotopy across a branch point, as shown in Figure \ref{fig:branch-point-loop}.
The constraint of twisted homotopy invariance says
\begin{equation} \label{eq:hom-inv-bp}
F(\wp, \vartheta) = F(\wp', \vartheta).
\end{equation}
Evaluating both sides of \eqref{eq:hom-inv-bp}, we obtain the explicit sums of paths shown in Figure
\ref{fig:branch-point-monodromy}.  Naively we
see $K+3$ terms in $F(\wp, \vartheta)$
and $K+1$ in $F(\wp', \vartheta)$.
The last two terms in $F(\wp,\vartheta)$, running from sheet $j$ to sheet $j$,
are of the form $X_a + X_{a'}$, where
$a$, $a'$ both project to the same charge $\bar a$, but differ by one unit of winding.
These two terms thus cancel one another thanks to our sign rule
\eqref{eq:winding-rule-2}.
The other $K+1$ terms precisely match the $K+1$ terms in $F(\wp', \vartheta)$.
\insfigscaled{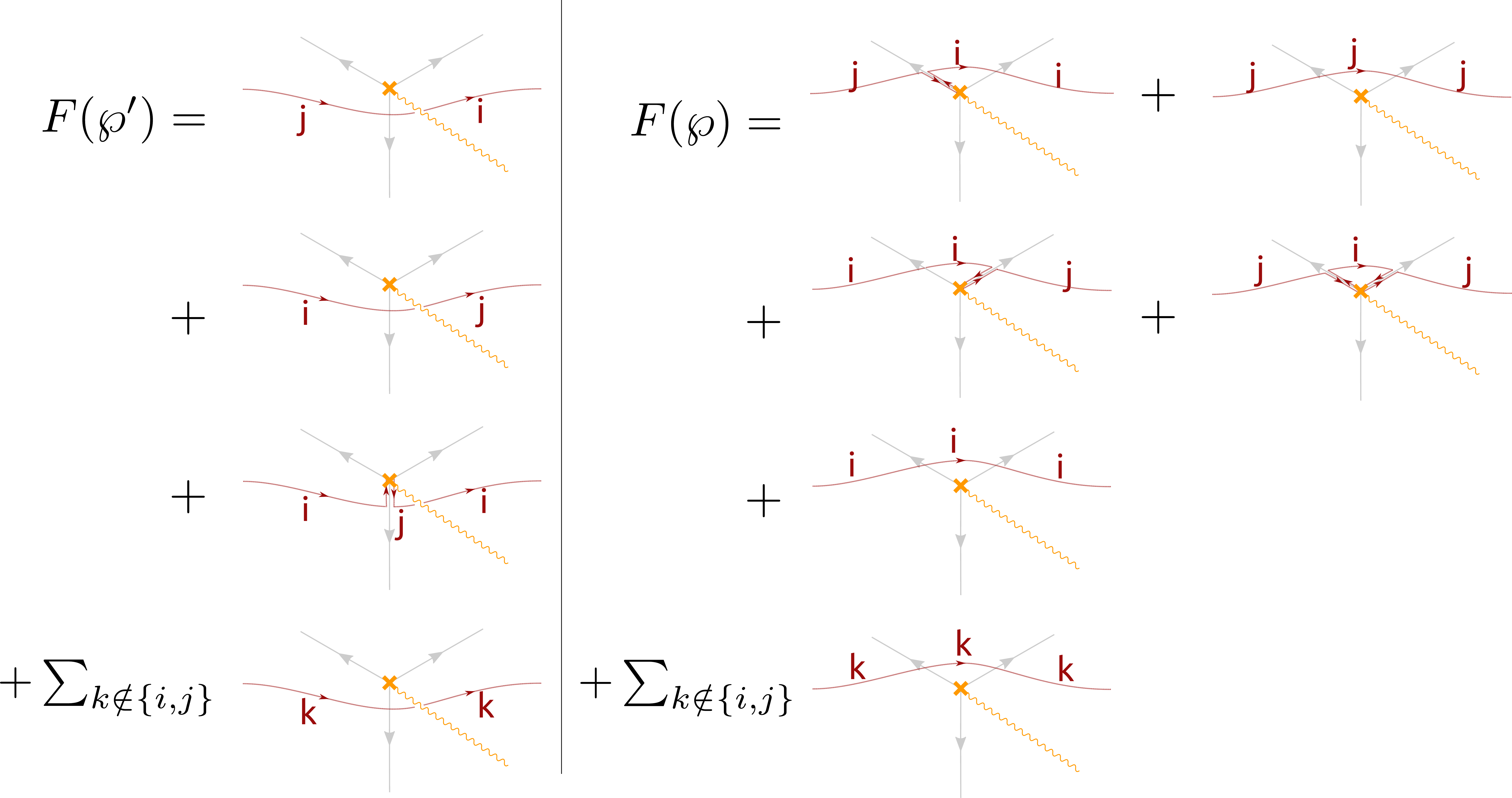}{0.19}{The explicit sums of paths which occur in \eqref{eq:hom-inv-bp}.  In
the equations shown in the figure, we represent the formal variable $X_a$ by an explicit
picture of a smooth path on $\Sigma$, whose canonical lift to $\tilde\Sigma$ represents the relative
homology class $a$.  The labels next to path segments show which sheet of the covering $\Sigma \to C$
the segments are on.  We have labeled the sheets in the same way as we did in Figure \ref{fig:branch-point-trajectories}.}  So the constraint \eqref{eq:hom-inv-bp} is indeed satisfied.

As an aside, note that this last
constraint would \ti{not} have worked out if there were no $\cS$-walls emerging from the branch point:  in other words, homotopy invariance of the framed 2d-4d BPS spectrum
really \ti{requires} that spectrum to undergo wall-crossing.  This is closely analogous to the
considerations of monodromy invariance that led Seiberg and Witten
to discover wall-crossing in the pure 4d BPS spectrum \cite{Seiberg:1994rs}.
The constraint of monodromy invariance
also would not have worked out if we had taken $\mu = -1$ instead of $\mu = +1$
in \eqref{eq:mu-near-bp}.

\subsection{$K=2$ theories and ideal triangulations} \label{sec:k2-examples}

\insfigscaled{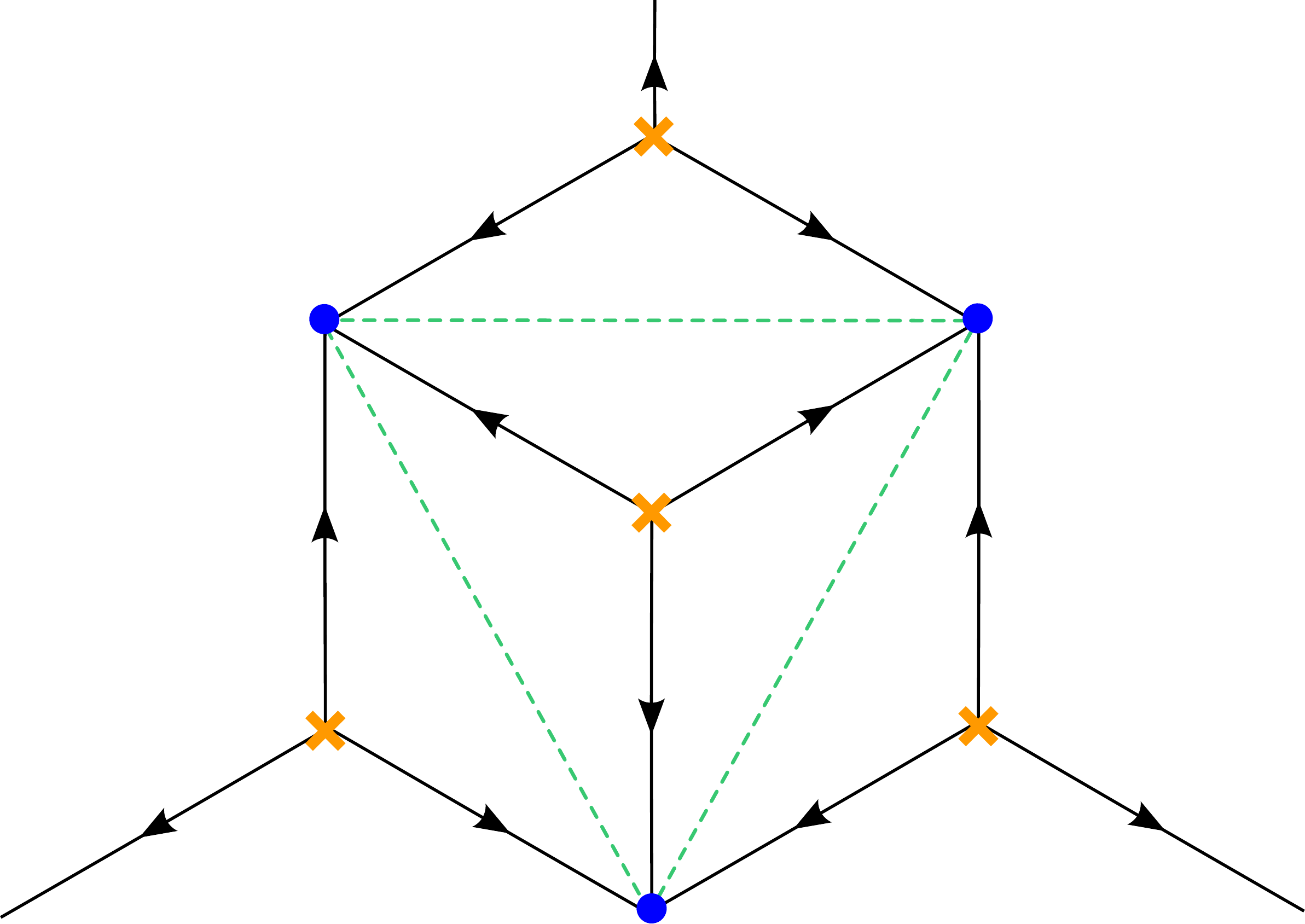}{0.28}{A portion of a spectral network $\wnet_\vartheta$
with $K=2$, and one face in
the corresponding ideal triangulation $T(\wnet_\vartheta)$ up to isotopy (green dashed lines).
Each wall of $\wnet_\vartheta$ begins on a branch point
(orange cross) and asymptotes to a defect (blue dot); these defects are
also the vertices of $T(\wnet_\vartheta)$.
Each branch point is contained in a unique face of $T(\wnet_\vartheta)$.}

In the special case $K=2$,
$\wnet_\vartheta$ is enormously simplified:
transverse intersections of $\cS$-walls as discussed in \S\ref{sec:wnet-basics}
require at least 3 distinct sheets $i$, $j$, $k$, and there is no room for this
if $K=2$.
So in this case there are no joints where new $\cS$-walls could be born, and so
$\wnet_\vartheta$ consists simply of three $\cS$-walls emerging from
each branch point.
Assuming that the parameters $m_a$ at the punctures $D_n$ are generic enough,
each of these $\cS$-walls asymptotes to one of the
punctures $D_n$.

When $K=2$, then, the dual to the network $\wnet_\vartheta$ is
an ideal triangulation $T(\wnet_\vartheta)$
of $C$ (determined up to isotopy).  See Figure \ref{fig:triangulation-vs-network}.
Combinatorially speaking,
the ideal triangulation $T(\wnet_\vartheta)$ and the spectral network $\wnet_\vartheta$
contain the same information.

It has been natural to wonder what is the appropriate higher-$K$
generalization of the notion of ideal triangulation.  In \S\ref{sec:spectral-networks} below we propose
a general definition of ``spectral network'' which we believe is the right answer to
this question.  The $\wnet_\vartheta$ which we have been discussing so far
are examples of spectral networks.

\section{Varying $\vartheta$} \label{sec:varying-theta}

In \S\ref{sec:wnet} above, we held $\vartheta$ fixed and studied the generating functions $F(\wp,\vartheta)$.  In this section we consider what happens when
$\vartheta$ is allowed to vary.  In so doing we will uncover new phenomena,
associated with the 4d BPS states in the theory $S[\fg,C,D]$.

$F(\wp,\vartheta)$ is piecewise constant as a function of $\vartheta \in \R$:
if the pair $(\wp,\vartheta)$ is generic and $\delta \vartheta$ small enough,
$F(\wp,\vartheta) = F(\wp,\vartheta+\delta\vartheta)$.
However, $F(\wp,\vartheta)$ jumps at some special values
$\vartheta = \vartheta_c$.
These special values arise for two distinct reasons,
to be described in the next two sections.

At the special values, we define for convenience\footnote{The meaning of the limit \eqref{eq:f-lim} is easy to understand if there are no other special values
$\vartheta_c'$ in some neighborhood of $\vartheta_c$.  However, it does sometimes happen that the
special values accumulate.  In that case we have to explain what kind of limit we mean in \eqref{eq:f-lim}.
One possibility would be to fix some $R>0$ and work with a
``$Z$-adic'' norm where $\norm{\sum c_a X_{a}} = \sum \abs{c_a} e^{-R \abs{Z_{\bar a}}}$.  We expect that
in this norm the limit \eqref{eq:f-lim} indeed exists.  At any rate, in what follows we work formally,
assuming that the limits make sense.}
\begin{equation} \label{eq:f-lim}
F(\wp, \vartheta_c^\pm) = \lim_{\vartheta \to \vartheta_c^\pm} F(\wp, \vartheta)
\end{equation}
(think of $\vartheta_c^\pm$ as representing $\vartheta_c \pm \varepsilon$ for
$\varepsilon \to 0^+$.)
What we will explain in the rest of this section is how to determine the relation
between $F(\wp, \vartheta_c^+)$ and $F(\wp, \vartheta_c^-)$, and how to extract
from this relation the 4d BPS degeneracies.

\subsection{Endpoints crossing $\cS$-walls}

First, there are ``simple'' jumps which occur when an $\cS$-wall
moves across one of the endpoints of $\wp$.  Namely:
$F(\wp,\vartheta)$ jumps whenever there is some
$\bar a \in \Gamma(z_1,z_1)$ or $\bar a \in \Gamma(z_2,z_2)$
with $e^{-\I \vartheta_c} Z_{\bar a} \in \R_-$ and $\mu(\bar a) \neq 0$.
If $\bar a \in \Gamma(z_1,z_1)$ and $\wp$ intersects the $\cS$-wall for
$\vartheta_c + \varepsilon$ then the jump is of the form
\begin{equation} \label{eq:s-jump-1}
 F(\wp,\vartheta_c^+) = (1 + \mu(a) X_{a_{\tilde z_1}}) F(\wp,\vartheta_c^-).
\end{equation}
Similarly,  if $\bar a \in \Gamma(z_2,z_2)$ and $\wp$ intersects the $\cS$-wall for
$\vartheta_c + \varepsilon$ then the jump is of the form
\begin{equation} \label{eq:s-jump-2}
 F(\wp,\vartheta_c^+) = F(\wp,\vartheta_c^-) (1 + \mu(a) X_{a_{\tilde z_2}}).
\end{equation}
The jumps \eqref{eq:s-jump-1}, \eqref{eq:s-jump-2} are easily
deduced from our rule \eqref{eq:S-factor-simple} for
computing $F(\wp,\vartheta)$ in \S\ref{subsec:Compute-FP}.

These equations can be compared to (2.27) of \cite{Gaiotto:2011tf}.
The signs appear different because in \cite{Gaiotto:2011tf} there is a cocycle $\sigma$
which was undetermined in that section.

\subsection{$\cK$-walls and the degenerate spectral networks $\wnet_{\vartheta_c}$}\label{sec:K-walls}

There is also a second, more interesting kind of jump which can occur at a critical phase
$\vartheta = \vartheta_c$.  In contrast to the previous case, these jumps do not arise because of
a change in the interaction between $\wnet_\vartheta$ and $\wp$.  Rather, they arise because of
a topology change in $\wnet_\vartheta$ itself.

\insfigscaled{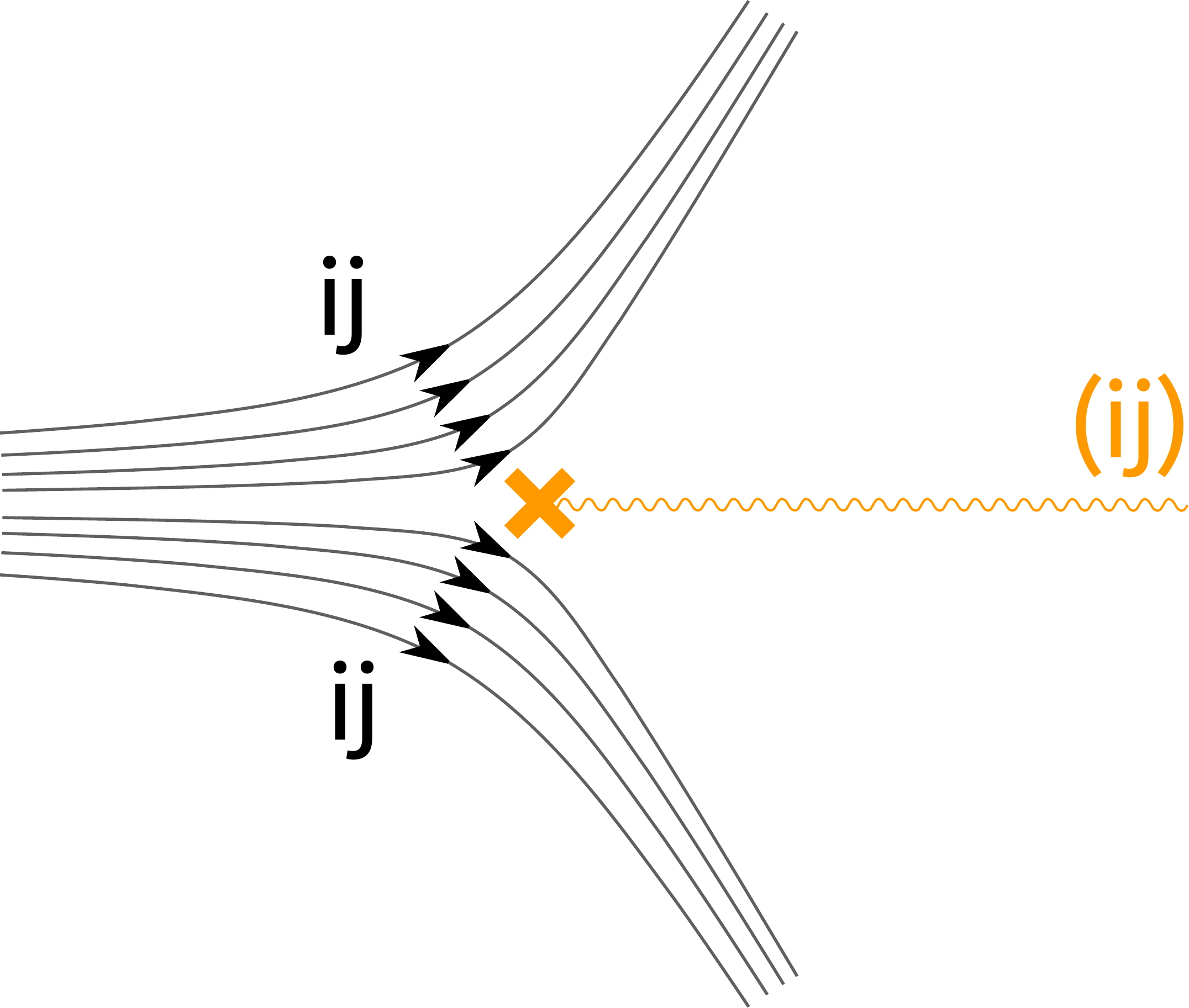}{0.18}{The behavior of $ij$-trajectories
approaching a branch point of type $(ij)$.}
The claim that a topology change in $\wnet_\vartheta$
could occur as we continuously vary $\vartheta$ might at
first seem strange.  After all, we have given a recipe for $\wnet_\vartheta$ in \S\ref{sec:wnet},
$\vartheta$ enters this recipe only
through the differential equation \eqref{eq:trajectory-de}, and this equation
evidently depends continuously on $\vartheta$.
The point is that the \ti{solutions} of \eqref{eq:trajectory-de}
can nevertheless exhibit discontinuous behavior,
because of the phenomenon of bifurcation near the branch points:  $ij$- or $ji$-trajectories coming close to
an $(ij)$ branch point can veer off in one of two directions, as indicated in Figure \ref{fig:near-branch-point}.
This figure can be considered to depict either a foliation of $C$ by $ij$-trajectories at fixed $\vartheta$
or the evolution of an $\cS$-wall in $\wnet_\vartheta$ as $\vartheta$ varies. From the latter
viewpoint it follows that the critical phases $\vartheta_c$ are those
at which some $\cS$-wall of type $ij$ or $ji$ runs directly into a branch point of type $(ij)$.
We call the $\vartheta_c$ where this occurs \ti{$\cK$-walls}.
A simple example is shown in Figure \ref{fig:wnet-jump}.

\insfigscaled{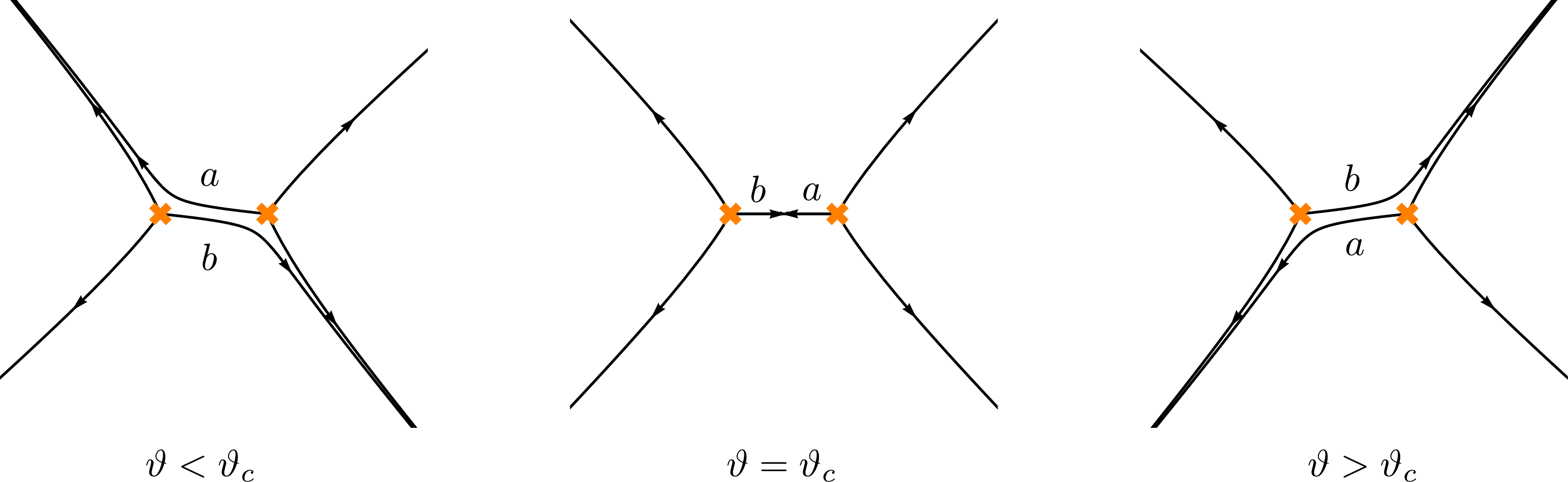}{0.41}{The simplest way in which $\wnet_\vartheta$ can jump at a $\cK$-wall.
As $\vartheta \to \vartheta_c$, two $\cS$-walls merge into a single two-way street
running between two branch points.  The two-way street supports two distinct charges $\bar a$, $\bar b$.}

To describe the relation between $F(\wp, \vartheta_c^+)$ and $F(\wp, \vartheta_c^-)$,
it is convenient to work directly with the limiting spectral network $\wnet_{\vartheta_c}$.
This limiting network has some features not seen for the generic $\wnet_\vartheta$.
In particular, several (possibly infinitely many) $\cS$-walls, supporting different
charges, might coalesce into a single segment as $\vartheta \to \vartheta_c$; hence a wall
$p$ of $\wnet_{\vartheta_c}$ might support several distinct charges.
Moreover, the $\cS$-walls coalescing onto $p$
might not all be oriented in the same direction; in this case we call $p$ a \ti{two-way street}.\footnote{We now
have two metaphors for the segments in
a spectral network.  They are \emph{walls} because they are the loci
where framed 2d-4d BPS degeneracies jump.  On the other hand
we will find that the metaphor of \emph{streets} is also very useful when tracking
the solitons.  Rather than insisting on one term, we will use them as synonyms.
We feel no inclination to go to wall-street.\label{fn:wall-street}}
Again see Figure \ref{fig:wnet-jump} for an example:  the saddle connection in the middle
of the figure is a two-way street.  Indeed, $\wnet_{\vartheta_c}$ always contains
at least one two-way street, since we get a two-way street whenever an $\cS$-wall
runs into a branch point.

\insfigscaled{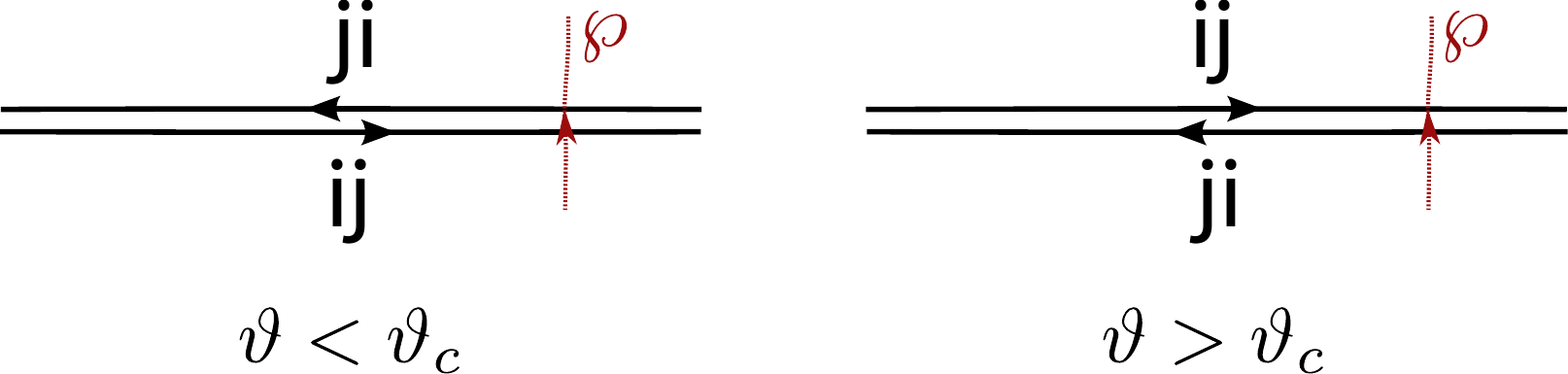}{0.45}{A path $\wp$ crossing a pair of nearly coincident
$\cS$-walls, which coalesce into a two-way street at $\vartheta = \vartheta_c$.}
Now let us consider a path $\wp$ which crosses a two-way street $p$.  A representative picture of the situation
when $\vartheta$ is near $\vartheta_c$ is indicated in Figure \ref{fig:crossing-two-way}.\footnote{In general the
picture might be a bit more complicated than Figure \ref{fig:crossing-two-way}, because there might be several walls
of type $ij$ and/or of type $ji$ which all coalesce at $\vartheta = \vartheta_c$.
An example appears in Figure \ref{fig:closed-trajectories} below.}
We define BPS soliton degeneracies $\mu^\pm(a,p)$ along the
walls of $\wnet_{\vartheta_c}$ by taking limits of $\mu(a,p)$
as $\vartheta \to \vartheta_c^\pm$.
Applying the rules of \S\ref{subsec:Compute-FP} and taking the limit $\vartheta \to \vartheta_c^\pm$,
we obtain formulas for $F(\wp, \vartheta_c^\pm)$:
\begin{align}
 F(\wp, \vartheta_c^-) &= \cD(\wp_+)
 \left(1 + \sum_{\bar b \in \Gamma^s_{ij}} \mu^-(b) X_{b_{\tilde z}}\right)
 \left(1 + \sum_{\bar a \in \Gamma^s_{ji}} \mu^-(a) X_{a_{\tilde z}}\right) \cD(\wp_-),\label{eq:two-way-crossing-2} \\
 F(\wp, \vartheta_c^+) &= \cD(\wp_+)
 \left(1 + \sum_{\bar a \in \Gamma^s_{ji}} \mu^+(a) X_{a_{\tilde z}}\right)
 \left(1 + \sum_{\bar b \in \Gamma^s_{ij}} \mu^+(b) X_{b_{\tilde z}}\right) \cD(\wp_-).\label{eq:two-way-crossing-1}
\end{align}
Here $\Gamma^s_{ij} \subset \Gamma_{ij}(z,z)$ denotes the set of charges supported along the two-way street.

Note that the order in these products really matters, since $X_{a_{\tilde z}}$ and $X_{b_{\tilde z}}$ do not
commute.
A convenient way to think about this is to regard the limiting network
$\wnet_{\vartheta_c}$ as equipped with a bit of extra structure:  each two-way street is
resolved into two infinitesimally separated and oppositely oriented ``lanes''.
For $\vartheta_c^-$ the division of lanes is according to the American rule (drive on the right),
while for $\vartheta_c^+$ it is according to the British rule (drive on the left).
We can determine $F(\wp, \vartheta_c^\pm)$ completely
from $\wnet_{\vartheta_c}$ and $\mu^\pm$,
using either \eqref{eq:two-way-crossing-2} or \eqref{eq:two-way-crossing-1} as appropriate.

\subsection{The jump of $F(\wp, \vartheta)$ at a $\cK$-wall} \label{sec:f-jump}

The jump of $F(\wp, \vartheta)$ at $\vartheta_c$ is given by a certain universal substitution $\cK$ acting on the formal variables $X_a$:
\begin{equation} \label{eq:k-jump}
F(\wp, \vartheta_c^+) = \cK\left(F(\wp, \vartheta_c^-)\right).
\end{equation}
$\cK$ is determined by the degenerate spectral network $\wnet_{\vartheta_c}$.
In this section we describe what $\cK$ is,
deferring the proof of \eqref{eq:k-jump} to \S\ref{sec:k-proof}.

We are going to
combine the soliton degeneracies on any two-way street $p$ into a new generating function $Q(p)$.
Unlike the generating functions $F(\wp, \vartheta)$ we have considered before, $Q(p)$ is
written in terms of formal variables $X_{\tilde \gamma}$
with $\tilde\gamma \in \tilde\Gamma$,
where $\tilde\Gamma = H_1(\tilde\Sigma; \IZ)$ modulo shifts by $2H$ (cf. the definition of
$\tilde\Gamma(\tz_1, \tz_2)$ in \S\ref{sec:tricky-sign}).
So these formal variables are associated to closed paths, rather than the open ones we have
encountered up to now.
We extend our multiplication rules,
\begin{equation}
 X_{\tilde \gamma} X_a = X_{\tilde \gamma + a}, \qquad X_{\tilde \gamma} X_{\tilde \gamma'} = X_{\tilde \gamma + \tilde \gamma'}
\end{equation}
(where the $+$ denotes the obvious action of $\tilde\Gamma$ on $\tilde\Gamma(\tz_1, \tz_2)$ or on $\tilde\Gamma$).
We also extend our sign rules by imposing
\begin{equation}
 X_H = -1,
\end{equation}
so the two $X_{\tilde\gamma}$ corresponding to different lifts of a single $\gamma \in \Gamma$ differ only by a sign.

One can build closed paths from open ones:  given $a \in \tilde \Gamma_{ii}(\tz,\tz)$,
there is a corresponding $\cl(a) \in \tilde \Gamma$ which is obtained just by forgetting the
basepoint $\tz^{(i)}$.
Using this we define
\begin{equation} \label{eq:def-Q}
Q(p) = 1 + \sum_{\bar a \in \Gamma_{ij}^s,\,\bar b \in \Gamma_{ji}^s} \mu^-(a,p) \mu^-(b,p) X_{\cl(a + b)}.
\end{equation}
(We will see later that we would have gotten the same $Q(p)$ if we had used
$\mu^+$ instead of $\mu^-$ on the right.)
$Q(p)$ is a power series in the variables $X_{\tilde\gamma}$, of a constrained sort: every
$\gamma$ that occurs is a sum $\bar a + \bar b$ of charges supported at $z$, and
hence has $e^{- \I \vartheta_c} Z_\gamma \in \R_-$.

Now we make a new genericity assumption.  Let $\Gamma_c \subset \Gamma$ be the set of all $\gamma \in \Gamma$
with $e^{- \I \vartheta_c} Z_\gamma \in \R_-$; we assume
that $\Gamma_c$ is generated by a single element $\gamma_0$.  (This condition holds automatically if our chosen
$u = (\phi_2, \cdots, \phi_K)$ is not on a wall of marginal stability in the
Coulomb branch.)  $Q(p)$ is then a power series in a single variable:
choosing a lift $\tilde\gamma_0$,
each $X_{\cl(a+b)} = \pm X_{\tilde \gamma_0}^n$
for some $n>0$.
Our next aim is to extract the BPS degeneracies by rewriting this power series as a product, \eqref{eq:Q-exp} below.

For each $\gamma \in \Gamma$ we can define a preferred lift $\tilde\gamma$ by the following rule:\footnote{In the first
preprint version of this paper we proposed a different lifting rule based on the principle that the product \eqref{eq:Q-exp}
should be finite.  More recently, in studying more complicated examples,
it has turned out that sometimes this finiteness is violated \cite{glmmn}.  For completeness
we have back-ported the corrected rule from \cite{glmmn} to here.}
represent $\gamma$ as a union of smooth closed curves $c_m$ on $\Sigma$; then $\tilde\gamma$ is the sum of
the canonical lifts of the $c_m$ to $\tilde\Sigma$, shifted by $\left( \sum_{m \le n} \delta_{mn} + \# (c_m \cap c_n) \right) H$
(of course since we work modulo $2H$, all that matters here is whether this sum is odd or even.)
One can check directly that $\tilde\gamma$ so defined is independent of the choice of how we represent $\gamma$
as a union of $c_m$ (this requirement is what forced us to add the tricky-looking shift.)
Moreover, with this definition one has
\begin{equation}
 X_{\tilde\gamma + \tilde\gamma'} = (-1)^{\inprod{\gamma, \gamma'}} X_{\widetilde{\gamma + \gamma'}}.
\end{equation}

We are ready to factorize $Q(p)$:
there exist exponents $\alpha_{\gamma}(p)\in\Z$ such that
\begin{equation} \label{eq:Q-exp}
Q(p) = \prod_{\gamma \in \Gamma_c} (1 - X_{\tilde \gamma})^{\alpha_\gamma(p)}.
\end{equation}
(Indeed, the equations determining the $\alpha_\gamma(p)$ from $Q(p)$ are
upper-triangular and hence can be solved.)
Then, for any $\gamma \in \Gamma_c$, define a 1-chain $L(\gamma)$ on $\Sigma$ by
\begin{equation} \label{eq:def-L}
L(\gamma) = \sum_p \alpha_\gamma(p)\,p_\Sigma,
\end{equation}
where $p$ runs over the walls in $\wnet_{\vartheta_c}$,
and $p_\Sigma$ is the oriented $1$-chain obtained by lifting $p$
as in \S\ref{subsec:4d-BPS-States}.

It is a crucial fact (proven in \S\ref{sec:k-proof}
below) that $L(\gamma)$ so defined is actually a 1-\ti{cycle}.
This allows us to define\footnote{Different representatives
of the class $\bar c$ differ by addition of 1-boundaries, and the 1-cycle $L(\gamma)$ has zero
intersection with any 1-boundary; this would not have worked if $L(\gamma)$ were merely
a 1-chain.}
an intersection number $\inprod{\bar c, L(\gamma)}$ for any $\bar c \in \Gamma(z_1, z_2)$.
Moreover, $e^{-\I \vartheta_c} \int_{p_\Sigma} \lambda \in \R_-$,
so $e^{-\I \vartheta_c} Z_{[L(\gamma)]} \in \R$;
under our genericity assumption this implies that the homology class
$[L(\gamma)]$ is a multiple of $\gamma$ (though not necessarily a positive multiple).
This fact will be useful below.

Finally we can define our universal substitution $\cK$:
\begin{equation} \label{eq:k-def}
 \cK(X_a) = \prod_{\gamma \in \Gamma_c} (1 - X_{\tilde\gamma})^{\inprod{\bar a,L(\gamma)}} X_a.
\end{equation}

\subsection{4d BPS degeneracies}\label{subsec:4d-degen}

According to the analysis of \cite{Gaiotto:2010be,Gaiotto:2011tf}, the jump $\cK$ of the framed 2d-4d degeneracies
captures the degeneracies of 4d
BPS particles.\footnote{This is possible because the jump occurs when framed 2d-4d BPS bound states
form/decay by binding/releasing 4d BPS particles.}  Since we have now given a formula for $\CK$,
it follows that we can use it to determine the 4d BPS degeneracies.  Comparing our formula \eqref{eq:k-def}
to those of \cite{Gaiotto:2010be,Gaiotto:2011tf}, we find that the enhanced BPS degeneracies reviewed in \S\ref{sec:enh-4d}
are here given by\footnote{More precisely, \eqref{eq:k-def} should be
compared with (2.30) of \cite{Gaiotto:2011tf}, except for the detail that in this paper we are working
with framed protected spin characters at $y=1$ rather than $y=-1$.  The needed modification for $y=1$ has not
quite appeared anywhere before, although
in (3.26) of \cite{Gaiotto:2010be} we did give the jump of the  framed degeneracies (without surface defects) at $y=1$.
We also noted there (in \S6.4)
that under the assumption that there are no ``exotic BPS states,'' the jump becomes somewhat simpler
(because then all of the factors $(-1)^m$ appearing in (3.26) are the same).
Strictly speaking then, \eqref{eq:k-def} should be compared with the most obvious combination of (2.30) of \cite{Gaiotto:2011tf} and (3.26) of \cite{Gaiotto:2010be},
taking into account this simplification.
A final detail:  in (2.30) of \cite{Gaiotto:2011tf} there appeared an undetermined $\pm 1$-valued cocycle $\sigma$;
in \eqref{eq:k-def} this sign is encoded in the choice of lift $\tilde\gamma$ of $\gamma$.}
\begin{equation} \label{eq:result-omega}
 \omega(\gamma, \bar a) = \inprod{L(\gamma), \bar a}.
\end{equation}

This result contains much more information
than the ordinary 4d BPS degeneracies:  it knows not only the total number of 4d BPS states but
also some local information about where they sit on $C$, as measured by their interaction with
surface defects.
Nevertheless it is also interesting to see how we can
recover simpler invariants:  using \eqref{eq:result-omega} and \eqref{eq:Omega-def},
we find that the 4d BPS degeneracy is
\begin{equation} \label{eq:result-Omega}
 \Omega(\gamma) = [L(\gamma)] / \gamma.
\end{equation}
In examples below, we will illustrate how this formula determines $\Omega(\gamma)$ in various
concrete situations.

\subsection{Wall-crossing formula}\label{subsec:2d4d-wcf}

In \S2.3 of \cite{Gaiotto:2011tf}, we wrote a \ti{2d-4d wall-crossing formula}
which should be obeyed by the degeneracies $\mu$ and $\omega$ in any coupled 2d-4d system.
We have given an explicit description of $\mu$ and $\omega$ in the theories we are now considering.
So it is natural to ask whether one can show directly that the $\mu$ and $\omega$ defined here indeed obey this formula.
In a sense, our definition of $\omega$ above was engineered so that this will be true.
We will not be able to give a complete proof here, but let us at least explain the idea.

The basic strategy of proof was already explained in \cite{Gaiotto:2011tf},
as follows.  Let $u = (\phi_2, \dots, \phi_K)$ denote a point of the Coulomb branch.
Let $\wp$ denote a path in $C$ which both begins and ends at
some point $z$.  As we vary the basepoint $z$ we can likewise
deform the path $\wp$ (in a unique way up to homotopy), giving a family of paths $\wp(z)$.
Now let $z$, $\vartheta$ and $u$ vary along small contractible loops $z(t)$, $\vartheta(t)$, $u(t)$
$(0 \le t \le 1)$ in parameter space, with $z(0) = z$,
and consider the corresponding generating functions
$F(\wp(z(t)), \vartheta(t), u(t))$,\footnote{Up until now we have usually represented
these generating functions as $F(\wp, \vartheta)$, holding $u$ fixed and implicit;
but to recover the usual statement of the wall-crossing
formula we have to let $u$ vary.} which we denote simply as $F(\wp, t)$.

Since $t$ is a closed contractible loop we have
\begin{equation} \label{eq:F-conserved}
 F(\wp, t=0)\ =\ F(\wp, t=1).
\end{equation}
On the other hand, as $t$ varies, $F(\wp, t)$ jumps at various critical $t_c$.
These jumps occur either when $z(t)$ crosses an $\cS$-wall in $\wnet_{\vartheta(t)}$
or when $\vartheta(t)$ crosses a $\cK$-wall.  At each critical $t_c$,
the jump of $F(\wp, t)$ is given by an explicit transformation $J(t_c)$ of the formal variables $X_a$.
Writing the total jump as ${\bf J} = \prod_{t_c} J(t_c)$ (with the product taken in order of increasing $t_c$),
\eqref{eq:F-conserved} thus says
\begin{equation} \label{eq:j-f}
 {\bf J}[ F(\wp, t=0) ]\ =\ F(\wp, t=0).
\end{equation}

Now, the statement of the 2d-4d wall-crossing formula is that  ${\bf J}$ is the identity, or more concretely,
\begin{equation} \label{eq:2d4d-wcf}
 {\bf J}[X_a] = X_a \text { for every } a \in \Gamma(z,z).
\end{equation}

The most direct way to obtain \eqref{eq:2d4d-wcf} from \eqref{eq:j-f} would be to show that
any $X_a$ for $a \in \Gamma(z,z)$
can be obtained as a linear combination of the $F(\wp, t=0)$ for various paths $\wp$
from $z$ to $z$.  We have not proven this, and indeed (as one sees by considering simple examples)
it cannot literally be true except under some
restrictions on the type of surface defects $D_n$ we allow; in general we expect to have to extend the set of
allowed $\wp$ to include some paths which run into the surface defects, as we did in \cite{Gaiotto:2010be}.

Even after extending the set of allowed $\wp$ appropriately, it does not appear to be straightforward
to show that any $X_a$ can be obtained as a linear combination of the $F(\wp, t=0)$.  We believe this is
an interesting and important question (related to conjectures of Fock-Goncharov on the relation between
universally positive Laurent polynomials and tropical points; we discussed the $K=2$ case of this connection
in \cite{Gaiotto:2010be}.)
However, we can also propose an alternative ``poor man's'' approach to proving \eqref{eq:2d4d-wcf}.
The idea is that even if we
cannot show that we can express $X_a$ literally as a function of the $F(\wp, t=0)$, we can at least
do so up to at most a finite ambiguity.  More precisely, we claim that
the group $G$ of automorphisms obeying the equation \eqref{eq:j-f} is \ti{finite}.
This point will be explained in \S\ref{sec:dimensions} below.  Assuming it for now, we conclude in particular that ${\bf J}$ is of finite order.
But the automorphisms $J(t_c)$ are all ``upper triangular'', in the sense that each $J(t_c)$
is of the form $1 + X$ where all of the $X$ belong to a common pronilpotent group.
It follows that ${\bf J}$ is also of this form; but then
${\bf J}$ cannot have finite order without being the identity.

\subsection{Proof of the $\cK$-wall formula} \label{sec:k-proof}

Here we provide the proofs omitted in \S\ref{sec:f-jump}.

First let us explain why $L(\gamma)$ as defined in \eqref{eq:def-L} is indeed a 1-cycle,
i.e. has no boundary.
For any two-way street $p$, the boundary of $p_\Sigma$ lies over the boundary of $p$.
More precisely, a boundary point of $p$ may be either a branch point or a joint;
$p_\Sigma$ has no boundary over a branch point, but does have a boundary over a joint.
So we need to check that $L(\gamma)$ has no boundary over a joint.
To establish this we consider Figure \ref{fig:six-way-junction}.
\insfig{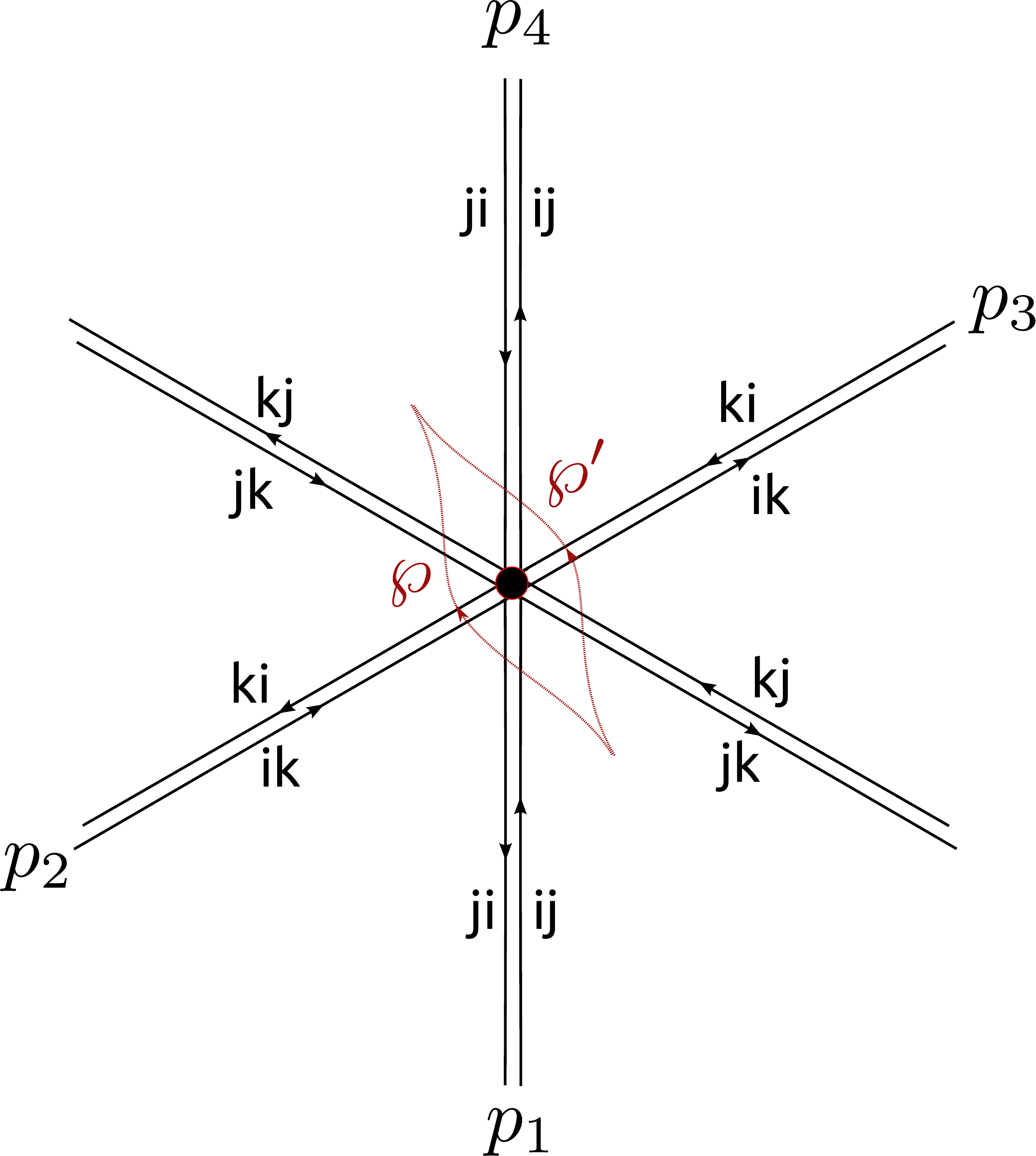}{A joint where two-way streets meet.  Each two-way street is shown
slightly ``resolved'', as explained at the end of \S\ref{sec:K-walls}.  The resolution here
is the ``American resolution,'' as appropriate since we are considering the limit
$\vartheta \to \vartheta_c^-$.}
Let $z$ denote the joint in the figure.
The coefficient of $z^{(i)}$ in $\partial L(\gamma)$ is
$\alpha_\gamma(p_1) + \alpha_\gamma(p_2) - \alpha_\gamma(p_3) - \alpha_\gamma(p_4)$, and
we would like to show that this vanishes.  Consider the open paths $\wp$ and $\wp'$.
The constraint of twisted homotopy invariance
says that $F(\wp, \vartheta_c^-) = F(\wp', \vartheta_c^-)$.
In particular, we can look at the pieces on both sides which involve paths which
both begin and end on sheet $i$:
\begin{equation}\label{eq:Fpm-equals}
F(\wp, \vartheta_c^-)_{ii} = F(\wp', \vartheta_c^-)_{ii}.
\end{equation}
Directly computing the two sides of \eqref{eq:Fpm-equals} we obtain
\begin{equation}
 Q(p_1) Q(p_2) = Q(p_3) Q(p_4).
\end{equation}
This amounts to
\begin{equation}
 \alpha_\gamma(p_1) + \alpha_\gamma(p_2) = \alpha_\gamma(p_3) + \alpha_\gamma(p_4),
\end{equation}
which is what we wanted to show.

Now let us explain why the jump formula \eqref{eq:k-jump} is true.
We will show that
the following hold both
for $\CF(\wp) = F(\wp, \vartheta_c^+)$ and $\CF(\wp) = \cK\left(F(\wp, \vartheta_c^-)\right)$:
\begin{description}
 \item[P1.] $\CF(\wp)$ is a twisted homotopy invariant of $\wp$ (in the sense explained
in \S\ref{sec:tricky-sign}).
 \item[P2.] If $\wp$ does not meet $\wnet_{\vartheta_c}$, then
\begin{equation}
 \CF(\wp) = \CD(\wp).
\end{equation}
 \item[P3.] If $\wp$ and $\wp'$ have endpoints off $\wnet_{\vartheta_c}$, $\CF$ obeys the composition law
\begin{equation}
 \CF(\wp) \CF(\wp') = \CF(\wp \wp').
\end{equation}
 \item[P4.] If $\wp$ crosses $\wnet_{\vartheta_c}$ exactly once at a point $z$ on a one-way street $p$, then $\CF(\wp)$ is of the form
\begin{equation} \label{eq:f-one-way}
 \CF(\wp) = \CD(\wp_+) \left(1 + \sum_{\bar a \in \Gamma_{ij}^s} \mu(a) X_{a_\tz}\right) \CD(\wp_-),
\end{equation}
for some $\mu(a) \in \Z$.
\item[P5.] If $\wp$ crosses $\wnet_{\vartheta_c}$ exactly once at a point $z$ on a two-way street $p$, and the intersection between $\wp$ and $p$
is positive (with respect to the orientation of $p$ as a $ji$ trajectory and the underlying orientation of $C$ as a complex curve),
then $\CF(\wp)$ is of the form
\begin{equation} \label{eq:f-two-way}
 \CF(\wp) = \CD(\wp_+) \left(1 + \sum_{\bar a \in \Gamma_{ji}^s} \mu(a) X_{a_\tz}\right) \left(1 + \sum_{\bar b \in \Gamma_{ij}^s} \mu(b) X_{b_\tz} \right) \CD(\wp_-)
\end{equation}
for some $\mu(a) \in \Z$ and $\mu(b) \in \Z$.
\end{description}

For $\CF(\wp) = F(\wp, \vartheta_c^+)$ these properties follow directly from what we have already said; the most nontrivial one is
P5, which is \eqref{eq:two-way-crossing-1}.  To prove them for
$\CF(\wp) = \CK(F(\wp, \vartheta_c^-))$ is slightly harder.
P1-3 are true of $F(\wp, \vartheta_c^-)$ and clearly preserved by $\CK$.
P4 is also true of $F(\wp, \vartheta_c^-)$, but we have to show it is preserved by $\CK$:
this follows simply from the fact that $\CK$ just multiplies each term by a function of the $X_{\tilde \gamma}$,
which indeed preserves the form \eqref{eq:f-one-way}.  P5 is the only really nontrivial one.  We begin with
the formula \eqref{eq:two-way-crossing-2} for $F(\wp, \vartheta_c^-)$, and need to show that the action of $\CK$
transforms it into the form \eqref{eq:f-two-way}.  Expanding out \eqref{eq:two-way-crossing-2} we find
various classes of terms:
\begin{align}
 F(\wp, \vartheta_c^-)_{ii} &= \wp_+^{(i)} \left( 1 + \sum_{\bar a, \bar b} \mu^-(a) \mu^-(b) X_b X_a \right) \wp_-^{(i)} = Q(p) \wp^{(i)}, \\
 F(\wp, \vartheta_c^-)_{ij} &= \wp_+^{(i)} \left( \sum_{\bar b} \mu^-(b) X_b \right) \wp_-^{(j)}, \\
 F(\wp, \vartheta_c^-)_{ji} &= \wp_+^{(j)} \left( \sum_{\bar a} \mu^-(a) X_a \right) \wp_-^{(i)}, \\
 F(\wp, \vartheta_c^-)_{jj} &= \wp^{(j)}, \\
 F(\wp, \vartheta_c^-)_{kk} &= \wp^{(k)} \quad \text{ for } k \notin \{i,j\}.
\end{align}
Within each class, the terms in the sum over $\bar a, \bar b$
differ only by multiplication by factors $X_{\tilde\gamma}$.  Since $L(\gamma)$ is a multiple of
$\gamma$, it follows that $\inprod{\gamma, L(\gamma)} = 0$.  Hence the action of $\CK$ on each class of terms is independent
of the particular term.  We consider them in turn.  For the $ii$ terms, $\CK$ acts by multiplication by
$\prod_{\gamma \in \Gamma_c} (1 + X_{\tilde\gamma})^{\inprod{\wp^{(i)}, L(\gamma)}}$, but using \eqref{eq:def-L}
we see that $\inprod{\wp^{(i)}, L(\gamma)} = -\alpha_\gamma(p)$
(since $p$ is the only edge of $\wnet$ crossed by $\wp$) and hence this reduces to
multiplication by $Q(p)^{-1}$.  On the $jj$ terms $\CK$ similarly acts by multiplication by $Q(p)$.
On the $ij$ terms the action of $\CK$ is more complicated:  it is multiplication by a new function $H = \prod_{\gamma \in \Gamma_c} (1 + X_{\tilde\gamma})^{\inprod{\wp_+^{(i)} + \bar b + \wp_-^{(j)}, L(\gamma)}}$.  But we have
$\inprod{\wp_+^{(i)} + \bar b + \wp_-^{(j)}, L(\gamma)} = - \inprod{\wp_+^{(j)} + \bar a + \wp_-^{(i)}, L(\gamma)}$ as one readily
sees using the facts that $\cl(\bar a+\bar b)$ is proportional to $L(\gamma)$ and
$\inprod{\wp^{(i)}, L(\gamma)} = - \inprod{\wp^{(j)}, L(\gamma)}$.
It follows that the action of $\CK$ on the $ji$ terms is multiplication by $H^{-1}$.
These facts together are sufficient to imply the desired P5.\footnote{Incidentally, they also imply
that $\CF(\wp)_{jj} = Q(p)$; this and the fact $\CF(\wp) = F(\wp, \vartheta_c^+)$ (which we are
in the process of showing) together prove our remark under \eqref{eq:def-Q}.}

Finally we must explain why these properties actually \ti{determine} $\CF(\wp)$.
As a warm-up, suppose that there are only one-way streets.  In that case we have shown in \S \ref{sec:F-basics}
and \S \ref{sec:wnet} above that the properties P1-P4 are sufficient to determine all of the $\mu(a)$,
and hence also to fix $\CF(\wp)$.
The same kind of argument can be made in the presence of two-way streets; the only new complication is that
the local structure around joints and branch points is more complicated.
Nevertheless, a direct computation
using only the fact that $\CF(\wp)$ obey P1-P5 determines the $\mu(a)$ appearing there
for all charges $a$ oriented out of a joint, in terms of the $\mu(a)$ for charges $a$
oriented into the same joint.
(The resulting explicit formulae are recorded in Appendix
\ref{app:joint-rules}, e.g. \eqref{eq:outgoing-2wayjoint}.)
Inductively we can thus determine the 2d-4d
degeneracies $\mu(a)$ everywhere on $\wnet_{\vartheta_c}$, just as we did in the (generic) case
with only one-way streets.
Having done so the $\CF(\wp)$ are also determined by P1-P5.  This completes the proof.

\section{Examples of $\CK$-walls}

In this section we illustrate the general discussion of \S\ref{sec:varying-theta} with some examples.

\subsection{Saddle connections}\label{sec:saddle-connection}

The simplest possibility
has already appeared in Figure
\ref{fig:wnet-jump} above.  It involves two $\cS$-walls,
colliding along a two-way street $p$ running
between two branch points.  We call such a two-way street
a ``saddle connection,'' following the standard terminology for trajectories of quadratic differentials
(which is literally what we are considering here in the case $K=2$).

The two $\cS$-walls support charges $\bar a$ and $\bar b$, with natural lifts $a$, $b$, and we have
\begin{equation}
\mu^-(a) = 1, \qquad \mu^-(b) = 1,
\end{equation}
with all other $\mu^-(\cdot)$ vanishing.
In this case \eqref{eq:def-Q} reads
\begin{equation}
Q(p) = 1 + X_{\cl(a+b)}.
\end{equation}
If we let $\gamma = \cl(\bar a + \bar b)$, the preferred lift is $\tilde\gamma = \cl(a+b) + H$;
we thus have
\begin{equation}
 Q(p) = 1 - X_{\tilde\gamma}.
\end{equation}
This can indeed be decomposed according to \eqref{eq:Q-exp}, taking $\alpha_\gamma(p) = 1$, and
$\alpha_{n \gamma}(p) = 0$ for $n \neq 1$.
We thus obtain the simple result
\begin{equation}
 L(\gamma) = p_\Sigma
\end{equation}
and $L(n\gamma) = 0$ for $n \neq 1$.
As we have seen above, this result completely determines the wall-crossing
at this $\cK$-wall, through the formula \eqref{eq:k-def}.

In particular, we have $[L(\gamma)] = \gamma$, so the 4d BPS degeneracy is
\begin{equation}
 \Omega(\gamma) = 1.
\end{equation}
We have thus found that a saddle connection represents a BPS hypermultiplet.
This recovers a result of \cite{Klemm:1996bj,Gaiotto:2009hg} for the case $K=2$, and extends it to arbitrary $K$.

\subsection{Closed loops} \label{sec:closed-loops}

\insfigscaled{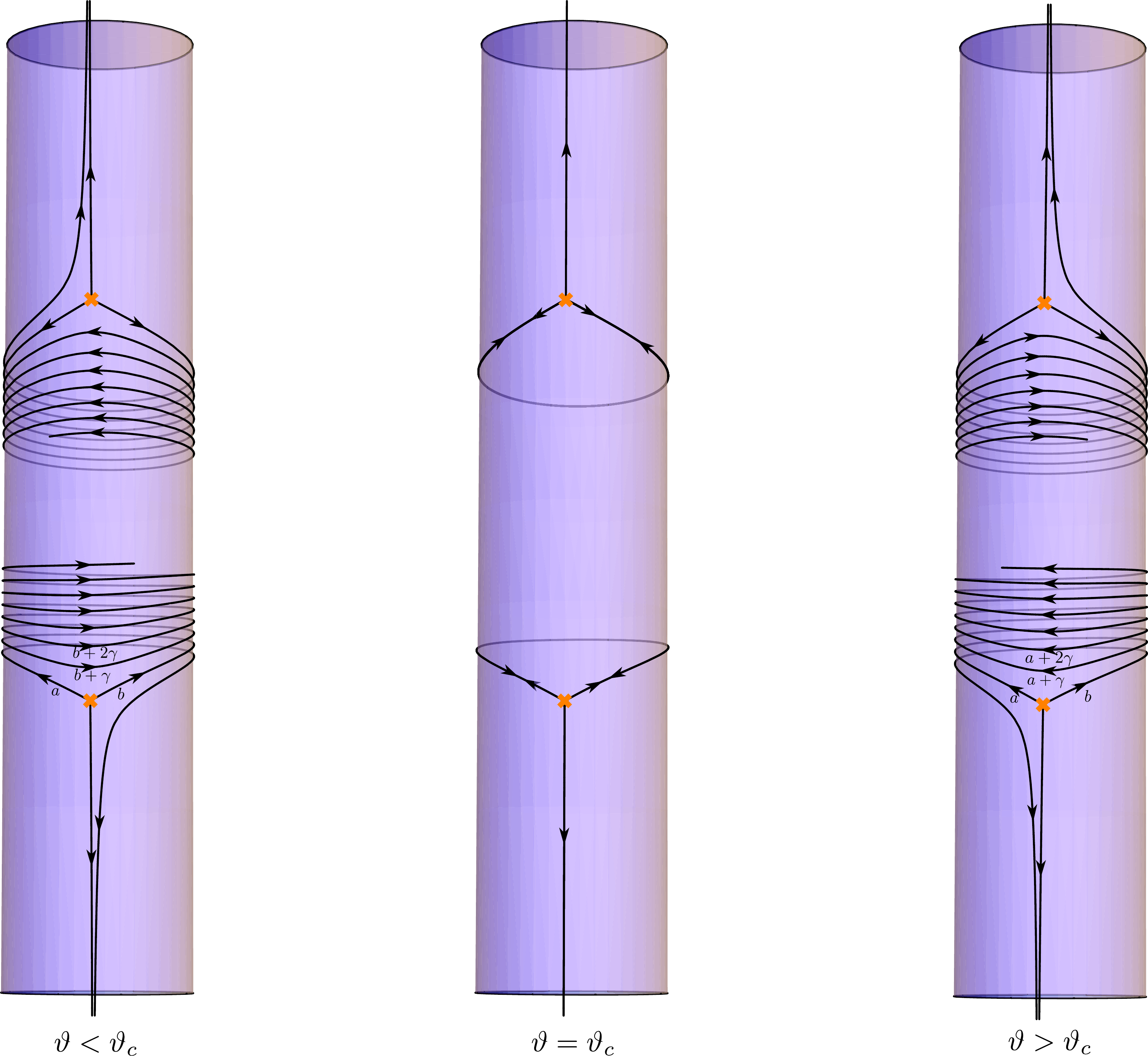}{0.45}{A more complicated way in which $\cS$-walls can collide head-on.  At $\vartheta < \vartheta_c$,
the $\cS$-wall supporting charge $a$
winds once clockwise around the cylinder and then disappears
toward the bottom, while the $\cS$-wall supporting charge $b$ winds many times counterclockwise
around the cylinder and eventually disappears toward the top.  In the figure, for clarity we truncate the
latter $\cS$-wall after it has wound around a few times.
At the critical $\vartheta = \vartheta_c$,
all of the windings of the $\cS$-wall supporting charge $b$
coalesce onto a single closed trajectory.  At $\vartheta > \vartheta_c$ these two $\cS$-walls exchange roles.}

A more interesting possibility is shown in Figure \ref{fig:closed-trajectories}.  This case involves
an $\cS$-wall which winds around a cylinder in $C$ many times, more and more tightly as
$\vartheta \to \vartheta_c$.  In this case two two-way streets appear simultaneously at
$\vartheta = \vartheta_c$; let $p$ denote one of them.
$p$ begins and ends at the \ti{same} branch point, and the soliton degeneracies
are a bit more interesting than the previous case:  there are infinitely many
nonvanishing $\mu^-(\cdot)$,
coming from the infinite set of windings of the $\cS$-wall around the cylinder, which have all
coalesced onto $p$.  Explicitly
\begin{equation}
 \mu^-(a) = 1, \qquad \mu^-(b + n(a+b)) = 1 \text{ for all } n \ge 0,
\end{equation}
with all other $\mu^-(\cdot)$ vanishing.
Hence by \eqref{eq:def-Q},
\begin{align}
Q(p) &= 1 + X_{\cl(a+b)} + X_{\cl(2(a+b))} + \cdots \\
&= (1 - X_{\cl(a+b)})^{-1}.
\end{align}
Define $\gamma = \cl(\bar a + \bar b)$.
The preferred lift of $\gamma$ is $\tilde\gamma = \cl(a+b)$, so we have
$Q(p) = (1 - X_{\tilde\gamma})^{-1}$, which means $\alpha_\gamma(p) = -1$.
Now, to compute $L(\gamma)$ we must sum the contributions from both two-way streets ---
call them $p^1$ and $p^2$ --- so we obtain
\begin{equation}
 L(\gamma) = - p^1_\Sigma - p^2_\Sigma.
\end{equation}

Both $p^1_\Sigma$ and $p^2_\Sigma$ are in the homology class $\gamma$,
so $[L(\gamma)] = -2\gamma$, or
\begin{equation}
\Omega(\gamma) = -2.
\end{equation}
Thus we have recovered the result
of \cite{Klemm:1996bj,Gaiotto:2009hg}:  this pair of closed trajectories
represents a BPS vectormultiplet.

In \cite{Gaiotto:2009hg} we derived the result $\Omega(\gamma) = -2$ in
this situation (in the special case $K=2$),
by a rather delicate analysis of the jumps of the vacuum expectation
values of line defects associated to \ti{closed} paths.
This analysis gave the right answer but depended on the assumption that
at $\vartheta = \vartheta_c$ the third trajectory emerging from the branch points in Figure \ref{fig:closed-trajectories}
ends on a puncture. As
Ivan Smith pointed out to us, this assumption can be violated for higher genus $C$:  the cylinder
of closed trajectories could be dividing $C$ into two pieces, one of which contains no puncture.
Our present analysis using line defects associated to \ti{open} paths is much simpler and is free of such extra assumptions.

Incidentally, from the point of view of the theory $S[\fg, C, D]$
these two closed trajectories would not seem to be the whole story.  Indeed, these two
$ij$-trajectories are actually the two ends of a one-parameter \ti{family} of closed
$ij$-trajectories, which sweep out a cylinder on $C$.  Any member of this family is
a BPS string.  So physically speaking we should regard this $\Omega(\gamma) = -2$ as giving
the contribution from this whole one-parameter family, not just from the two ends of the family.
From our current point of view though, the ends play a privileged role, in
that they are $\CS$-walls, while the trajectories in the interior of the cylinder are not.

Finally, although we have hidden it up to this point, the behavior of the networks
$\wnet_\vartheta$ as $\vartheta \to \vartheta_c$ is actually rather complicated.  In Figure
\ref{fig:closed-trajectories}, at $\vartheta < \vartheta_c$, we see two $\cS$-walls which wind
many times around the cylinder.  There we truncated them at some finite distance in order not
to make the figure too confusing.  In Figure \ref{fig:cylinder-colored} we show the full $\cS$-walls
at some particular $\vartheta < \vartheta_c$.  Note that the two winding $\cS$-walls thread
past one another many times before they escape the cylinder.  As we vary
$\vartheta$ slightly, the angle at which they are threaded changes, and there are infinitely many
critical phases $\vartheta_n$ at which they actually collide head-to-head.  The critical phases
$\vartheta_n$ accumulate at the value $\vartheta_c$.  At each of these
critical phases we have a saddle connection and the network $\wnet_\vartheta$ jumps,
just as we discussed in \S\ref{sec:saddle-connection}.
\insfigscaled{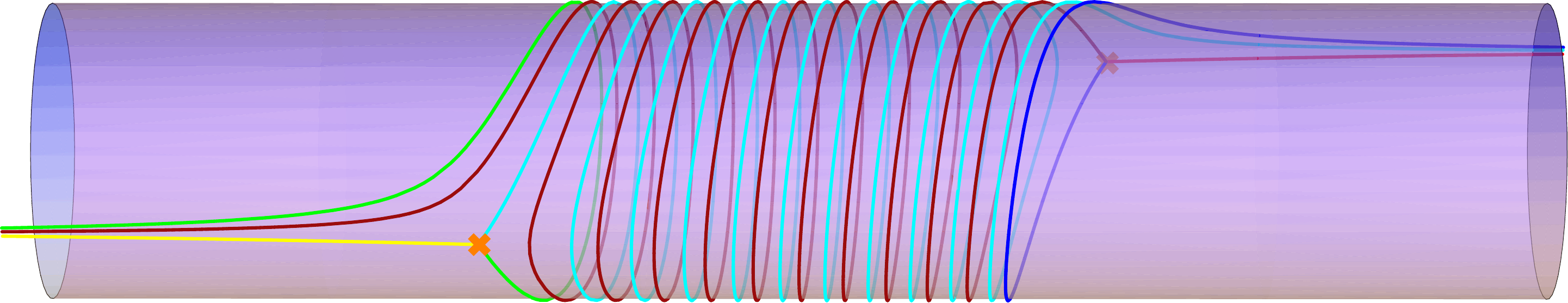}{0.45}{Another picture of the $\CS$-walls on the cylinder for $\vartheta$
near the critical phase $\vartheta_c$.  Here we do \ti{not} truncate the walls, and we distinguish
different walls by giving them different colors.}
We analyzed this limiting process rather closely in \cite{Gaiotto:2009hg}; but we emphasize
that in our current analysis, if all we want to know is the BPS degeneracy $\Omega(\gamma)$ which
appears exactly at the critical phase $\vartheta_c$,
it is \ti{not} necessary to study this infinite sequence of jumps.

\subsection{Three-string webs} \label{sec:three-string}

\insfigscaled{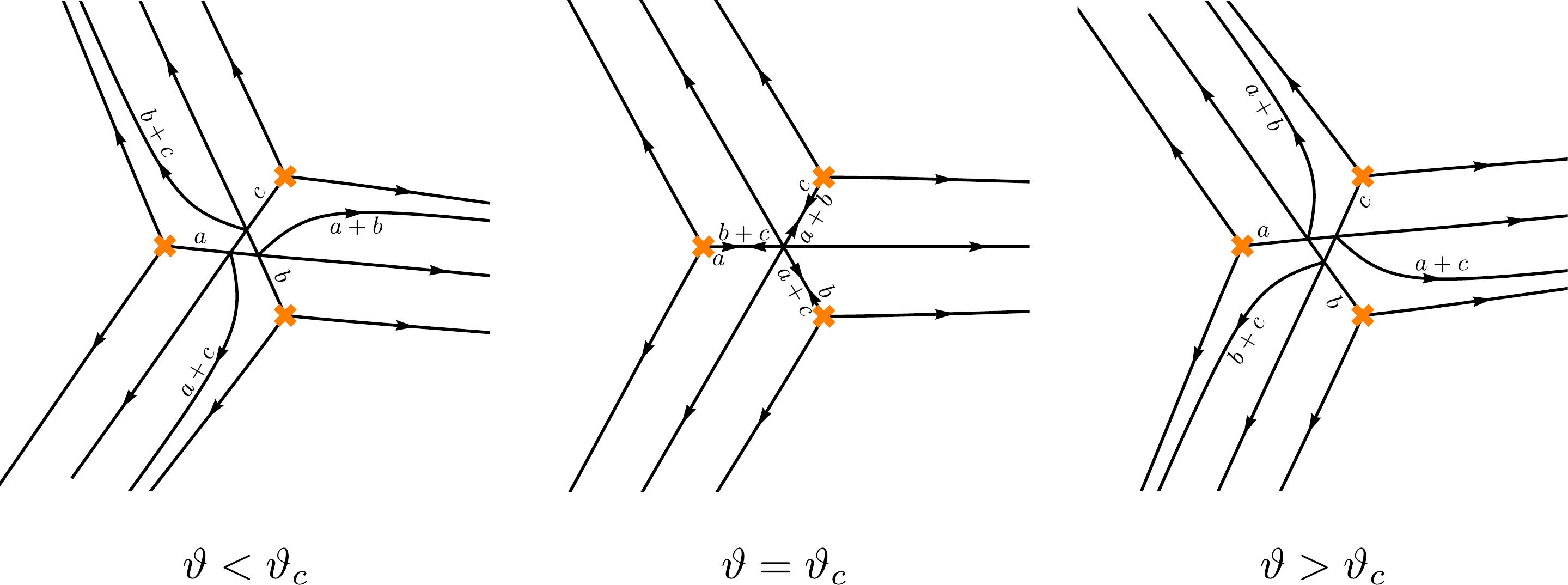}{0.52}{A new way in which $\cS$-walls can collide, which appears only in
theories with $K > 2$.  The picture shows a patch of $C$ containing three branch points.  $\wnet_\vartheta$
involves both ``primary'' $\cS$-walls born from the branch points (supporting charges $a$, $b$, $c$) and ``secondary'' $\cS$-walls
born from intersections between primary
$\cS$-walls (supporting charges $a+b$, $b+c$, $a+c$.)
At the critical phase $\vartheta = \vartheta_c$, primary and secondary walls collide head-on, along a locus which forms
a three-string finite web.}

When we go beyond $K=2$, we encounter many new and varied phenomena.

The simplest new possibility is shown in Figure \ref{fig:three-pronged-network}.
Near the critical phase $\vartheta = \vartheta_c$, we see \ti{three} pairs of $\cS$-walls nearly colliding head-on.
Each pair consists of one ``primary'' $\cS$-wall, born at a branch point, and one ``secondary'' $\cS$-wall, born at the intersection
between two primary walls.  At the critical phase $\vartheta = \vartheta_c$ each such pair of walls collides along
a two-way street $p$.
Each of the three two-way streets $p$
supports a different pair of two 2d-4d charges, with
all three pairs summing to $\gamma = \cl(\bar a+\bar b+\bar c)$.

We would like to compute the 1-cycle $L(\gamma)$.
Let us focus attention on the two-way street $p$
supporting the 2d-4d charges $\bar a$ and $\bar b+\bar c$.
At $\vartheta < \vartheta_c$,
the $\cS$-wall supporting charge $\bar a$ is emerging directly from a branch point,
and thus has a natural lift $a$ with $\mu(a) = 1$.  The
$\cS$-wall supporting $\bar b+\bar c$, on the other hand, is emerging from a joint where walls supporting $\bar b$ and $\bar c$ intersect.  These two walls
in turn emerge from branch points, so at the joint $\mu^-(b) = 1$, $\mu^-(c) = 1$.
It follows from \eqref{eq:cv-wcf} that the emerging wall has
$\mu^-(b+c) = 1$.  Since these are the only two $\cS$-walls
which collide along $p$, all other $\mu^-(\cdot)$ vanish along $p$.
Plugging into \eqref{eq:def-Q} we obtain
\begin{equation}
 Q(p) = 1 + X_{\cl(a+b+c)}
\end{equation}
and hence, letting $\gamma = \cl(\bar a+\bar b+\bar c)$ with preferred lift $\tilde\gamma = \cl(a+b+c) + H$, for suitable $a,b,c$,
we find $\alpha_\gamma(p) = 1$.  The same result holds for the other two two-way streets $p^n$,
so $L(p)$ is the sum:
\begin{equation}
 L(\gamma) = p^1_\Sigma + p^2_\Sigma + p^3_\Sigma.
\end{equation}
In other words, letting $N$ denote the finite web made up of the three two-way streets $p^n$,
we have
\begin{equation}
 L(\gamma) = N_\Sigma.
\end{equation}
Moreover, $[N_\Sigma] = \gamma$, so the 4d BPS degeneracy here is
\begin{equation}
\Omega(\gamma) = 1.
\end{equation}
So we have found that the finite web $N$, made up of three strings which meet
at a junction, corresponds to a BPS hypermultiplet.

\subsection{The setting sun} \label{sec:setting-sun}

Next we briefly consider a more complicated example, in which several overlapping finite webs
appear simultaneously.
Unlike the previous examples, we just draw the degenerate network of interest, and not the nondegenerate ones
at nearby phases (which would be terribly cluttered in this example).  See Figure \ref{fig:setting-sun}.
\insfig{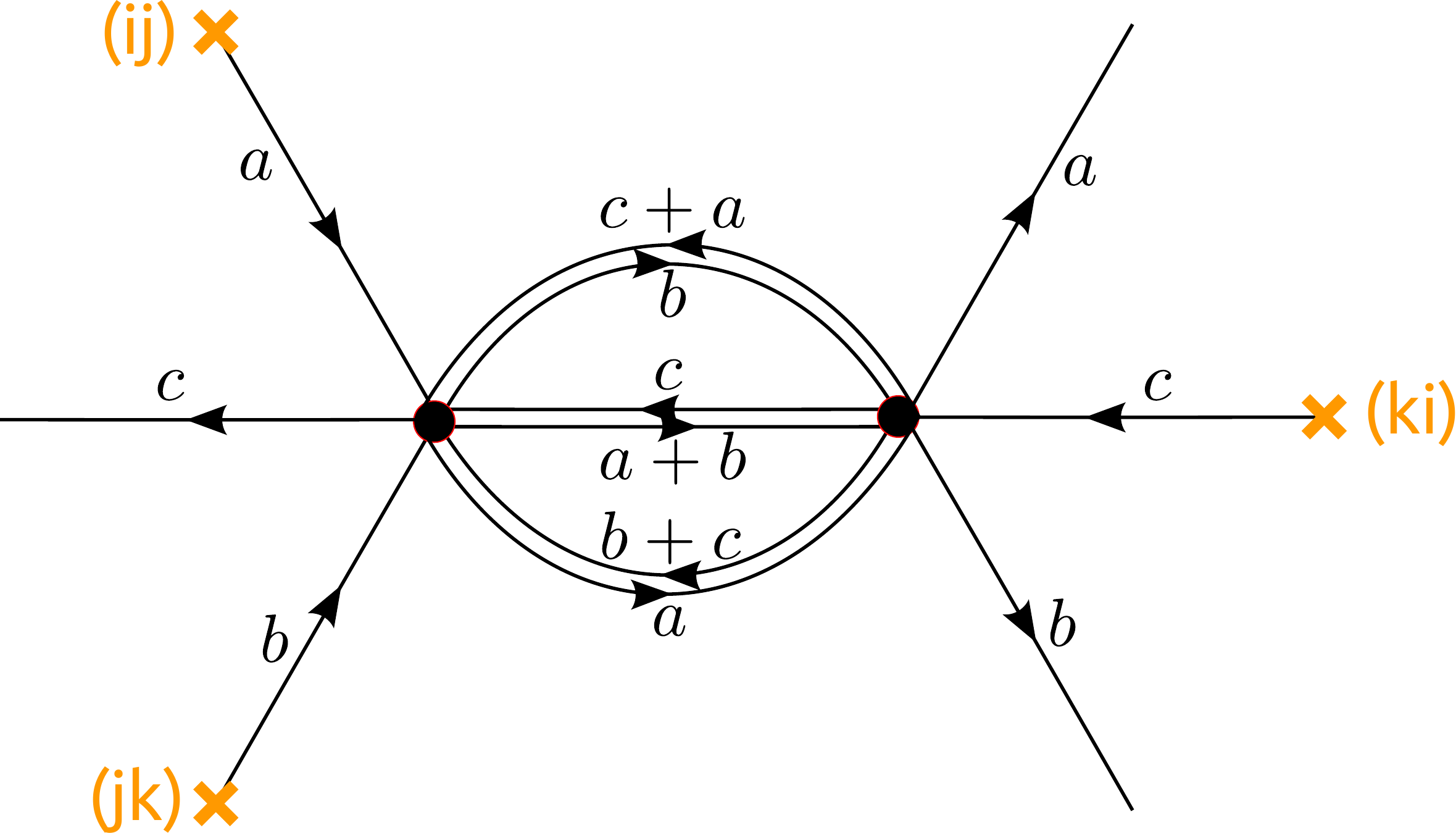}{The ``setting sun'':  a picture which could occur as part of a degenerate spectral network
at some $\vartheta = \vartheta_c$.  We show the two-way streets slightly resolved, with the ``American resolution'',
which would occur in the limit $\vartheta \to \vartheta_c^-$.  Each street supports the charges indicated as
well as their shifts by positive multiples of $\gamma = \cl(\bar a + \bar b + \bar c)$.}

Label the three two-way streets in this figure as $p^1$, $p^2$, $p^3$ from top to bottom.  Using the rules of
Appendix \ref{app:joint-rules}, we can directly compute the soliton degeneracies $\mu(\cdot, \cdot)$
on all the streets in the figure.  The nonzero degeneracies on $p^1$, $p^2$, $p^3$ come out to
\begin{align}
 \mu (b, p^1) = 1, & \quad \mu(c + a + n (b + c + a), p^1) = 1 \text{ for } n \ge 0, \\
 \mu (c, p^2) = 1, & \quad \mu(a+b, p^2) = 1, \\
 \mu (a + n (b + c + a), p^3) = 1 \text{ for } n \ge 0, & \quad \mu(b+c, p^3) = 1.
\end{align}
(All other streets turn out to be one-way, as anticipated in Figure \ref{fig:setting-sun}, and so do not contribute to $L$ below.)
Define $\gamma = \cl(\bar a + \bar b + \bar c)$, and let $\tilde\gamma$ be its preferred lift.
Plugging into \eqref{eq:def-Q} and keeping careful track of windings, we find
\begin{align}
 Q(p^1) = Q(p^3) &= (1 - X_{\tilde \gamma})^{-1}, \\
 Q(p^2) &= 1 - X_{\tilde \gamma}.
\end{align}
This gives
\begin{equation}
 L(\gamma) = -p^1_\Sigma - p^3_\Sigma + p^2_\Sigma.
\end{equation}
This class is homologically trivial, $[L(\gamma)] = 0$.  (It could hardly be otherwise,
since the projection of $L(\gamma)$ to $C$ lies in a contractible region containing no branch points, which means
$L(\gamma)$ itself lies in the disjoint union of three contractible open sets on $\Sigma$.)
So in particular this degenerate network does not contribute to the 4d BPS spectrum:
\begin{equation} \label{eq:zero-contrib}
 \Omega(\gamma) = 0.
\end{equation}
Nevertheless the enhanced degeneracies $\omega(\gamma, \cdot)$ are certainly \ti{not} zero in this example:
they are given as usual by \eqref{eq:result-omega}.

\insfigscaled{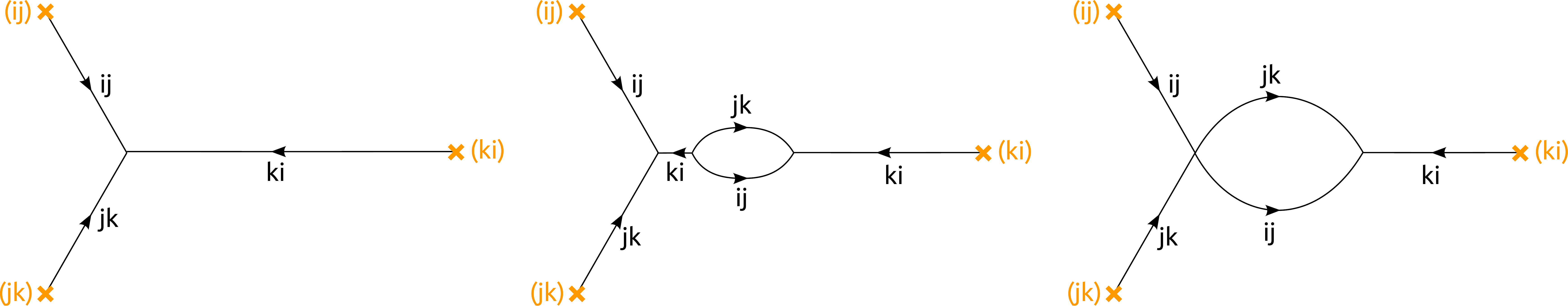}{0.21}{Three finite webs, belonging to a 1-parameter family of finite webs
parameterized by an interval:
as we move along the interval, the ``bubble'' in the middle of the web expands from zero size (left) to a finite maximum size (right).}

How should we understand the result \eqref{eq:zero-contrib}?  First note that
there is a 1-parameter family of finite BPS webs here, parameterized by an interval,
as indicated in Figure \ref{fig:two-finite-webs}.  The two ends of the family are built from the $\cS$-walls appearing in
the degenerate network $\wnet_{\vartheta_c}$, while the other finite webs involve a ``bubble'' made out of
BPS strings which are not
$\cS$-walls.  This is the same phenomenon we had in \S\ref{sec:closed-loops} above, where we considered a cylinder swept out
by BPS strings; the boundaries of the cylinder were $\cS$-walls, but the generic BPS strings inside the cylinder were not.

In the present case
one end of the family (at left) looks like the ``three-string web'' we encountered in \S\ref{sec:three-string}
above.  If this three-string web occurred
in isolation it would give rise to a BPS hypermultiplet, with $\Omega(\gamma) = +1$.  However, when it sits in this
1-parameter family its contribution cannot be evaluated in isolation:  rather we must quantize the whole family at once.
What we have seen is that the contribution to $\Omega(\gamma)$ from this family vanishes.

\section{Some BPS spectra}\label{sec:Examples}

In this section we finally show how the spectral networks $\wnet_\vartheta$ can be used
to determine the 4d BPS spectrum in some simple examples of theories $S[\fg = A_{K-1}, C, D]$.

A first comment is that all of the examples described in Section 9 of \cite{Gaiotto:2009hg} are also examples
of the structures considered here:  more precisely they are examples of the case $K=2$, where (as explained in \S\ref{sec:k2-examples})
studying the spectral networks $\wnet_\vartheta$ is equivalent to studying some special ideal triangulations
of $C$.  We therefore regard those examples
as incorporated here by reference, and move on to the really new phenomena.

\subsection{The pentagon theory revisited}

The first new example we consider is obtained by taking $K=3$ and $C = \IC\IP^1$,
with a single defect at $z = \infty$, imposing the boundary conditions
that $\phi_2$ has a pole of order 4 and $\phi_3$ one of order 8.
These conditions imply that after rescaling and shifting the coordinate $z$, one can put
$\phi_2$ and $\phi_3$ in the form
\begin{equation}
 \phi_2 = 3 \Lambda^2 \, \de z^{2}, \qquad \phi_3 = (z^2 + u) \, \de z^{3}.
\end{equation}
Here $\Lambda$ is a parameter and $u$ parameterizes the 1-dimensional Coulomb branch $\CB$.
For any particular $u$, we have a corresponding $3$-fold cover of $C$ (Seiberg-Witten curve) given by \eqref{eq:sw-curve},
\begin{equation}
 \Sigma = \{ \lambda^3 + (3 \Lambda^2 \, \de z^{2}) \lambda + (z^2 + u) \, \de z^{3} = 0 \} \subset T^* C,
\end{equation}
or if we write more concretely $\lambda = x \, \de z$,
\begin{equation} \label{eq:sw-pentagon-1}
 \Sigma = \{ x^3 + 3 \Lambda^2 x + z^2 + u = 0 \}.
\end{equation}

We can now study the BPS spectrum, by scanning through the spectral networks $\wnet_\vartheta$ as $\vartheta$ varies
between $0$ and $\pi$ and looking
for critical phases where the topology of $\wnet_\vartheta$ jumps.

\insfig{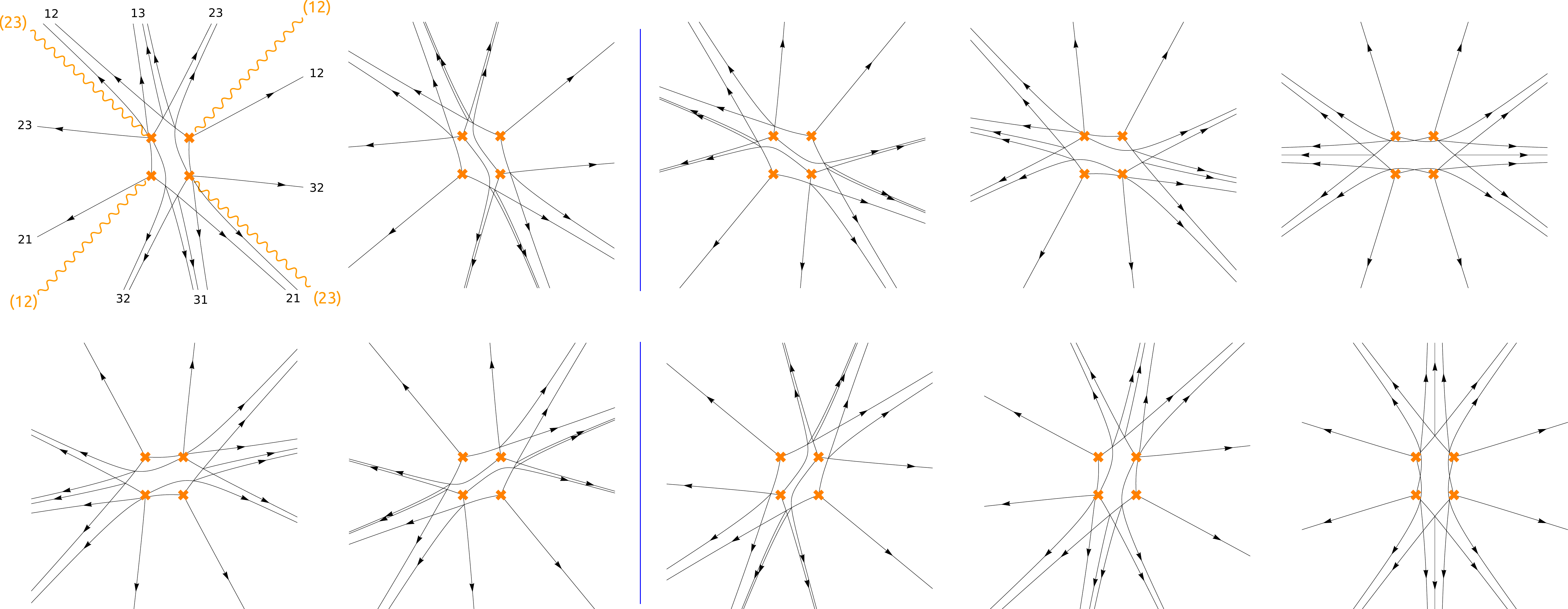}{The network $\wnet_\vartheta$ when $u=0$, at the phases $\vartheta = \pi n / 10$, with
$1 \le n \le 10$; the first row begins with $n=1$ and the second with $n=6$.  We indicate the labelings of the $\cS$-walls
in the $n=1$ figure:  all walls in each indicated group carry the same label (e.g. there are three walls carrying the label $13$
exiting the figure to the north.)  The two critical phases at which a $\cK$-wall occurs
are indicated by blue lines; one is between $n=2$ and $n=3$, the other between $n=7$ and $n=8$.  See \cite{spectral-network-movies}
for an animated version of this figure.}
\insfigscaled{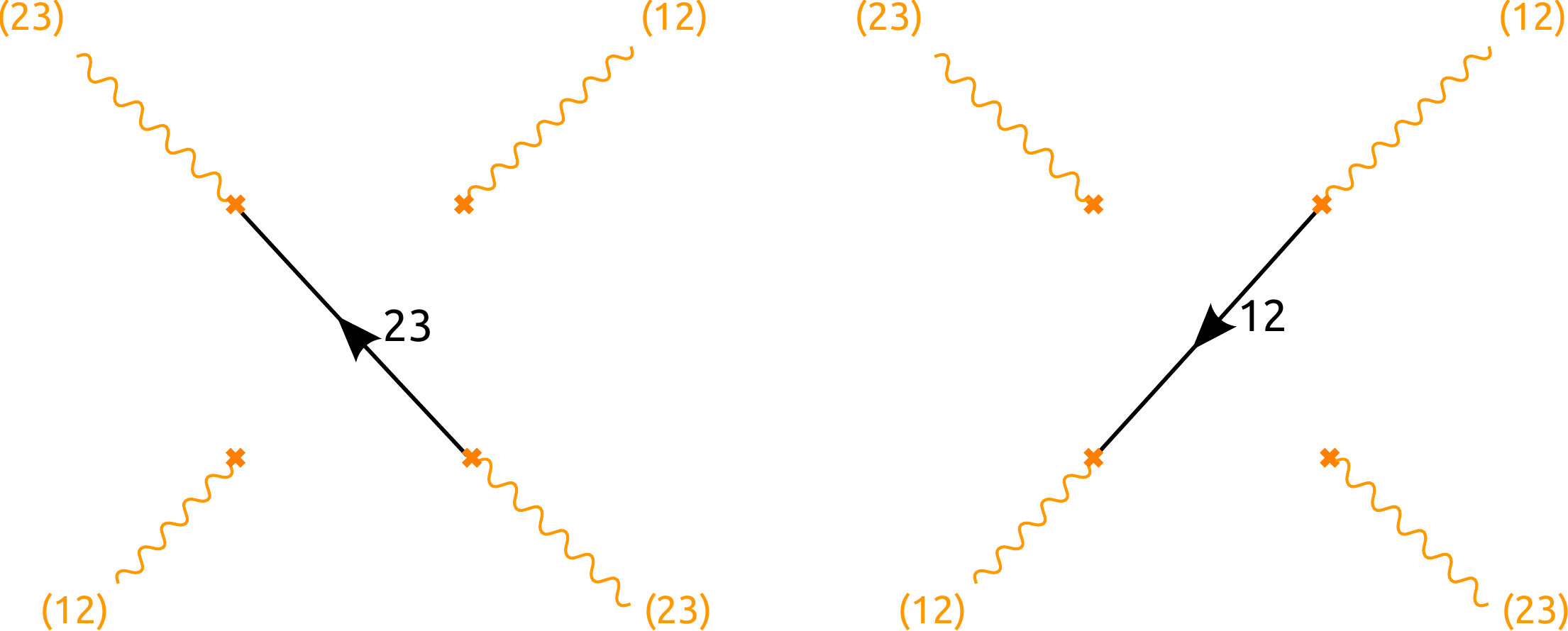}{0.45}{The two saddle connections which appear at the critical phases, when $u=0$.  They could equally
well have been represented by segments with the opposite orientation and the labels $ij$ transposed.}
For example, suppose $u = 0$.  In this case
$\wnet_\vartheta$ jumps 2 times as we vary the phase, as we show in Figure \ref{fig:pentagon-evolution-strong}.
Both of these jumps are of the type we discussed in \S\ref{sec:saddle-connection}:
at the critical phase there is a saddle connection, i.e. an $ij$-trajectory
running between two $(ij)$ branch points.
We depict these two saddle connections in Figure \ref{fig:pentagon-bps-states-strong}.  Following the recipe of \S\ref{sec:f-jump} we see that
they correspond to closed loops $L_1$, $L_2$ on $\Sigma$ (just obtained by lifting the saddle connections to $\Sigma$) with corresponding
charges $\gamma_1 = [L_1]$, $\gamma_2 = [L_2]$, and the 4d BPS degeneracies are
\begin{equation}
\Omega(\gamma_1) = 1, \qquad \Omega(\gamma_2) = 1.
\end{equation}
(We also have $\Omega(-\gamma_1) = \Omega(-\gamma_2) = 1$; these other two BPS multiplets would be encountered in varying
$\vartheta$ between $\pi$ and $2 \pi$.  The network $\wnet_{\vartheta+\pi}$ is obtained from $\wnet_\vartheta$ just by
transposing the labels $ij$ on all walls, so we do not need to draw new figures for this range of phases.)
The intersection pairing between these charges is $\inprod{\gamma_2, \gamma_1} = 1$;
this reflects the fact that the two saddle connections cross at a single point, at which they have a single sheet in common
(sheet $2$ in the notation of Figure \ref{fig:pentagon-bps-states-strong}.)

On the other hand, suppose we take $u/\Lambda = 5$.  In this case the picture looks somewhat different:  $\wnet_\vartheta$ jumps 3 times
as we vary $\vartheta$ from $0$ to $\pi$, as shown in Figure \ref{fig:pentagon-evolution-weak}.  These three jumps correspond to three finite
webs as indicated in Figure \ref{fig:pentagon-bps-states-weak}.
\insfig{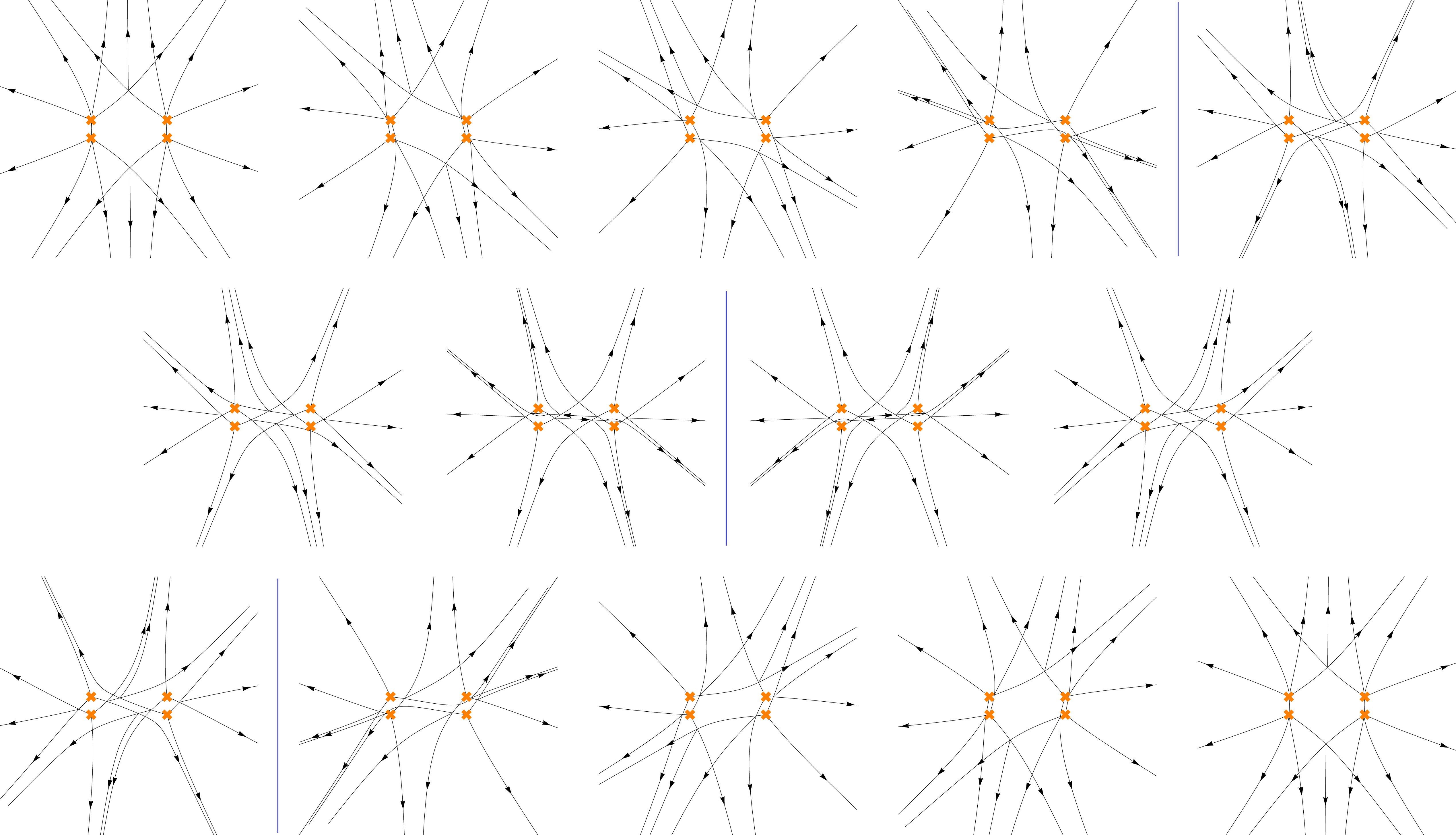}{The network $\wnet_\vartheta$ when $u=4$ and $\Lambda=1$, at the phases $\vartheta = \pi n / 200$, with
(reading from left to right and top to bottom)
$n = 1, 23, 45, 68, 82, 90, 97, 103, 110, 118, 132, 155, 177, 199$.  The three critical phases at which a $\cK$-wall occurs
are indicated by blue lines.  See \cite{spectral-network-movies}
for an animated version of this figure.}
\insfigscaled{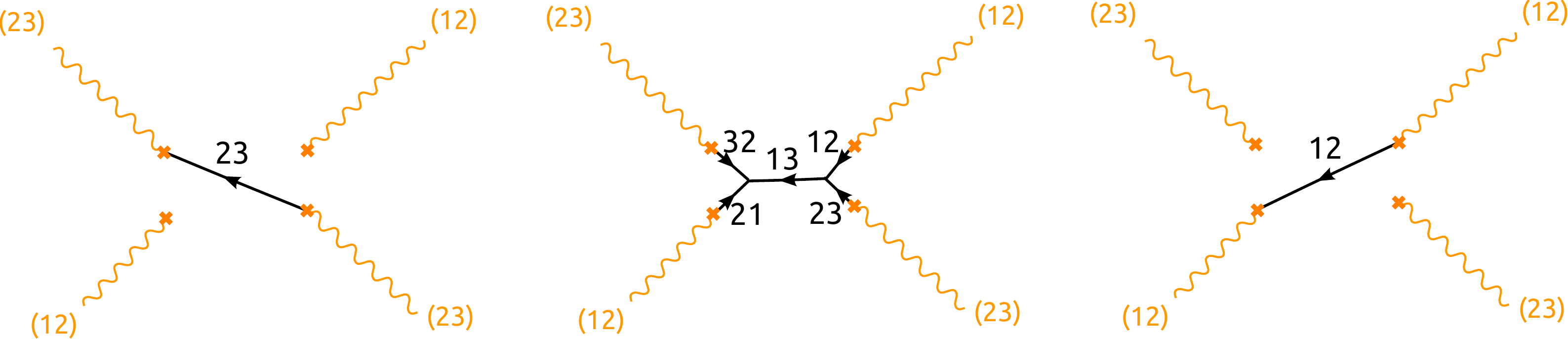}{0.45}{The two saddle connections and one more complicated finite web
which appear at the critical phases, when $u=4$
and $\Lambda = 1$.  Each segment could equally well be replaced
by one with the opposite orientation and the labels $ij$ transposed.}

Again following \S\ref{sec:f-jump} we find that all three of these lift to closed loops on $\Sigma$, now with corresponding charges
$\gamma_2$, $\gamma_1 + \gamma_2$ and $\gamma_1$ in order, and we have
\begin{equation}
 \Omega(\gamma_1) = 1, \qquad \Omega(\gamma_2) = 1, \qquad \Omega(\gamma_1 + \gamma_2) = 1
\end{equation}
(along with $\Omega(-\gamma_1) = \Omega(-\gamma_2) = \Omega(-\gamma_1-\gamma_2) = 1$, as before).

So the 4d BPS spectrum changes as we vary $u$.  Of course, this is not unexpected:
it is the wall-crossing phenomenon, and occurs exactly as
predicted by the wall-crossing formula \cite{Denef:2007vg,ks1,Gaiotto:2008cd,Gaiotto:2009hg}.
At large $\abs{u}$ the two BPS multiplets of charges $\gamma_1$ and $\gamma_2$ form
a bound multiplet of charge $\gamma_1 + \gamma_2$.

The BPS spectrum here (two hypermultiplets with symplectic product $1$ at small $\abs{u}$,
three hypermultiplets at large $\abs{u}$) might look familiar to the reader:
it is just the same structure one meets in the first nontrivial Argyres-Douglas theory.
Wall-crossing in Argyres-Douglas theories was first studied in \cite{Shapere:1999xr};
we also studied this particular theory in
Section 9.4.4 of \cite{Gaiotto:2009hg}.  There
we took $K=2$ and $\phi_2 = z^3 - 3 \Lambda^2 z + u$,
thus obtaining the Seiberg-Witten curve
\begin{equation}
 \Sigma = \{ \lambda^2 + (z^3 - 3 \Lambda^2 z + u) dz^{2} = 0 \},
\end{equation}
or writing $\lambda = x\, \de z$,
\begin{equation}
 \Sigma = \{ x^2 + z^3 - 3 \Lambda^2 z + u = 0 \}.
\end{equation}
Now comes the point:  the change of variables $x \to -z$, $z \to x$ transforms this into \eqref{eq:sw-pentagon-1}!
This change of variables
does not quite preserve $\lambda$, but it takes $\lambda \to \lambda - \de(x z)$,
so as far as the periods of $\lambda$ over closed cycles
are concerned, these two Seiberg-Witten curves
are fully equivalent.  This constitutes strong evidence that the two 4d
theories $S[A_1, C, D]$ and $S[A_2, C, D']$ which we have considered
are actually the same.
These two descriptions of the theory however
privilege different classes of surface
defect, one with $K=2$ vacua and one with $K=3$.
Moreover they lead to rather different-looking representations of the BPS states:
in the $K=2$ picture, all three BPS hypermultiplets at large $u$ are represented by saddle
connections, quite unlike the situation depicted in Figure \ref{fig:pentagon-bps-states-weak}.  We regard the fact that the BPS degeneracies nevertheless
agree as a useful consistency check of our story.

This ``duality'' between two different descriptions of the same theory is an example
of a more general phenomenon.  There is a class of theories discussed in \cite{Cecotti:2010fi}, labeled by pairs
$(G,G')$ of Dynkin diagrams, and the theory with diagram $(G, G')$ is the same as the one with diagram $(G', G)$.
The example we considered here is the case of $(A_1, A_2)$.

\subsection{The pure $SU(3)$ theory at strong coupling}\label{sec:su3-strong}

For our next example we take again $K=3$ and $C = \IC\IP^1$,
but this time with defects both at $z = 0$ and $z = \infty$, with each defect
imposing the boundary conditions that $\phi_2$ has a pole of order at most 2 and $\phi_3$ one of order 4.
These conditions imply that after rescaling the coordinate $z$, one can put
$\phi_2$ and $\phi_3$ in the form
\begin{equation}
 \phi_2 = -3 u_2 \, \left( \frac{\de z}{z} \right)^2, \qquad \phi_3 = \left( \frac{\Lambda}{z} + u_3 + \Lambda z \right) \left(\frac{\de z}{z}\right)^{3}.
\end{equation}
This corresponds to the pure $SU(3)$ theory \cite{Gaiotto:2009hg}.
Here $\Lambda$ is a parameter (the dynamical scale)
and $(u_3,u_2)$ parameterize the Coulomb branch.
In this paper will not attempt a complete study of the BPS spectrum in this theory:
we just describe what happens at the locus where $u_3 = 0$ and $\abs{u_2} \ll \abs{\Lambda}^{\frac23}$.
This locus is in the ``strongly coupled'' region of the theory.

We could proceed immediately to the pictures of $\wnet_\vartheta$ as $\vartheta$ varies,
but to calibrate our expectations, it is useful to make
some preliminary exploration.
At a generic point $(u_3,u_2)$, the Seiberg-Witten curve $\Sigma$ is a 3-fold covering of $C$
with $4$ simple branch points.  If $u_2 = 0$ then the $4$ branch points coalesce in pairs.
Each pair consists of an $(ij)$ and a $(jk)$ branch point, which at $u_2=0$ coalesce to a single
branch point, with cyclic monodromy $(ijk)$.  There is nothing singular about this situation from the point of view of the IR 4-dimensional
physics (in contrast with the case where two $(ij)$ branch points coalesce, in which case the mass of a BPS hypermultiplet goes to zero.)
For small $u_2$ and generic $\vartheta$, we can work out (and verify by computer calculation) what
the spectral network $\wnet_\vartheta$ around such a pair looks like:
see Figure \ref{fig:two-colliding-bp}.
Taking the limit $u_2 \to 0$ we find that the branch points with monodromy $(ijk)$ emit $8$ walls.
As $\vartheta$ is increased continuously through an angle
$\pi/3$, these $8$ walls rotate by one unit counterclockwise.
\insfig{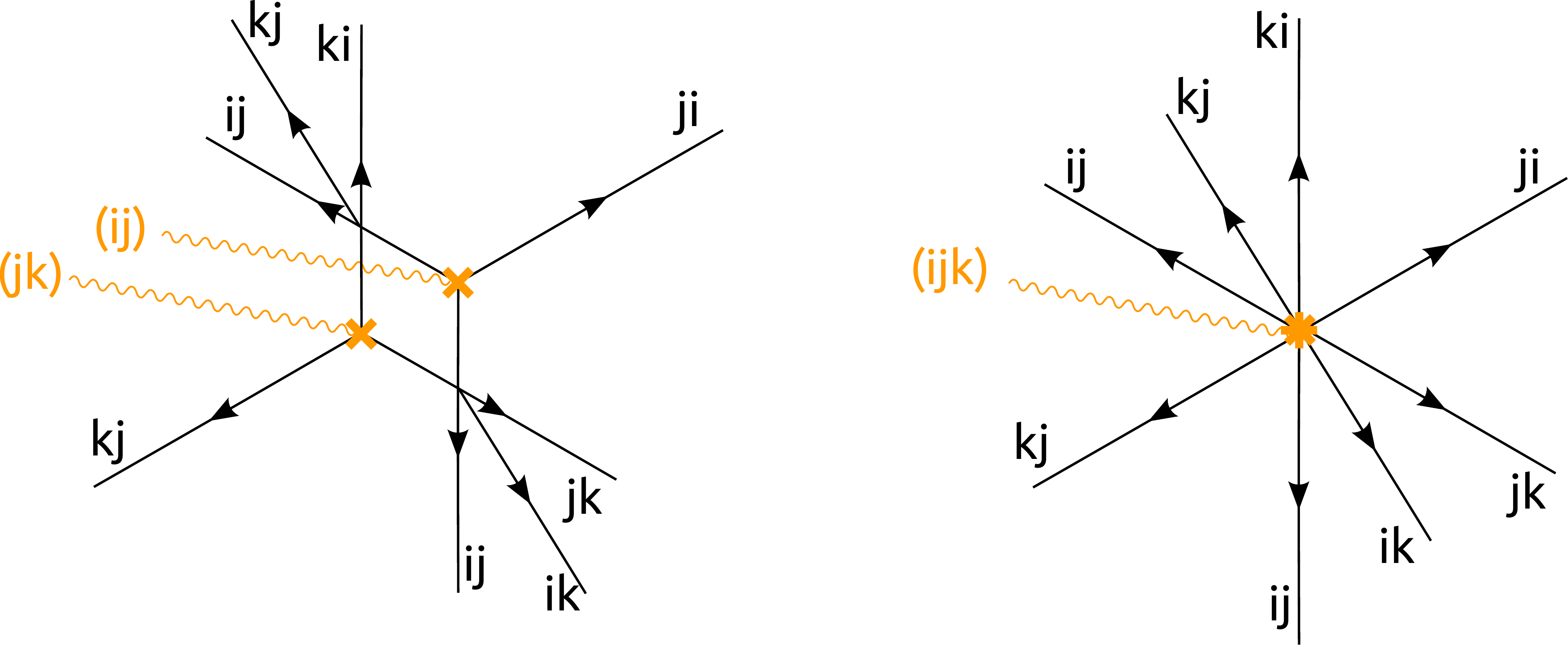}{Left: the spectral network $\wnet_\vartheta$ in a neighborhood of a pair of nearby branch points
of types $(ij)$ and $(jk)$; right: the limit where the branch points coalesce to form a branch point of type $(ijk)$.
This situation occurs in the pure $SU(3)$ theory as $u_2 \to 0$.}

Now fixing $u_3 = 0$ and $u_2$ small (or even $u_2 = 0$) we can draw $\wnet_\vartheta$ as $\vartheta$ varies from $0$ to $\pi$.
In fact, the picture has a further approximate
symmetry under a shift $\vartheta \mapsto \vartheta + \frac{\pi}{3}$ (which becomes exact at $u_2=0$), so we only need
to look at the variation over some range of length $\frac{\pi}{3}$.  We show this variation in Figure \ref{fig:su3-evolution}.
\insfigscaled{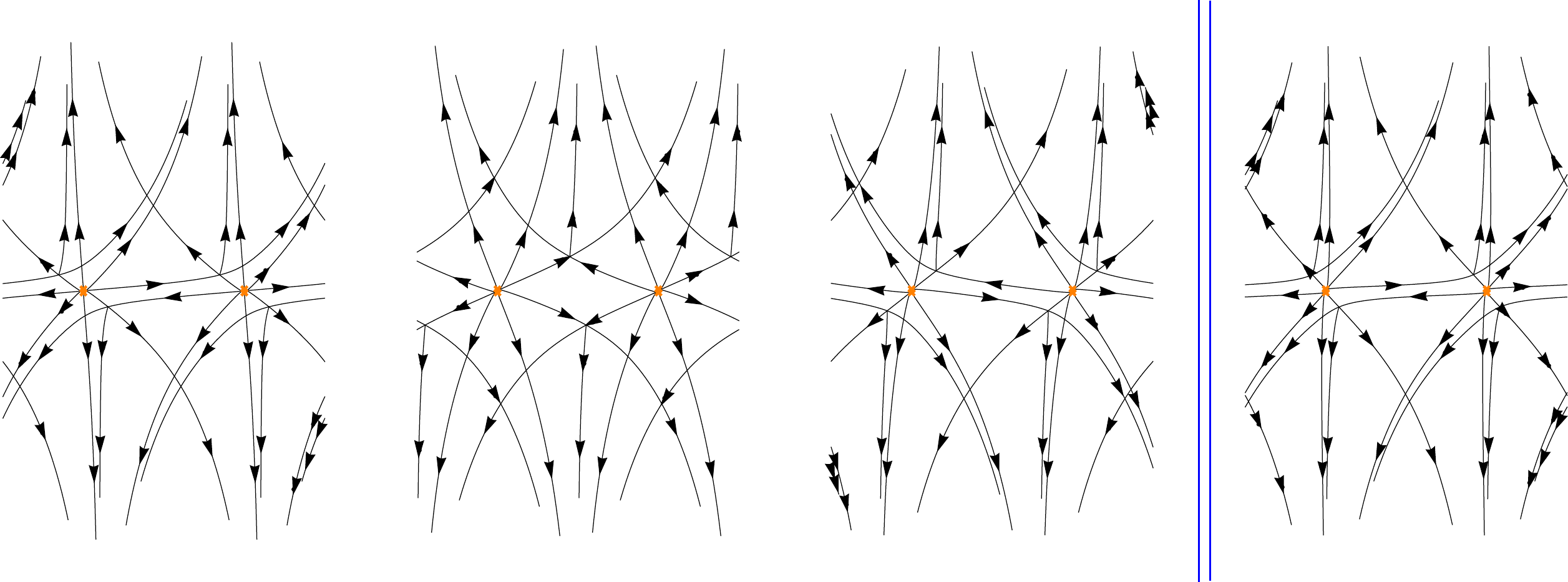}{0.6}{The evolution of $\wnet_\vartheta$ from $\vartheta = \frac{\pi}{30}$ to $\vartheta = \frac{11 \pi}{30}$,
at $u_3 = 0$ and very small $u_2$.  Each figure is drawn on a flattened-out cylinder, so the left and right sides should be identified.
Note that the last figure looks essentially identical to the first.
There are two critical phases $\vartheta_c$ in this range, very close together,
marked by the two blue lines.  See \cite{spectral-network-movies}
for an animated version of this figure.}
We find two critical phases $\vartheta_c$ very close together.  At each critical phase
a BPS hypermultiplet
appears, represented by a two-way street connecting the two pairs of branch points.
(One of these is evident in the figure; the other is harder to spot since it exits the right side and re-appears on the left.)
As $\vartheta$ varies from $0$ to $\pi$ this picture is repeated twice more, giving $2$ more BPS states each time.
Thus altogether we find that
the strong-coupling spectrum of the $SU(3)$ theory consists of $6$ distinct BPS hypermultiplets
(or $12$ if we include the antiparticles).  This agrees with the recent
result of \cite{Alim:2011ae,Alim2011a}
where the same spectrum is obtained using quiver representations.

\subsection{The theory of $9$ free hypermultiplets}\label{sec:nine-hypers}

Next let us consider $K=3$ and $C = \IC\IP^1$, with three defects, two ``full'' (at $z = \pm 1$)
and one ``simple'' (at $z = 0$).  This means that at $z = \pm 1$ we impose the condition
that $\phi_2$ has a pole of order $2$ and $\phi_3$ one of order $3$, while
at $z = 0$ we require that the discriminant $\Delta = 27 \phi_3^2 - 4 \phi_2^3$ has a pole
of order $4$.
Concretely this means we take
\begin{equation}
 \phi_2 = \frac{- 3 m^2 + a z + b z^2}{z^2 (z+1)^2 (z-1)^2} \de z^{2}, \qquad \phi_3 = \frac{2 m^3 - a m z - c z^2 + d z^3}{z^3 (z+1)^3 (z-1)^3} \de z^{3}.
\end{equation}
Here $m, a, b, c, d$ are complex parameters, related to the flavor masses which will appear below.

According to \cite{Gaiotto:2009we}
the corresponding theory $S[A_2, C, D]$ is
a theory of $9$ free hypermultiplets, transforming in the $(\mathbf{3}, \mathbf{3}, +1)$ of an
$SU(3) \times SU(3) \times U(1)$ flavor symmetry.  The mass parameter for the $U(1)$ flavor symmetry is $m$, while
those for the two $SU(3)$ are the residues $r_1^{(i)}$, $r_{-1}^{(i)}$ of $\lambda^{(i)}$ at $1$, $-1$ respectively
(these are some functions of $m, a, b, c, d$).
As a test of this statement (and of our whole picture) one
can choose some arbitrary values for $a$, $b$, $c$, $d$, $m$ and study the BPS spectrum.
We should expect to find $9$ BPS multiplets,
corresponding to the quanta of the elementary hypermultiplet fields,
 carrying charges $\gamma(i,j)$ for $1 \le i,j \le 3$.
The expected central charges of these BPS multiplets are determined by the flavor mass parameters:
\begin{equation}
 Z_{\gamma(i,j)} = 2 \I (-m + r_1^{(i)} + r_{-1}^{(j)}).
\end{equation}

We found it simplest to study the spectrum in the regime where $\abs{m} \gg \abs{r_{\pm 1}^{(i)}}$.
In this regime the $4$ branch points coalesce into two pairs,
each pair sitting very close to one of the two full punctures.
For several chosen values of parameters in this regime,
we indeed found $9$ BPS multiplets with exactly the predicted central charges.
See Figure \ref{fig:nine-hypers} for an example.
\insfigscaled{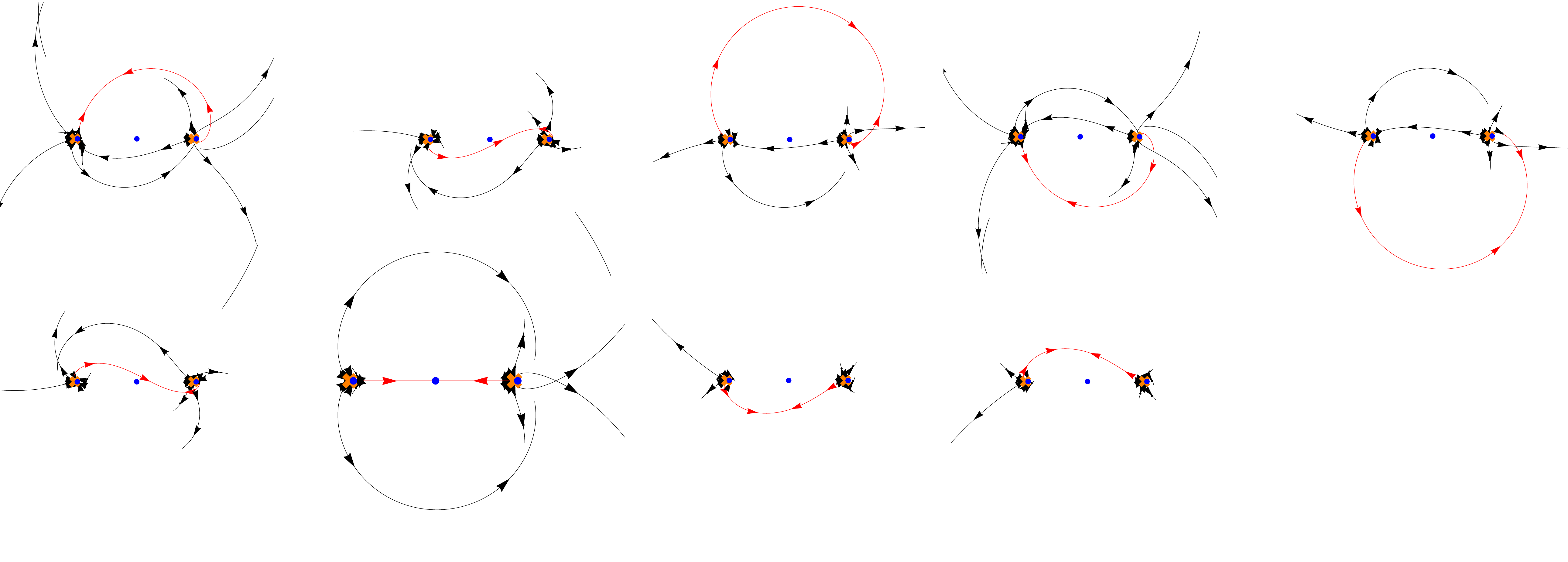}{0.28}{The cutoff spectral networks $\wnet_\vartheta[\Lambda]$ where $\Lambda e^{\I \vartheta}$ runs over the $9$ values $\half Z_{\gamma(i,j)}$, and we
have fixed parameters such that $\abs{m} \gg \abs{r_{\pm 1}^{(i)}}$.  In each network
we see a pair of $\cS$-walls meeting head-to-head, thus combining into a single path,
highlighted in red;
this corresponds to a BPS hypermultiplet with central charge $Z_{\gamma(i,j)}$.}

In addition we found $6$ extra BPS multiplets, $3$ associated to each of the $2$ full punctures,
with central charges
\begin{equation}
 Z_{\gamma_{\pm 1}(i,j)} = 2 \I (r_{\pm 1}^{(i)} - r_{\pm 1}^{(j)}) \qquad (i > j).
\end{equation}
(Taking $i < j$ in this formula gives the central charges for the corresponding antiparticles.)
In the regime we are considering, these multiplets are represented by small loops
around the punctures.
They might at first seem unexpected, but they have a natural explanation:
they do not represent 4d particles at all but rather 2d particles living on the surface defect
$\bS_z$.  Indeed,
as we emphasized in \cite{Gaiotto:2011tf}, the quantity $\omega(\gamma, a)$
in general must be interpreted as a sum of contributions from 4d particles carrying charge
$\gamma$ and 2d particles carrying the same charge. The phenomenon that a closed loop around
a puncture represents a 2d particle carrying flavor charge
also arose there, in the context of the $\IC\IP^1$ sigma model.

\section{General spectral networks and path lifting}\label{sec:spectral-networks}

As we have remarked in \S\ref{sec:k2-examples},
in the theories $S[A_1, C, D]$, each spectral network $\wnet_\vartheta$
corresponds naturally to an ideal triangulation of $C$.  Ideal triangulations are
rather flexible objects and one might study them without regard for whether they
arise from any $\wnet_\vartheta$.
In an analogous way, we now generalize the $\wnet_\vartheta$
we have worked with thus far to some purely topological objects $\wnet$,
which we will refer to as \ti{spectral networks.}

A spectral network $\wnet$ is associated to a branched cover
$\Sigma \to C$.  In contrast to the case of $\wnet = \wnet_\vartheta$ constructed in
\S\ref{sec:wnet}, now we do not require that $\Sigma \subset T^*C$.
In particular, we do not have the canonical 1-form $\lambda$ on $\Sigma$ anymore.
In consequence the $\cS$-walls making up a spectral network are not solutions
to any differential equation; rather, locally they are arbitrary paths.  We thus
gain some flexibility in the $\cS$-walls, but at the same time we lose some
topological data which were previously induced by the 1-form $\lambda$
(Seiberg-Witten differential).  We will build substitutes for those
data explicitly into our definition of ``spectral network.''

\subsection{General spectral networks}\label{subsec:GenSpecNet}

Let $C$ be an oriented real surface with a (perhaps empty) boundary.
Each connected component of the boundary is a copy of $S^1$;
on each such component fix a (nonempty) set of marked points.  Also fix a (perhaps empty) set of marked
points in the interior of $C$. We refer to the interior marked points as \emph{punctures} and
we refer to all the marked points as \ti{singular points} (although they are not singularities of $C$)
and denote them as $\mathfrak{s}_n$.  We require that there is at least one singular point.

Let $\Sigma \to C$ be a $K$-fold branched covering which is unramified
over the boundaries and the singular points.
Let $C'$ be $C$ minus the branch points.
For simplicity we assume the branch points are all simple,
so the monodromy around each branch point just exchanges
two sheets of $\Sigma$. (This condition can likely be relaxed at the price of a
more cumbersome definition.)

A \emph{spectral network subordinate to
the covering $\Sigma$} is a collection
\begin{equation}
\wnet = ( o(\mathfrak{s}_n), \{z_\mu\}, \{p_c\} )
\end{equation}
where the symbols refer to the following data:

\begin{description}
\item[D1.] For each singular point $\mathfrak{s}_n$, $o(\mathfrak{s}_n)$
is a \ti{partially ordered subset} of the set of sheets of $\Sigma$
over a neighborhood of $\mathfrak{s}_n$.  $o(\mathfrak{s}_n)$ must contain
at least two elements, and if $\fs_n$ is a puncture, $o(\mathfrak{s}_n)$ must contain
\ti{all} of the sheets over a neighborhood of $\fs_n$.

\item[D2.] $\{z_\mu\}$ is a locally finite collection of points on $C'$, called \emph{joints}.

\item[D3.] $\{p_c\}$ is a finite or countable collection of
closed segments (i.e. images of
\emph{embeddings} of $[0,1]$ into $C$), called \ti{walls} or \emph{streets} (depending
which metaphor is more useful in a given context, cf. footnote \ref{fn:wall-street}).
For each orientation
$o$ of the street $p_c$, $p_c$ is labeled with an ordered pair of distinct sheets of the
covering $\Sigma \to C$ over $p_c$.  Reversing the orientation reverses this ordered pair of sheets.
So $p_c$ comes with two labels which we could
write as $(o, ij)$ and $(-o, ji)$.

\end{description}

The data must satisfy the following conditions:

\begin{description}

\item[C1.] The segments $p_c$ cannot cross one another (but they are allowed to
have common tangents).  Each $p_c$ must begin on a branch point or a joint,
and must end on a joint or a singular point.  Any compact subset of $C'$
intersects only finitely many segments.

\item[C2.] Around each branch point $\fb$ there is a neighborhood where $\wnet$
looks like Figure \ref{fig:branch-point-trajectories}.  That is,
each branch point of type $(ij)$ is an endpoint of three
streets which carry labels $(o,ij)$ or $(o,ji)$, and
the streets encountered consecutively traveling around a loop around $\fb$
have oppositely ordered sheets.

\item[C3.] Around each joint $z_\mu$ there is a neighborhood where
$\wnet$ looks like Figure \ref{fig:trajectories-meeting-commuting},
Figure \ref{fig:ijkl-collisions}, or Figure \ref{fig:ijkl-collision-birth}.

\item[C4.] If a segment with label $ij$ ends at a singular point $\mathfrak{s}_n$,
then $i$ and $j$ lie in the ordered subset $o(\mathfrak{s}_n)$,
and with respect to the ordering of $o(\mathfrak{s}_n)$ we have $i < j$.

\end{description}

Our definition of ``spectral network'' is somewhat provisional.
With an eye toward the future let us mention one natural
generalization:  we could have relaxed the requirement that there is at least
one singular point $\fs_n$, and allowed
the segments to be infinite in one direction.  This generalization would be needed
if we want our definition to encompass the networks $\wnet_\vartheta$ in theories $S[\fg, C]$
where the set of defects is actually \ti{empty}.
The resulting spectral networks would be expected to look much more complicated --- e.g.
the streets may well be dense on $C$.  Nevertheless, as we described in the introduction,
we think that it should be possible to extend everything described in this paper to this case.

\subsection{Canonical examples:  the $\wnet_\vartheta$}

The $\wnet_{\vartheta}$ discussed in previous sections, which arose naturally from the
physics of theories of class $S$, are essentially examples of spectral networks.

The data D2 and D3 above appeared in \S\ref{sec:wnet}.
The datum D1 is determined by the behavior of the $\lambda^{(i)}$ near $\fs_n$.  We will not describe it
explicitly here, except in the basic case where $\fs_n$ is a full defect:  in that
case $o(\fs_n)$ consists of all the sheets,
ordered by $\re(e^{-\I \vartheta} m^{(i)})$.  As we saw in \S\ref{sec:falling-down-holes},
the $\wnet_\vartheta$ indeed obey our condition C4 with this
ordering.

We say $\wnet_\vartheta$ are ``essentially'' spectral networks
because of two technical points:

\begin{enumerate}
 \item In the networks
$\wnet_\vartheta$ as we defined them, it is possible for a wall to ``die''
at a joint, as indicated in Figure \ref{fig:ijkl-collision-death}.
In contrast, our present definition of spectral network
we do not allow this:  the walls always continue through joints.

This difference arises because in $\wnet_\vartheta$
we included a wall $p$ only if it supports a charge $a$ with $\mu(a,p) \neq 0$.
We could have dropped the requirement $\mu(a,p) \neq 0$, thus including some additional
``invisible'' walls.  This would not have changed anything in previous
sections, except to make the notation a bit more cumbersome.

\item $\wnet_\vartheta$ can fail to be a spectral network because it has a wall which
is not of type $ij$ or $ji$ but accidentally runs into a branch point of type $(ij)$.  It would be
possible to extend the definition of spectral network to include this situation, but for simplicity
we have avoided it.
\end{enumerate}

\subsection{Soliton content}\label{subsec:Soliton-content}

To a spectral network we can associate some additional data
having to do with \ti{solitons}, which we now define.

A \emph{soliton} $s(z)$, where $z\in p_c$, is an immersion of $[0,1]$ into $\Sigma$,
which begins and ends on preimages of $z$, and such that its projection to $C$ lies
in the spectral network $\wnet$ (or more precisely in a very small neighborhood of $\wnet$;
this correction is necessary because at joints we smooth out
the sharp corners.)  If $p_c$ carries the label $(o, ij)$,
then we say $s(z)$ is \emph{compatible with $o$}
if it begins on $z^{(i)}$ and ends on $z^{(j)}$, and the projection of
$s(z)$ to $C$ begins with orientation $-o$ and ends with orientation $o$.
The soliton content is, for each street $p_c$ and each point $z \in p_c$, a pair of sets of
solitons $\CS^o_c(z)$, $\CS^{-o}_c(z)$, such that
the solitons in $\CS^{o}_c(z)$ are compatible with $o$ and
those in $\cS^{-o}_c(z)$ are compatible with $-o$.

The soliton sets must satisfy some rules, which we will refer to
as \emph{solitonic traffic rules}.  Actually the rules come in two variants:  either
\ti{American} or \ti{British}.  In order to state the rules we introduce a
resolution of the streets of the spectral network, regarding each street as resolved into
two oriented ``lanes'' infinitesimally displaced from one another, either in the ``American'' or the ``British'' fashion.  See Figure \ref{fig:resolutions}.
\insfig{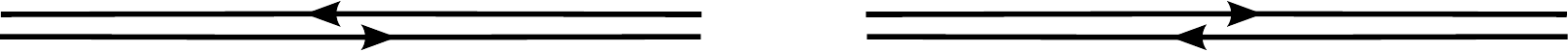}{The American (left) and British (right) resolutions of a street.}
Note that this definition uses the
orientation of $C$.  The American and British traffic rules are thus related to one another by a
reflection in the plane.  In what follows we show the British rules only; to get the American
rules simply requires a little reflection.

\begin{description}
\item[ST1.] As $z$ moves continuously along a street $p_c$, the soliton sets
$\CS_c^{\pm o}(z)$ evolve continuously, by the natural parallel transport; in
other words, the soliton sets do not ``jump.''
With this in mind, we abuse notation by writing the soliton sets
simply as $\CS_c^{\pm o}$, suppressing the trivial $z$ dependence.

\insfig{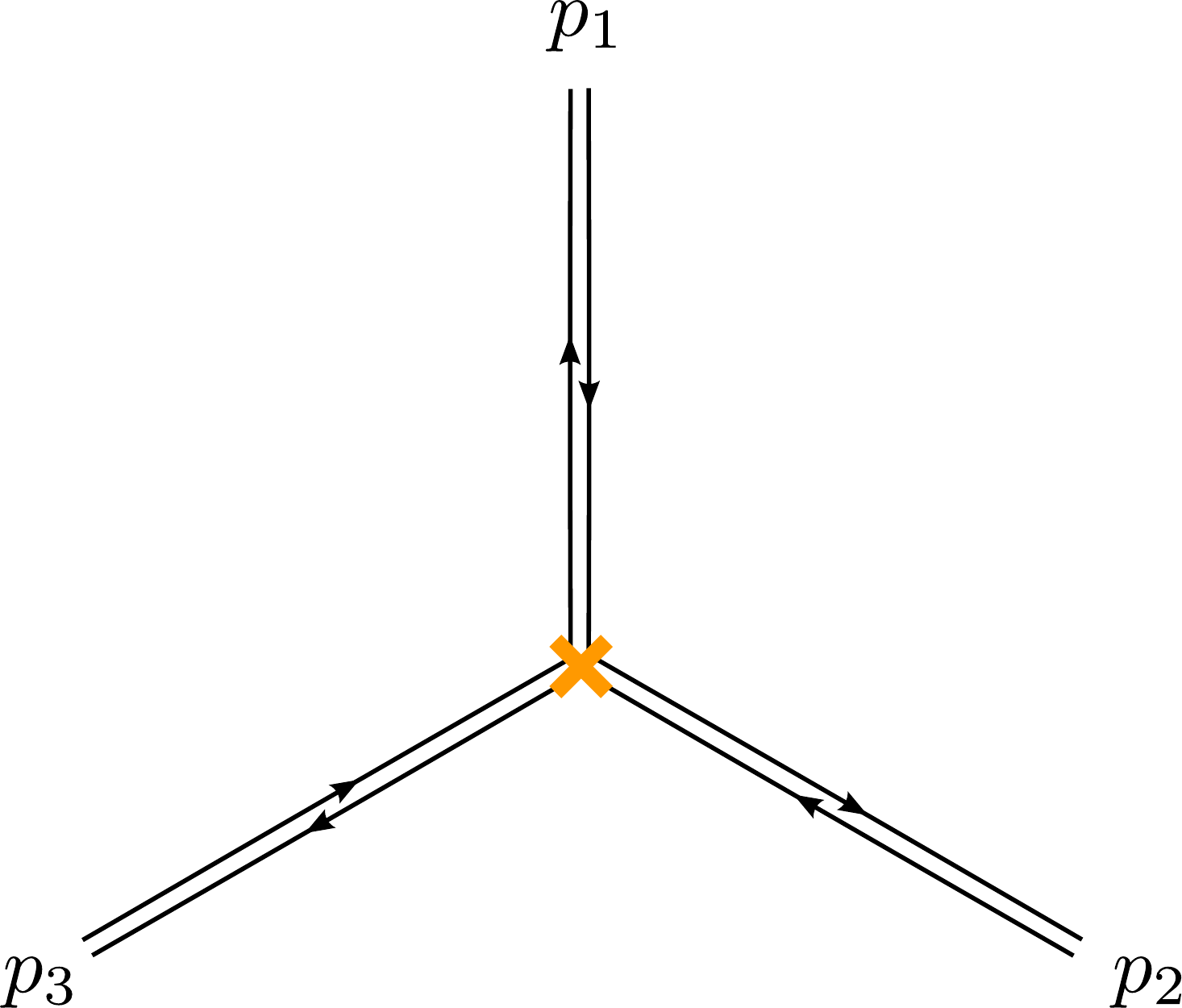}{A spectral network in the vicinity of a branch point.}
\item[ST2.]
Let $\fb$ be a branch point.  The network $\wnet$ then looks like Figure
\ref{fig:soliton-rules-2} in the vicinity of $\fb$.
There are three streets $p_c$ emerging from $\fb$.
We denote the set of orientations of each street by $\{\iin, \out\}$.
For each $c$, define the \emph{light soliton} $s_c$ as follows:
if $p_c$ carries the label $(\out,ij)$,
then $s_c(z)$ begins on $z^{(i)}$, travels along $p_c^{(i)}$
back to the ramification
point on $\Sigma$ covering $\fb$, and returns along $p_c^{(j)}$ to
$z^{(j)}$.  Then,
the soliton sets are related by
\begin{equation}
\CS_1^{\out} = \CS_3^\iin \cup \{ s_1 \}, \quad
\CS_2^{\out} = \CS_1^\iin \cup \{ s_2 \}, \quad
\CS_3^{\out} = \CS_2^\iin \cup \{ s_3 \}.
\end{equation}
(In this equation the solitons are understood to be evolved continuously from
one street to the other.  Since there is a branch point at $\fb$, this continuous
evolution depends on which path we follow to go from one street to the next;
we follow the \ti{short} path, i.e. we go around an arc of length $2 \pi / 3$,
not $4 \pi / 3$.)

\insfigscaled{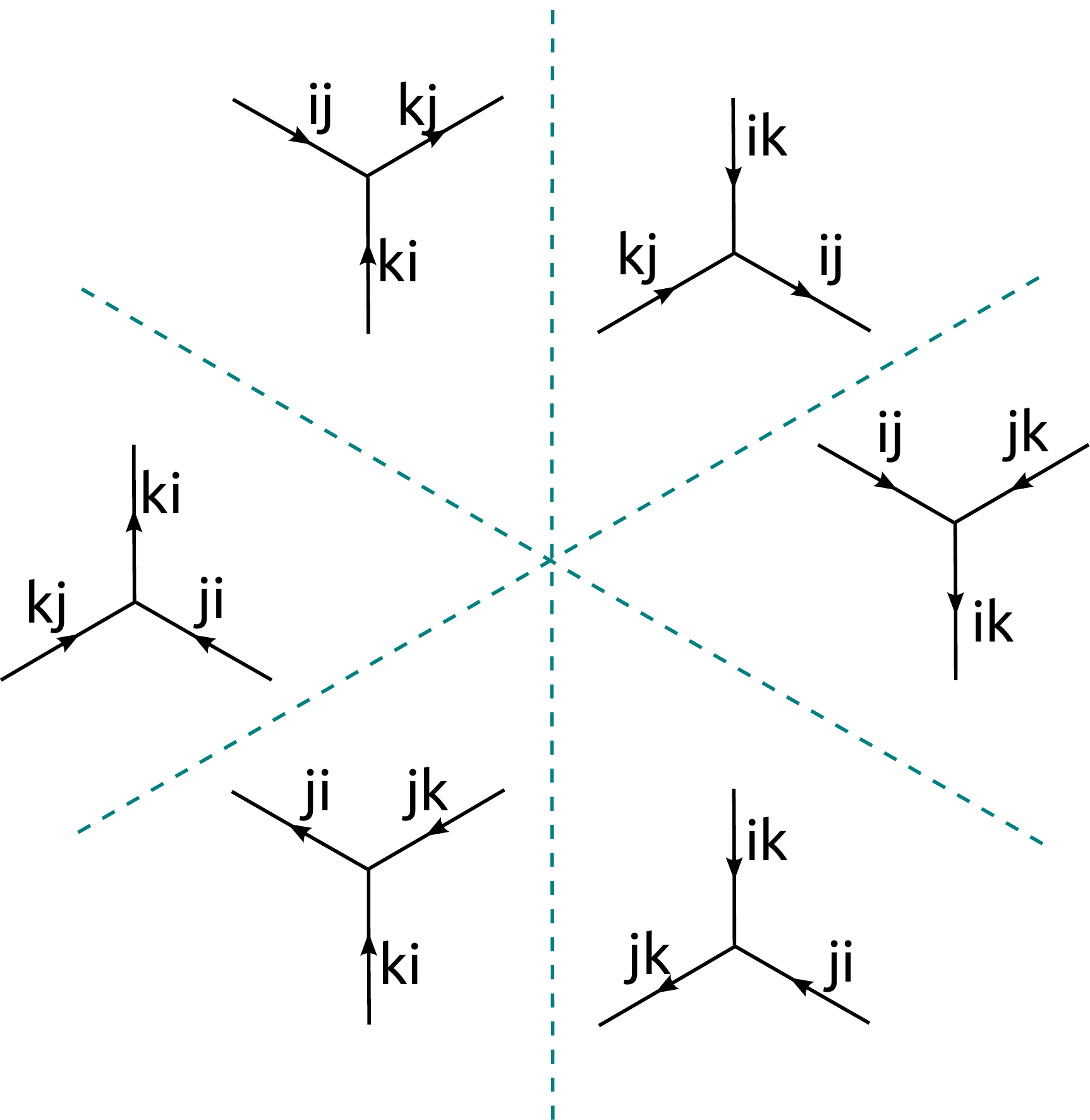}{0.34}{The rule for constructing outgoing solitons from a
six-way junction, with the British resolution.
We show six allowed types of junction,
one in each ``Weyl chamber'' around the joint.  The most general outgoing soliton
is constructed by a diagrammatic prescription
as follows.  We fix any set of incoming solitons; for each one we draw
an incoming line on the diagram, coming in along the appropriate direction,
and on the left side of the median.
We then combine these incoming solitons with junctions of the allowed types,
to make a single outgoing line.  Every diagram so obtained determines an outgoing
soliton in a natural way, by concatenation of the incoming solitons.}

\insfigscaled{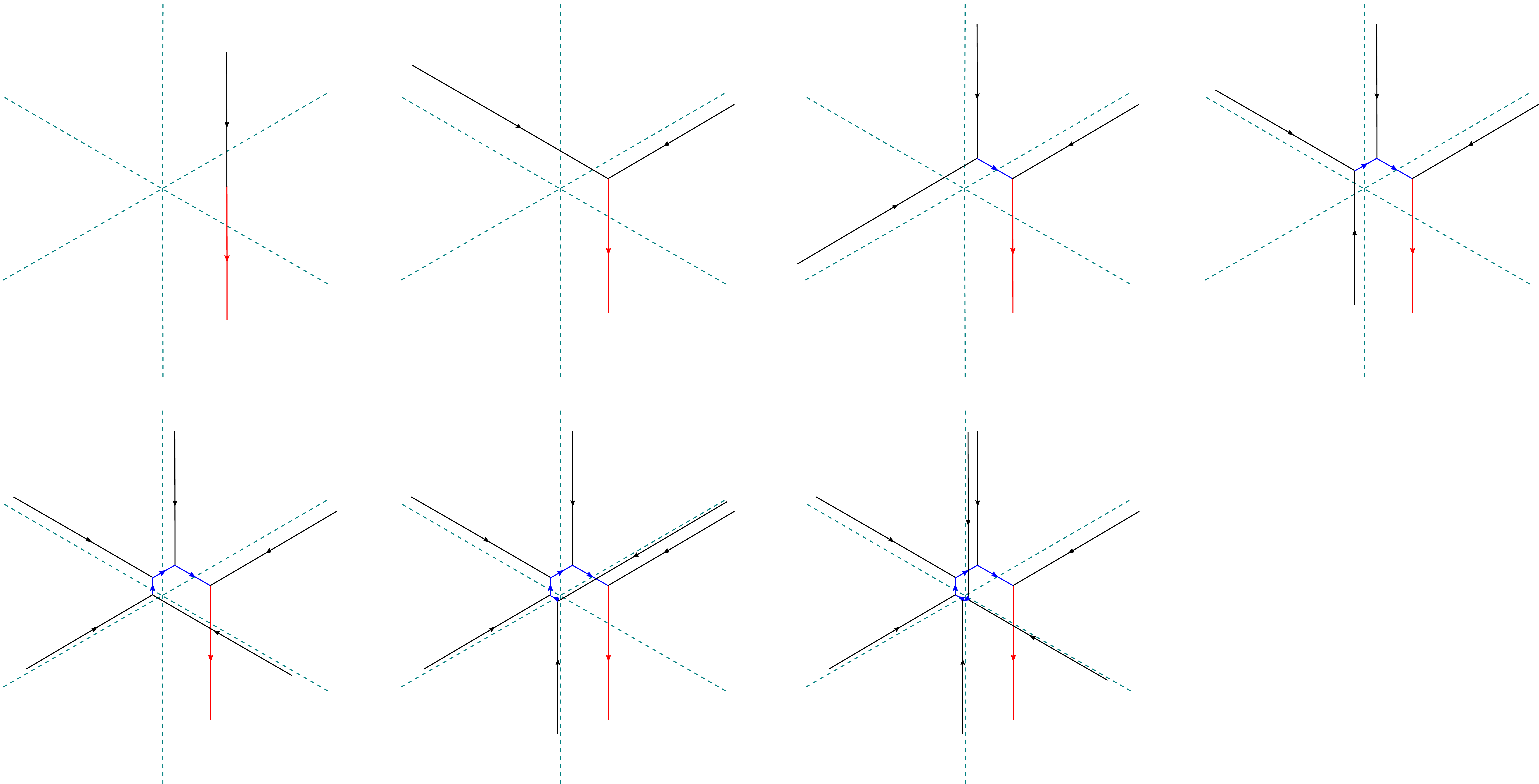}{0.15}{Some examples of diagrams representing outgoing solitons.
Incoming solitons are represented by black lines, the outgoing path by a red line.
(The first example is exceptional in that the incoming line is the same as
the outgoing one.)}

\item[ST3.] In a sufficiently small neighborhood of
a joint of the ``four-way junction''
type shown in Figure \ref{fig:trajectories-meeting-commuting} (where $ij$ and $kl$ walls meet),
the soliton sets vary continuously, i.e. the outgoing soliton sets are equal to the
incoming ones.
In a sufficiently small neighborhood of a joint of the ``six-way junction'' type
(where $ij$, $jk$ and $ki$ walls meet),
the outgoing soliton sets are
determined by the incoming ones according to the
rules shown in Figures \ref{fig:joint-rule}, \ref{fig:joint-rule-examples}.
(A motivation for this peculiar-looking rule
is explained in Appendix \ref{app:joint-rules}.)

\end{description}

Importantly, for any spectral network $\wnet$, the
solitonic traffic rules (including the choice of American or British)
uniquely determine the soliton content.
To prove this claim one can consider a discrete
version of the mass filtration of the spectral networks $\wnet_\vartheta$:
just define the \ti{length} of a soliton to be the number of walls in its
projection to $C$, and construct the soliton content by induction on length
using the solitonic traffic rules.

It is convenient to distinguish two different possibilities for the soliton content.
Consider the two soliton sets $\CS_c^o$, $\CS_c^{-o}$ on a street $p_c$.
If both are nonempty, we refer to $p_c$ as a
\emph{two-way street}.  If exactly one is empty, $p_c$
is a \emph{one-way street}, and in that case we write $\CS_c$ for the single
nonempty soliton set.  (It never happens that both are empty.)
If there are no two-way streets
we say that $\wnet$ is a \emph{nondegenerate} spectral network; otherwise we
say it is \emph{degenerate}.

If $\wnet$ is nondegenerate, the solitonic traffic rules
simplify considerably, to the following.
Consider the joint of Figure \ref{fig:ijkl-collisions}.  We continue the $ij$ and
$jk$ soliton sets continuously through the joint.
For the $ik$ wall, after passing through the joint we add to the set
of $ik$ solitons all new solitons of the
following description:  the new soliton begins at $z^{(i)}$,
projects on $C$ to the path going back to the joint,
then follows one of the the $ij$ solitons from
sheet $i$ to sheet $j$ above the joint, then follows one of the $jk$ solitons from
sheet $j$ to sheet $k$ above the joint, then returns on sheet $k$ on the
trajectory projecting to the $ik$ wall ending at $z^{(k)}$.
In particular, these simplified rules do not depend on whether we started out with
American or British traffic rules (as we should expect since we are
considering the case where there are no two-way streets.)

So for nondegenerate spectral networks the soliton content
is unique, while for degenerate ones there are two possible soliton contents,
one following American traffic rules and one British.

In previous sections we considered integers
$\mu(p,a)$ attached to the spectral networks $\wnet_\vartheta$.  These integers are determined
by the soliton content of $\wnet_\vartheta$ --- indeed they are simply
counts of the solitons with appropriate signs:  $\mu(p_c, a) X_a = \sum_{\nu \in \CS_c} X_{[\tilde s_\nu]}$.
So the soliton content is a slight extension of the $\mu(p,a)$ to keep track of the actual solitons,
not only their number.  This extension will actually not be used for anything in this paper (all
of our constructions really depend only on $\mu$)
but we believe it may be useful in the future.

\subsection{Path lifting}\label{subsec:Path-Lifting}

In \S\S\ref{sec:F-basics}-\ref{sec:wnet}
we studied at great length the generating functions $F(\wp, \vartheta)$ of
framed 2d-4d BPS degeneracies.
Let us set aside the physical meaning of these functions for a moment and just
think of them as some interesting mathematical objects.
We found in \S\ref{subsec:Compute-FP} that $F(\wp, \vartheta)$ could
be completely constructed from the datum of the spectral network $\wnet_\vartheta$
together with its soliton content.
One of the main motivations of our definition of spectral network is that,
given \ti{any} spectral network, we can make a very similar construction.
In this section we describe that construction.  For notational convenience we
consider only the case of a nondegenerate spectral network, but what we write
has an obvious extension to the degenerate case.

We will be rather brief since
everything is parallel to what we did in \S\ref{subsec:Compute-FP}.
However, we slightly modify
\S\ref{subsec:Compute-FP}, in two respects.  First,
with an eye toward future applications,
instead of \ti{homology} classes of open paths
on $\tilde\Sigma$, we will keep track of \ti{homotopy} classes.
This turns out to be no more difficult.
We will generally denote the homotopy objects with a bold letter to distinguish
them from their homology cousins.  Second, instead of considering smooth paths $\wp$ in $C$
and using their canonical lift to $\tilde C$, it will be convenient to
work with arbitrary paths $\twp$ on $\tilde C$ from the beginning.
This will involve lifting many objects from $C$ to $\tilde C$ or from $\Sigma$
to $\tilde \Sigma$; we always use a tilde
to denote the lifted objects.

For any homotopy class $\ba$ of open paths on $\tilde\Sigma$, we introduce a corresponding
formal variable $\bX_\ba$,
subject to the relation that if $\ba$ and $\ba'$ project to the same homotopy class on
$\Sigma$
then we have
\begin{equation} \label{eq:bold-winding-rule-2}
 \bX_\ba / \bX_{\ba'} = (-1)^{w(\ba,\ba')}.
\end{equation}
Now, given a spectral network $\wnet$, let $\tilde \wnet$ be the preimage of $\wnet$ on
$\tilde C$.  Given a
path $\twp$ on $\tilde C$ whose endpoints are not on $\tilde \wnet$,
we are going to define a formal sum of these variables, with integer coefficients:
\begin{equation} \label{eq:F-expand}
 \bF(\tilde \wp, \wnet) = \sum_{\ba} \fro'(\tilde \wp, \wnet, \ba) \bX_{\ba}.
\end{equation}
(In \S\ref{sec:gen-func} we met a similar expansion for which the coefficients $\fro'$
were interpreted as framed 2d-4d BPS degeneracies.  In the more general setting of this section,
we do not know what the physical interpretation of $\fro'$ should be.)

The assignment $\bF(\cdot, \wnet)$ will obey two important properties:
\begin{itemize}
 \item For two concatenatable paths $\twp$, $\twp'$ on $\tilde C$,
\begin{equation} \label{eq:composition-bold}
\bF(\twp \twp', \wnet) = \bF(\twp, \wnet) \bF(\twp', \wnet).
\end{equation}

\item If $\twp$ and $\twp'$ are two paths on $\tilde C$ which project to the same
homotopy class on $C$,
\begin{equation} \label{eq:twisted-homotopy}
\bF(\twp, \wnet) = (-1)^{w(\twp, \twp')} \bF(\twp', \wnet).
\end{equation}
(In particular, if $\twp$ and $\twp'$ are homotopic,
then $\bF(\twp, \wnet) = \bF(\twp', \wnet)$.)
\end{itemize}
Given any path $\twp$ on $\tilde C'$ we first define
\begin{equation}
 \bD(\twp) = \sum_{i=1}^K \bX_{\twp^{(i)}}.
\end{equation}
where  $\twp^{(i)}$ is the open path given by
lifting the initial point of $\twp$ to the $i$-th sheet and then using the
canonical connection on $\tilde \Sigma \to \tilde C$ to lift the path.
For any $\twp$ which does not cross $\tilde \wnet$ we have simply
\begin{equation} \label{eq:F-no-cross}
 \bF(\twp, \tilde\wnet) = \bD(\twp).
\end{equation}
For any $\twp$ which crosses $\tilde \wnet$ exactly once at a wall $p_c$, we have
\begin{equation} \label{eq:F-one-cross}
 \bF(\twp, \tilde\wnet) = \bD(\twp_+) \left(1 + \sum_{\nu \in \CS_c} \bX_{\ba(\nu,\twp)}\right) \bD(\twp_-),
\end{equation}
where $\ba(\nu,\twp)$ is a particular lift of the soliton $s_\nu$
to $\tilde\Sigma$, described in the next paragraph.

Let $\wp$ be the projection of $\twp$ to $C$.  Let $z$ be the point of intersection between $\wp$ and $\wnet$.
Let $\tz$ be the point of $\twp$ lying over $z$, dividing $\twp$
into $\twp_+ \twp_-$.
Let $o_+$ be a tangent vector at $z$ oriented along $\wnet$,
and $o_-$ a tangent vector at $z$ oriented oppositely to $\wnet$.
Let $A_+$ be some
path from $\tz$ to $o_+$ in the circle fiber $\tilde C_z$,
and $A_-$ a path from $o_-$ to $\tz$ in $\tilde C_z$.  The
concatenation $A = A_- + A_+$ is thus a path from $o_-$ to its antipode
$o_+$ in $\tilde C_z$;
we fix our choices of $A_\pm$ so that $A$ is homotopic to
a simple arc which covers half of $\tilde C_z$, and this arc includes
the tangent vector $\wp'(z)$.
Each soliton $s_\nu(z)$ has a canonical lift to a smooth path
$\tilde s_\nu(z)$ on $\tilde \Sigma$ from $o_+^{(i)}$ to $o_-^{(j)}$.
The path $\ba(\nu,\twp)$ on $\tilde \Sigma$ from $\tz^{(i)}$ to $\tz^{(j)}$ is defined by
\begin{equation}
 \ba(\nu,\twp) = A^{(i)}_+ + \tilde s_\nu(z) + A^{(j)}_-.
\end{equation}

Finally, to define $\bF(\twp, \wnet)$ for an arbitrary path $\twp$ on $\tilde C$ whose endpoints are
not on $\tilde\wnet$, we perturb $\twp$ slightly
so that its intersections with $\tilde \wnet$ are all transverse, break it into
pieces which meet $\tilde \wnet$ at most once, and use \eqref{eq:F-no-cross},
\eqref{eq:F-one-cross} and the composition property
\eqref{eq:composition-bold}.

Our construction makes the composition property \eqref{eq:composition-bold} manifest.
The same arguments given in \S\ref{sec:homotopy-invariance} apply
to our current construction and show the twisted homotopy invariance
\eqref{eq:twisted-homotopy}.

The path-lifting rule $\bF$ will be put to some mathematical use in
\S \ref{sec:moduli} below.
As mentioned in \S \ref{sec:Introduction}, we are hopeful that it will also have
some interesting physical applications, as well as connections to
upcoming work of Goncharov and Kontsevich.

\section{Coordinates for moduli of flat connections}\label{sec:moduli}

In this last section we discuss a mathematical application of spectral networks.
Given a spectral network $\wnet$ subordinate to a $K$-fold covering $\Sigma \to C$, we will construct
a map
\begin{equation}
 \Psi_\wnet: \CM(\Sigma, GL(1); \fm) \to \CM_F(C, GL(K); \fm)
\end{equation}
where $\CM(\Sigma, GL(1); \fm)$ is a moduli space of twisted flat $GL(1)$-connections $\nabla^\ab$ on $\Sigma$
and $\CM_F(C, GL(K); \fm)$ is a moduli space of twisted flat $GL(K)$-connections $\nabla$ on $C$, decorated
by some ``flag data'' as we explain below.
This map is a local symplectomorphism, and conjecturally 1-1 onto its image.

Roughly speaking,
$\nabla = \Psi_\wnet(\nabla^\ab)$ is obtained by a two-step process.  We first push forward the $GL(1)$ connection $\nabla^\ab$ from
$\Sigma$ to $C$.  This gives a flat $GL(K)$ connection on the complement of the branch locus in $C$, which is
everywhere diagonal.  This connection however cannot be extended over the branch points, because it has
monodromy around them.  To deal with this problem we cut $C$ into pieces along the network $\wnet$,
and then reglue the connection with a nontrivial (and non-diagonal) transition function, controlled by
the soliton content of $\wnet$.
This process eliminates the monodromy around branch points, while not introducing unwanted
monodromy anywhere else.  We call this construction ``nonabelianization.''

The space $\CM(\Sigma, GL(1); \fm)$ is a torsor for $(\IC^\times)^{2 g_{\bar\Sigma}}$,
where $\bar\Sigma$ is the closure of $\Sigma$,  so another way to
read $\Psi_\wnet$ is as a \ti{local coordinate system} on $\CM_F(C, GL(K); \fm)$.
These coordinate systems are closely related to the ``cluster coordinates''
introduced by Fock and Goncharov \cite{MR2233852}.  In the case $K=2$, it is straightforward to see that
our coordinates in fact coincide with the cluster coordinates.  In the case $K>2$ a fully explicit description
of the most general cluster coordinate system is not available in the literature; we conjecture that
the $\Psi_\wnet$ actually give the most general cluster coordinates, thus filling this gap.

One point in favor of this conjecture is that when
$\wnet$ is of a special form, $\Psi_\wnet$ yields a coordinate system which was explicitly described by
Fock and Goncharov; this will be described in \cite{gmn6-to-appear}.

Although we are emphasizing its mathematical content in this section, this
construction also has a natural physical meaning, which appears most clearly
when the theory is compactified from four to three dimensions on $S^1$.
Indeed, the expansion \eqref{eq:pt-exp} of the parallel transport of $\nabla$
as a linear combination of parallel transports $\bX_a(\nabla^\ab)$ is precisely the
``Darboux expansion'' of the vacuum expectation value of the supersymmetric interface
associated with $\tilde \wp$.  This point of view was described at length in \cite{Gaiotto:2011tf}.

\subsection{Nonabelianization}\label{subsec:PushAbelian}

We now explain how, given a nondegenerate spectral network $\wnet$ subordinate
to a covering $\Sigma \to C$, and given also a
twisted flat rank 1 connection over $\Sigma$, we
construct a corresponding twisted flat rank $K$ connection over $C$.

By a \ti{twisted flat rank $K$ connection}
over a real surface $S$ we mean a pair of a complex rank $K$ vector bundle
on $\tilde S$ and a flat connection therein, such that the
holonomy around each fiber of $\tilde S$ is $-1$.
This is a very small modification of the usual notion of a rank $K$ connection over $S$.
In particular, given a spin structure on $S$, there is a canonical
isomorphism between the moduli space of twisted flat connections and that
of ordinary connections.\footnote{To obtain this isomorphism we use the fact that
a spin structure is the same thing as a fiberwise double covering of $\tilde S$;
then given a twisted flat connection, pulling back to this double cover gives a new flat
connection with holonomy $+1$ around the fibers, which then descends to a flat connection
over $S$.}
Nevertheless, this slight twisting is important
in the construction we are about to describe.

Now suppose given a rank 1 twisted flat connection $(\CL,\nabla^\ab)$ over $\Sigma$.
Recall that $C'$ denotes $C$ minus the branch points of the covering $\Sigma \to C$.  Let
$\pi: \tilde \Sigma \to \tilde C'$ be the projection.  Then define the complex
rank $K$ vector bundle
\begin{equation}
 E = \pi_*(\CL).
\end{equation}
The fiber of $E$ at any $\tz \in \tilde C'$ is simply
\begin{equation}\label{eq:simple-pushforward}
E_{\tz} = \bigoplus_i \CL_{\tz^{(i)}}.
\end{equation}

Given the flat rank 1 connection $\nabla^\ab$ on $\CL$,
there is a canonical ``pushforward'' flat connection $\pi_*(\nabla^\ab)$ on $E$.
But the reader should beware that $\pi_*(\nabla^\ab)$  is \ti{not} the connection
we are trying to construct.  Indeed, while $\pi_*(\nabla^\ab)$ is flat on $\tilde C'$,
it has nontrivial holonomy around any small loop $\ell_\fb$ in $\tilde C'$ linking
the fiber over a branch point $\fb$; this monodromy is
induced by the permutation of sheets attached to $\fb$.
It follows that $\pi_*(\nabla^\ab)$ cannot be extended
to a flat connection over $\tilde C$, since on $\tilde C$ the loop $\ell_\fb$ would
be contractible.
Our construction will modify $(E, \pi_*(\nabla^\ab))$ in a way
which eliminates this holonomy.

The key is the path-lifting rule $\bF$ defined in \S \ref{subsec:Path-Lifting}.
Let $C'_\wnet := C' \setminus \wnet$.
Given any path $\tilde \wp$ from $\tz_1 \in \tilde C'_\wnet$ to $\tz_2 \in \tilde C'_\wnet$
we have built a formal sum of paths on $\tilde\Sigma$, given by \eqref{eq:F-expand}:
\begin{equation} \label{eq:F-expand-repeat}
 \bF(\tilde \wp, \wnet) = \sum_{\ba} \fro'(\tilde \wp, \wnet, \ba) \bX_{\ba}.
\end{equation}
Each $\ba$ in the sum \eqref{eq:F-expand-repeat}
is a homotopy class of paths on $\tilde\Sigma$ from $\tz_1^{(i)}$
to $\tz_2^{(j)}$ for some $i$, $j$.
Let
\begin{equation}
\bX_\ba(\nabla^\ab) \in \Hom(\CL_{\tz_1^{(i)}}, \CL_{\tz_2^{(j)}})
\end{equation}
denote the parallel transport of $\nabla^\ab$ along any path in the class $\ba$.
The fact that $\nabla^\ab$ is a flat connection implies that $\bX_\ba(\nabla^\ab)$ depends
only on the class $\ba$
and not on the particular path, and the fact that $\nabla^\ab$ has holonomy $-1$
around the fibers of $\tilde \Sigma$ implies that $\bX_\ba(\nabla^\ab)$
obeys the relation \eqref{eq:bold-winding-rule-2}.
Hence the map $\bX_\ba \to \bX_\ba(\nabla^\ab)$ is well defined.

We now define a
new operator $\bF(\tilde \wp, \wnet, \nabla^\ab)$
just by replacing $\bX_\ba \to \bX_\ba(\nabla^\ab)$
in \eqref{eq:F-expand-repeat}:
\begin{equation} \label{eq:pt-exp}
 \bF(\tilde \wp, \wnet, \nabla^\ab) := \sum_{\ba} \fro'(\tilde \wp, \wnet, \ba) \bX_\ba(\nabla^\ab) \in \Hom(E_{\tz_1}, E_{\tz_2}).
\end{equation}
$\bF(\tilde \wp, \wnet, \nabla^\ab)$ has precisely
the formal properties one would expect
if it were the parallel transport operator in some flat vector bundle:  it is homotopy invariant
and behaves properly under composition.
The key to our construction is that indeed
\ti{$\bF(\tilde \wp, \wnet, \nabla^\ab)$ is the parallel
transport operator from $\tz_1$ to $\tz_2$ for a twisted rank $K$
flat connection $({\hat E}, \nabla)$ over $C$.}
$({\hat E}, \nabla)$ will be constructed in an essentially
tautological way to make this sentence true.

The most direct way of building $({\hat E}, \nabla)$ is to construct its
sheaf ${\mathbb F}$ of flat sections, as follows.  For
any open set $\CU \subset \tilde C$, ${\mathbb F}(\CU)$ consists of all sections $\psi$
of $E$ over $\CU \cap \tilde C'_\wnet$ such that
for any path $\tilde \wp \subset \CU$
with endpoints $\tz_1, \tz_2 \in \tilde C'_\wnet$, we have
\begin{equation}
\psi(\tz_1) \bF(\tilde \wp, \wnet, \nabla^\ab) = \psi(\tz_2).
\end{equation}
(Note that the parallel transport acts from the right, in accordance
with Appendix \ref{app:Conv-conv}.)
Any $\psi \in {\mathbb F}(\CU)$ is obviously determined by its value at a single
$\tz_0 \in \CU \cap \tilde C'_\wnet$.  Conversely,
if $\CU$ is contractible, then any chosen $\psi(\tz_0) \in E_{\tz_0}$ can
be extended to $\psi \in {\mathbb F}(\CU)$ (to prove this one uses
the invariance of $\bF(\tilde \wp, \wnet, \nabla^\ab)$ under homotopies of $\tilde \wp$
and the composition law \eqref{eq:composition-bold}.)
So $\mathbb F$ is a locally constant sheaf of $K$-dimensional vector spaces.
Giving such a sheaf is equivalent to giving a rank $K$ vector bundle with
connection; this is our desired $({\hat E}, \nabla)$.

It will be useful in what follows to have a more concrete description of $({\hat E}, \nabla)$.
The stalk of $\mathbb F$ at any $\tz \in \tilde C'_\wnet$ is just $E_{\tz}$, so
on $\tilde C'_\wnet$ we have a canonical isomorphism ${\hat E} \simeq E$.  Moreover, for
paths $\tilde \wp \subset \tilde C'_\wnet$ we have $\bF(\tilde \wp, \wnet) = \bD(\tilde \wp)$,
from which it follows that $\bF(\tilde \wp, \wnet, \nabla^\ab)$ is just the
parallel transport of $\pi_*(\nabla^\ab)$.  So on $\tilde C'_\wnet$ we have
simply $({\hat E}, \nabla) \simeq (E, \pi_* \nabla^\ab)$.
Now, $\tilde C'_\wnet$ is divided into various connected components $\CU_{\alpha}$,
separated from one another by the lifts of walls of $\wnet$, which are topologically cylinders in $\tilde C$.
We want to know how to patch together the bundle ${\hat E}$ across these cylinders.
Given a wall $p_c$ whose lift $\tilde p_c$ separates components $\CU_{\alpha}$ and $\CU_{\beta}$, we define
a section of $\End(E) \vert_{\tilde p_c}$ by
\begin{equation}\label{eq:tran-fn}
 \CT_c^{\alpha,\beta}(\tz) = 1 + \sum_{\nu \in {\CS}_c(z)} \bX_{\ba(\nu, \twp)}(\nabla^\ab)
\end{equation}
where $\twp$ denotes a short path which crosses from
$\tz_1 \in \CU_{\alpha}$ to $\tz_2 \in \CU_{\beta}$,
and we take the limit
as $\tz_1$ and $\tz_2$ approach the common point $\tz$ of $\tilde p_c$.
Recall here that $\CS_c(z)$ is the soliton set of $\wnet$ at $z$ ($\wnet$
is nondegenerate, so only one of the soliton sets is nonempty.)
The bundle ${\hat E}$ can be realized by extending $E$ to the closure of each component
and then gluing along the boundary cylinders:
\begin{equation}
 {\hat E} = \left( \bigsqcup_\alpha E \vert_{\bar \CU_{\alpha}} \right) / \sim
\end{equation}
where the equivalence relation $\sim$ identifies $(\psi, \tz) \in E \vert_{\bar \CU_{\alpha}}$
with $(\CT_{c}^{\alpha,\beta} \psi, \tz) \in E \vert_{\bar \CU_{\beta}}$ whenever $\tz$ lies on the cylinder
$\tilde p_c$.

\subsection{Singularities and flags} \label{sec:flags}

The twisted flat connection $({\hat E}, \nabla)$
produced by our nonabelianization construction carries a
bit of extra structure which will be important in what follows.
In this section we briefly describe it.

Let us consider
$({\hat E}, \nabla)$ in a neighborhood $\CU \subset \tilde C$ of one of the singular points
$\mathfrak{s}_n$.  On $\CU$ we can trivialize the covering $\tilde\Sigma \to \tilde C$ and hence
there is a decomposition into line bundles (cf. \eqref{eq:simple-pushforward}),
\begin{equation} \label{eq:e-simp-decomp}
E = \bigoplus_i \CL^{(i)},
\end{equation}
preserved by the connection $\pi_* \nabla^\ab$.
We do not generally have such a decomposition for
${\hat E}$, because the gluing transformations $\CT^{\alpha,\beta}_c$ mix the
different $\CL^{(i)}$.  Still, the $\CT^{\alpha,\beta}_c$ do preserve some structure, as follows.
Recall that the walls $p_c$ of $\wnet$ which end at $\fs_n$ carry labels $ij$ which
obey a constraint:  both $i$ and $j$ must belong to $o(\fs_n)$ and we must
have $i < j$.  Using \eqref{eq:tran-fn} it then follows that all the
$\CT^{\alpha,\beta}_c$ are ``upper triangular'':
for $i \notin o(\fs_n)$ they preserve $\CL^{(i)}$, and for $i \in o(\fs_n)$ they
preserve the subspace
\begin{equation}
 F^{(i)} = \bigoplus_{j \ge i} \CL^{(j)}.
\end{equation}
So we find that ${\hat E}$ decomposes
over $\CU$ as
\begin{equation} \label{eq:e-decomp}
{\hat E} = \left( \bigoplus_{i \notin o(\fs_n)} \CL^{(i)} \right) \oplus \hat E_{\fs_n}
\end{equation}
where each $\CL^{(i)}$ is $\nabla$-invariant, and
$\hat E_{\fs_n}$ has rank $\abs{o(\fs_n)}$ and
carries various $\nabla$-invariant subbundles $F^{(i)}$.  The ranks of these subbundles
and of their intersections are determined by the structure of the partial ordering in $o(\fs_n)$.

In the simplest case $o(\fs_n)$ is a totally ordered set containing all $K$ sheets.
In that case we may as well identify the labels with integers $1 \le i \le K$,
and write the ordering as $1 < 2 < \cdots < K$.  Then the
extra structure just discussed is simply a $\nabla$-invariant filtration of ${\hat E}$,
\begin{equation}
 0 = F^{(K)}  \subset F^{(K-1)} \subset F^{(K-2)} \subset \cdots \subset F^{(0)} = {\hat E},
\end{equation}
where $F^{(i)}$ has dimension $K-i$.  If the monodromy of $\nabla$ around $\fs_n$ is diagonalizable
with all eigenvalues distinct, then there exist $K!$ such invariant filtrations, and
our construction picks one out of those $K!$.

What we have found is that our construction naturally produces not just $({\hat E}, \nabla)$, but a bit of extra structure around each $\fs_n$, namely the decomposition \eqref{eq:e-decomp} and
the subbundles $F^{(i)}$.  We refer to this extra structure as \ti{flag data}.

\subsection{The nonabelianization map}

Given a nondegenerate spectral network $\wnet$ subordinate to a covering
$\Sigma \to C$, we have defined a nonabelianization operation which takes
a twisted flat rank $1$ connection over $\Sigma$ to a twisted flat rank $K$
connection over $C$ with flag data.  In this section we discuss the corresponding
map $\Psi_{\wnet}$ between \ti{moduli spaces} of twisted flat connections.
For simplicity we take the case where $C$ only has punctures, not boundaries,
and restrict attention to $\wnet$ for which each $o(\fs_n)$ is a total ordering
of the set of all sheets.\footnote{Our discussion can be generalized
to include marked points on boundaries; in that case the relevant moduli spaces
are spaces of connections with \ti{irregular} singularities, on the surface obtained
by shrinking each boundary component of $C$ to a point.
In the special case $K=2$ this was discussed in \cite{Gaiotto:2009hg}.
Examples with irregular singularities and $K>2$ are discussed in
\cite{gmn6-to-appear}.}

Let us first describe the relevant moduli space of rank $1$ connections $\nabla^\ab$.
The covering surface
$\Sigma$ has punctures $\mathfrak{s}_n^{(i)}$, preimages of the $\fs_n$ on $C$.
Fix a complex line bundle $\CL$ over $\tilde{\Sigma}$;
we will consider flat connections
on $\CL$ with fixed holonomy around the punctures.
Namely, fix parameters $\fm^{(i)}_n \in \R / \Z$, with $\fm^{(i)}_n \neq \fm^{(j)}_n$
for $i \neq j$.
Let $\ell_{n^{(i)}}$ be a small counterclockwise
loop around $\fs_n^{(i)}$, and $\tilde \ell_{n^{(i)}}$ its canonical
lift to $\tilde \Sigma$.  We require that the holonomy of $\nabla^\ab$ around
$\tilde \ell_{\fs_n^{(i)}}$ is $\exp (2 \pi \I \mathfrak{m}_n^{(i)})$.
As usual, we consider connections $\nabla^\ab$
only up to gauge equivalence;
our gauge group is the group of smooth
maps $\bar \Sigma \to \IC^\times$,
where $\bar \Sigma$ is the closure
of $\Sigma$.  Let $\CM(\Sigma, GL(1); \mathfrak{m})$ be the resulting
moduli space of twisted rank 1 flat connections.
$\CM(\Sigma, GL(1); \mathfrak{m})$ is a torsor for
$H^1(\bar \Sigma, \IC^\times) \simeq (\IC^\times)^{2g}$, where $g$
is the genus of $\bar\Sigma$.

Next, what can we say about the rank $K$ flat connections $\nabla$ we obtain
by nonabelianization?
Fix a small clockwise loop $\ell_n$ around $\fs_n$, and let $\tilde \ell_n$
be its canonical lift to $\tilde C$.
The holonomy of $\nabla$ around $\ell_n$
is the product of several factors:  the holonomy of $\nabla^\ab$ along the
pieces of $\ell_n$ running between walls of $\wnet$, and the
transformations $\CT_p$ attached to the walls.  With respect to the decomposition
\eqref{eq:e-simp-decomp}, the former are diagonal matrices whose product is
$\diag \{\exp[ 2\pi \I \fm_n^{(i)}] \}$, while the latter are unipotent
upper triangular.  Their product is thus an upper-triangular matrix, with
diagonal elements $\exp[ 2\pi \I \fm_n^{(i)} ]$.
Since we have assumed these values
distinct, it follows that the monodromy of $\nabla$ around $\tilde \ell_n$
is semisimple, with eigenvalues $\exp[ 2\pi \I \fm_n^{(i)} ]$.
So, let $\CM(C, GL(K);\mathfrak{m})$
denote the moduli space of twisted rank $K$ flat connections on $C$
(with fixed topology of the principal bundle),
such that the monodromy around $\tilde \ell_n$ is conjugate to
\begin{equation}\label{eq:diag-elts-diag}
\diag \{ \exp[ 2\pi \I \fm_n^{(1)}], \dots,\exp[2\pi \I \fm_n^{(K)}] \}.
\end{equation}

As we have noted, nonabelianization produces not only a twisted flat connection over $C$,
but a twisted flat connection over $C$
equipped with flag data.  Thus we also define an extended moduli
space $\CM_F(C, GL(K);\mathfrak{m})$, consisting of elements of
$\CM(C, GL(K);\mathfrak{m})$ with flag data.
$\CM_F(C, GL(K);\mathfrak{m})$ is a finite cover of $\CM(C, GL(K);\mathfrak{m})$,
as explained at the end of \S\ref{sec:flags}.
Nonabelianization gives a map
\begin{equation}\label{eq:Psi-MAP}
\Psi_{\wnet}: \CM(\Sigma, GL(1); \mathfrak{m} )\to \CM_F(C, GL(K); \mathfrak{m}).
\end{equation}

One simple property of $\Psi_\wnet$ is worth noting at once.  Given a spectral network $\wnet$,
let $\wnet^*$ be obtained by reversing the labels on all walls and the orderings at all punctures.
Then we have
\begin{equation}
 \Psi_{\wnet^*}(\CL, \nabla^\ab) = \Psi_{\wnet}(\CL^*, (\nabla^\ab)^*)^*.
\end{equation}
This fact may be of some practical use:  the reason is that $\wnet_\vartheta^* = \wnet_{\vartheta + \pi}$,
and determining the relation between $\Psi_{\wnet_\vartheta}$ and $\Psi_{\wnet_{\vartheta+\pi}}$ is equivalent to
determining the ``spectrum generator'' which captures the full BPS spectrum.

For the rest of this section we will study $\Psi_\wnet$ as defined above.
However, we must note that in the physical applications the actual
moduli space $\CM$ of the physical theory $S[\fg, C, D]$ compactified on $S^1$
is not quite $\CM(C, GL(K); \fm)$;
it is more closely related to the space of $SL(K)$ connections (but even
this is a slight oversimplification, see \cite{Gaiotto:2010be}).
We expect that there is a variant of our nonabelianization map
for which the codomain is precisely $\CM$.
The domain of this variant map should not be quite $\CM(\Sigma, GL(1); \fm)$;
rather it should be a space
of twisted rank 1 connections on $\Sigma$ subject to some additional constraints and/or
carrying some additional structure.  This is related to the fact that
the IR charge lattice of $S[\fg, C, D]$ is not $H_1(\Sigma;\Z)$, but rather an appropriate
subquotient of it \cite{Gaiotto:2009hg}.  (See \cite{MR1397059,dongaits}
for a careful discussion of subtle
issues of this sort that arise
in abelianization of Higgs bundles for a general group $G$; we expect that, as far as
these topological issues are concerned, the story for moduli of twisted
flat connections will be similar.) We leave the proper extension of our construction
to account for these variations as an open problem. It is related to open problem 2
of \S \ref{sec:Introduction}.

\subsection{Holomorphic symplectic structures} \label{sec:hol-symp}

In this section we explain one of the important properties of $\Psi_\wnet$:
both its domain and codomain are holomorphic symplectic, and $\Psi_\wnet$ is a
holomorphic symplectic map,
\begin{equation}\label{eq:preserve-sf}
\Psi_{\wnet}^*(\varpi_C) = \varpi_{\Sigma}.
\end{equation}

We begin with $\CM(\Sigma, GL(1); \mathfrak{m})$.  The easiest way to understand
its holomorphic symplectic form is to recall that on the space of \ti{untwisted} connections
there is a standard such form \cite{MR85k:14006,boalch-thesis}.  This form, formally speaking, is
obtained by symplectic quotient from
\begin{equation} \label{eq:alpha-int}
\varpi_{\Sigma} = \int_{\Sigma} \delta \alpha \wedge \delta \alpha
\end{equation}
where $\alpha$ denotes the abelian connection 1-form.\footnote{Here and below,
we use the freedom to fix a gauge
so that we consider only variations $\delta \alpha$ which vanish near the punctures.
This makes  the integral \eqref{eq:alpha-int}
  convergent, despite the fact that the connections we consider are singular.}
Using a choice of spin structure on $\Sigma$ to relate twisted and untwisted connections
we can transfer this holomorphic symplectic structure to
$\CM(\Sigma, GL(1); \mathfrak{m})$; by abuse of notation we denote that structure
also as $\varpi_\Sigma$.

Similarly, $\CM_F(C, GL(K); \mathfrak{m})$ has symplectic form obtained
by symplectic quotient from
\begin{equation}
\varpi_{C} = \int_{C}{\rm Tr}\,\delta \CA  \wedge \delta \CA
\end{equation}
with $\CA$ the nonabelian connection 1-form.  Again we use a spin structure on $C$
to pass between twisted and untwisted connections.

To see why \eqref{eq:preserve-sf} is true,
first consider variations $\delta \alpha$ which have support away
from the spectral network $\wnet$.  In this case the corresponding
variation of $\CA$ is simply
$\delta \CA = \pi_* \delta \alpha$.
It follows easily that for such variations we have $\varpi_\Sigma(\delta \alpha, \delta \alpha') = \varpi_C(\delta \CA, \delta \CA')$.
If $\delta \alpha$ has support intersecting $\wnet$ then the situation is slightly more involved:
there is a distribution-valued contribution to $\delta \CA$ supported on $\wnet$.
By a gauge transformation $\delta \alpha \to \delta \alpha + \de \chi$
we may assume however that $\delta \alpha$ vanishes in neighborhoods of joints.
So it is enough to consider what happens in a patch intersecting only a single wall $p_c$ of $\wnet$.
Taking local coordinates $(x,y)$ on $C$
where $p_c$ is the locus $y = 0$, one finds the form
\begin{equation} \label{eq:delta-a-sing}
 \delta \CA = \pi_* \delta \alpha + s(x) \delta(y) \de y,
\end{equation}
where in the decomposition
\eqref{eq:e-simp-decomp} $s(x)$ is off-diagonal and $\pi_* \delta \alpha$ diagonal.
One thus has ${\rm{Tr}} (s(x) \de y \wedge \pi_* \delta \alpha') = 0$,
so the last term in \eqref{eq:delta-a-sing}
does not contribute to $\varpi$; we get once again
$\varpi_\Sigma(\delta \alpha, \delta \alpha') = \varpi_C(\delta \CA, \delta \CA')$.

\subsection{Dimension counts} \label{sec:dimensions}

Now let us compare the dimensions of the moduli spaces related by $\Psi_\wnet$.

First we consider
$\CM(C, GL(K);\mathfrak{m})$.  This space can be represented
as the space of $GL(K,\IC)$ matrices $A_1, \dots, A_{g_C}$, $B_1, \dots, B_{g_C}$,
$C_1, \dots, C_s$, subject to some relations:  first, the eigenvalues of the $C_n$
are fixed; second, we impose
\begin{equation} \label{eq:fund-rel}
 \prod_i [A_i,B_i] \prod_n C_n = 1;
\end{equation}
third, we divide out by overall conjugation.
The condition on the eigenvalues of the $C_n$ eliminates $K s$ degrees of freedom.
The equation \eqref{eq:fund-rel} imposes only
$K^2-1$ independent conditions:  if $\prod_n \det C_n\not=1$
there are no solutions, but if $\prod_n \det C_n = 1$, while there are solutions,
the determinant equation is redundant.  Similarly, the center of the
gauge group acts trivially, so the quotient by the gauge group reduces the dimension
by $K^2-1$.  Altogether we get
\begin{equation}
\dim_\IC \CM(C, GL(K);\mathfrak{m}) = (2g_C + s) K^2 - K s - 2(K^2-1).
\end{equation}
On the other hand, using the Riemann-Hurwitz relation
\begin{equation}
\dim_\IC \CM(\Sigma, GL(1);\mathfrak{m}) = 2g_\Sigma = K(2g_C-2) + B + 2,
\end{equation}
where $B$ is the branching number (which is simply the number of branch points,
since we assume all branch points are simple.)
It follows that
\begin{equation} \label{eq:diff-dim}
\dim_\IC \CM(C, GL(K);\mathfrak{m}) - \dim_\IC \CM(\Sigma, GL(1);\mathfrak{m}) = (K^2-K)(2g_C + s - 2) - B.
\end{equation}
Since $\Psi_{\wnet}$ is symplectic it is in particular locally injective,
so the quantity in \eqref{eq:diff-dim} must be nonnegative.
We thus get a topological restriction
on the branching number of a branched cover $\Sigma \to C$ of degree $K$
admitting a spectral network:
\begin{equation}\label{eq:Branch-bound}
B \leq (K^2-K)(2g_C + s - 2).
\end{equation}
If this bound is saturated, $\Psi_{\wnet}$ is a local
symplectomorphism.  Note that this does happen for the
spectral curves \eqref{eq:sw-curve}, to which the spectral networks
$\wnet_\vartheta$ are subordinate.

Equation \eqref{eq:Branch-bound} is a
purely topological statement.  It says that there cannot
be too many branch points in a
spectral network.  It would be interesting to give a
more direct topological proof of this assertion.

In the case where \eqref{eq:Branch-bound} is saturated and $\Psi_\wnet$ is a covering
map, we have constructed not only a local isomorphism between $\CM(C)$ and $\CM(\Sigma)$,
but also a local identification between $(\pi_* \CL, \pi_* \nabla^\ab)$ and $(\hat E, \nabla)$.
We can use this to tie up a loose end from \S\ref{subsec:2d4d-wcf}.  Letting $(\CL,\nabla)$
vary we get a \ti{universal bundle} $\pi_*(\CL)$ over $\CM(\Sigma) \times C$.
Restricting this to some point
$\tilde{z} \in \tilde{C}$ gives a line bundle over $\CM(\Sigma)$, which we call $\pi_*(\CL)_{\tilde{z}}$.
Similarly we have a universal bundle $\hat E_{\tilde{z}}$ over $C$.\footnote{Actually this universal bundle does not quite exist owing
to subtleties involving the center of $GL(K)$; but its bundle of endomorphisms does exist, which is all we will use below.}
Now let $A(\Sigma)$ be the algebra of
global rational sections of $\End(\pi_*(\CL)_{\tilde{z}})$ (i.e. the algebra of sections over the generic point),
and similarly $A(C)$ the algebra of global rational sections
of $\End(\hat E_{\tilde{z}})$.  $A(\Sigma)$ is a central simple algebra of degree $K$ over
$\CR(\CM(\Sigma))$, the field of rational functions on $\CM(\Sigma)$,
and similarly $A(C)$ is a central simple algebra of degree $K$ over $\CR(\CM(C))$.
Our nonabelianization construction gives an embedding of
$A(C)$ in $A(\Sigma)$ and a compatible embedding of $\CR(\CM(C))$ in $\CR(\CM(\Sigma))$.  Now, we have just
shown that this embedding realizes $\CM(\Sigma)$ as a \ti{cover} of $\CM(C)$.  It follows that
$\CR(\CM(\Sigma))$ is a finite extension of $\CR(\CM(C))$ (since it is an extension and the two have the same
dimension, hence the same transcendence degree over $\IC$, and are both finitely generated).
Thus the group $Aut(\CR(\CM(\Sigma)):\CR(\CM(C)))$ is finite.
But now it follows by dimension counting that $Aut(A(\Sigma):A(C))$ is also finite.

\subsection{Equivalence of spectral networks}\label{subsubsec:EqivalentGSN}

Given a nondegenerate spectral network $\wnet$
we have constructed a corresponding nonabelianization map $\Psi_\wnet$.
It is natural to ask how this map depends on $\wnet$.

We can get some intuition by remembering previous results about the case $K=2$.  In that case, as already mentioned
in \S\ref{sec:k2-examples}, $\wnet$ determines
an ideal triangulation $T(\wnet)$ of $C$ up to isotopy.  $T(\wnet)$ does not change if we vary $\wnet$ by an isotopy (while holding $\Sigma$ fixed).
Moreover, the map $\Psi_\wnet$ in this case only depends on $T(\wnet)$.  So we find that $\Psi_\wnet$ is an isotopy invariant of $\wnet$.

We now extend this discussion to general $K$.  We begin by defining a notion of equivalence between
nondegenerate spectral networks.  Our experience from the case $K=2$ suggests that equivalence should at least include isotopy.
In fact, for $K > 2$ it will be natural to
allow even some moves which are \ti{not} isotopies.  Given two
nondegenerate spectral networks $\wnet$ and $\wnet'$, an
\ti{equivalence} between $\wnet$ and $\wnet'$ is a one-parameter family of path-lifting rules
$\bF(\cdot, t)$ to coverings $\Sigma(t)$, such that:

\begin{description}
\item[E1.] $\bF(\cdot, 0) = \bF(\cdot, \wnet)$ and $\bF(\cdot, 1) = \bF(\cdot, \wnet')$.

\item[E2.] For any point $\tz$ on $\tilde{C}$ and pair $0 \le t_1 \le t_2 \le 1$, there exists a formal sum of paths
$R(\tz, t_1, t_2)$, such that for any path $\wp$ from $\tz_1$ to $\tz_2$ we have
\begin{equation} \label{eq:equalpt}
 \bF(\wp, t_2) = R(\tz_1, t_1, t_2) \bF(\wp, t_1) R(\tz_2, t_1, t_2)^{-1}.
\end{equation}
\end{description}
The point of this definition is that if $\wnet$ and $\wnet'$ are connected by an equivalence then we have
\begin{equation} \label{eq:equalpsi}
\Psi_\wnet = \Psi_{\wnet'}.
\end{equation}
(The meaning of the equal signs in \eqref{eq:equalpt}, \eqref{eq:equalpsi} has
to be clarified in case the coverings $\Sigma(t)$ vary with $t$:  in this case
we identify the different coverings using the ``Gauss-Manin'' parallel transport induced by
the family $\{\Sigma(t)\}$.)

One natural way of getting an equivalence is to consider a 1-parameter family of nondegenerate spectral networks $\wnet(t)$
and take $\bF(\cdot, t) = \bF(\cdot, \wnet(t))$.  In this case $R(\tz, t_1, t_2)$ is a product,
with one factor of the form $(1 + \sum_{\nu \in \cS_c(t)} \bX_{\ba(\nu)})$ for each $t$ with $t_1 \le t \le t_2$
such that $z$ lies on an $\cS$-wall $p_c(t)$ in $\wnet(t)$.

Here are some examples of equivalences:
\begin{description}
\item[M1.] An \ti{isotopy}.  This is a one-parameter family of spectral networks $\wnet(t)$ subordinate to coverings
$\Sigma(t)$, such that the $\Sigma(t)$ vary continuously in the obvious sense (in particular the branch points move continuously),
each wall $p_c$ varies by isotopy, and the joints $z_\mu$ move continuously. The combinatorics of $\wnet(t)$ do not change during
an isotopy.

\insfig{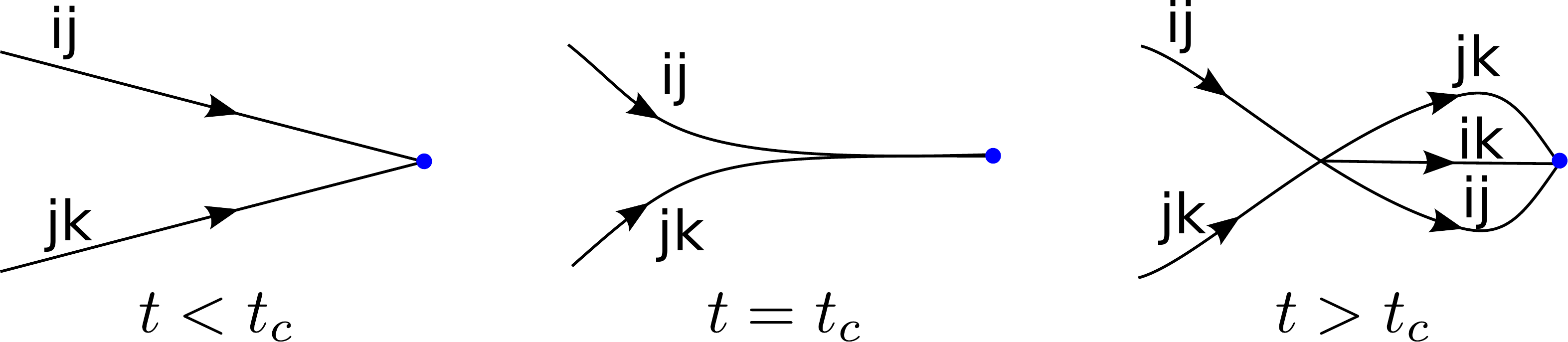}{Crossing at a defect:  an equivalence which relates two non-isotopic spectral networks.}
\item[M2.] A \ti{crossing at a defect}.  This is a family $\wnet(t)$ which implements a
move as illustrated in Figure \ref{fig:infinity-crossing}.  Here the combinatorics of the network
do change at the critical value $t = t_c$.  Nevertheless the family gives an equivalence.

\insfig{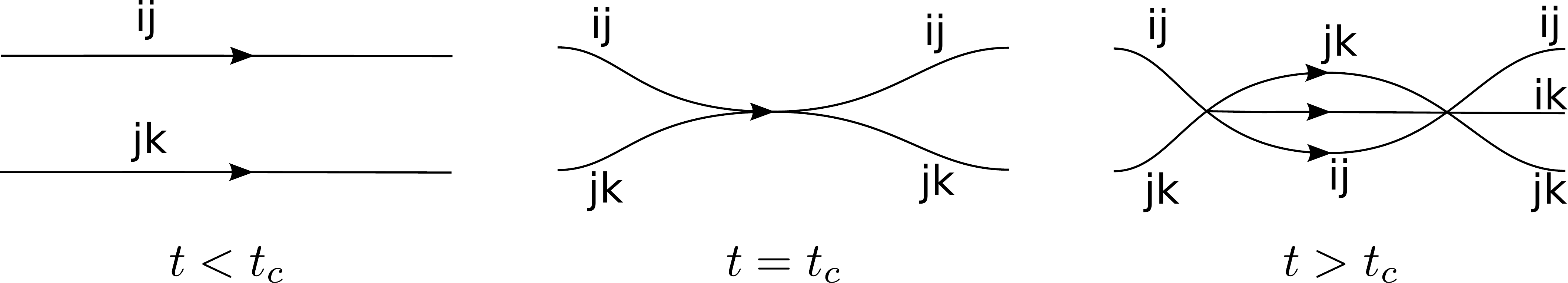}{Bubble:  an equivalence which relates two non-isotopic spectral networks.}
\item[M3.] A \ti{bubble}.  This is a family $\wnet(t)$ which implements a
move as illustrated in Figure \ref{fig:bubble}.

\insfig{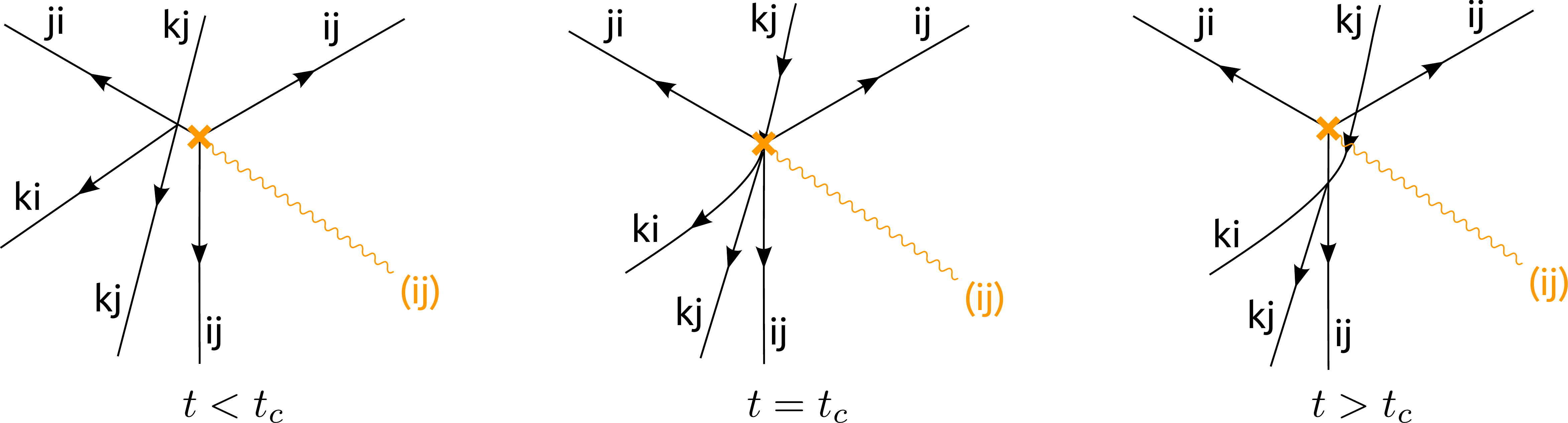}{Branch point traversal:  an equivalence which relates two non-isotopic spectral networks.}
\item[M4.] A \ti{traversal of a branch point}.  This is a family $\wnet(t)$ which implements a
move as illustrated in Figure \ref{fig:branch-point-traversal}.  The member $\wnet(t_c)$ strictly speaking is not
quite a spectral network according to our definition, because of the extra $kj$ lines passing through the branch point;
nevertheless one can define $\bF(\cdot, t_c)$ by a straightforward extension of our rules, and after so doing,
E1, E2 above are obeyed.

\end{description}

All of these equivalences really occur in practice, e.g. in families $\wnet_\vartheta$ as $\vartheta$
varies.  We emphasize that these equivalences are \ti{not} the jumps of $\wnet_\vartheta$ which we studied in
\S\ref{sec:K-walls}:  those jumps generally connect spectral networks which are \ti{inequivalent} in our sense.

\subsection{A picture of the set of all spectral networks}\label{subsec:PictureSN}

Fix a curve $C$ and singularities $\fs_n$ and orderings $o(\fs_n)$.
Exploration of some examples leads to a rough conjecture about the set $X = X(C, \fs_n,o(\fs_n))$ of all
spectral networks with these data fixed.

There should be a natural topology on $X$, such that $X$ is decomposed into connected cells,
separated by codimension-1 loci where the spectral networks become degenerate.  (Ideally the
cells should even be contractible, but this might require extending our definition of spectral
network to allow more non-generic phenomena which occur in codimension greater than 1,
e.g. allowing several joints to coalesce.)
Moving around in a single cell corresponds to varying $\wnet$ by
equivalences in the
sense of \S\ref{subsubsec:EqivalentGSN}.  Crossing one of the boundaries between
cells corresponds to one of the more interesting ``$\CK$-wall'' jumps that occur when
the network degenerates, which we discusssed in \S\ref{sec:varying-theta}.

We leave it as an interesting open problem to formulate this picture
on a more rigorous and precise basis.

\subsection{Coordinate systems and a cluster conjecture}\label{subsec:Coord-System}

Let us consider a spectral network $\wnet$ for which \eqref{eq:Branch-bound}
is saturated.  In this case, as we have noted,
$\Psi_\wnet$ is at least locally one-to-one.
Thus the holonomies $\tilde\CY_{\tilde\gamma}$ of $\nabla^\ab = \Psi_\wnet^{-1}(\nabla)$,
where $\tilde\gamma$ runs over a basis of $H_1(\tilde\Sigma;\IZ)$,
provide \ti{local coordinates} on $\CM_F(C, GL(K);\mathfrak{m})$.

It is natural to ask whether $\Psi_\wnet$ might be even globally one-to-one,
so that it would give global coordinates on its whole image $\CU_\wnet$.
We conjecture that this is indeed the case.\footnote{This is the moment where it is important
that we use the moduli space $\CM_F(C, GL(K);\mathfrak{m})$ with flag data rather than just $\CM(C, GL(K);\mathfrak{m})$.  If we used $\CM(C, GL(K);\mathfrak{m})$, then
$\Psi_\wnet$ would be only finite-to-one.}
At least in the case $K=2$ this conjecture
is true:  in this case the coordinate system in
question is essentially the Fock-Goncharov coordinate system \cite{MR2233852,Gaiotto:2009hg}.
Some further special cases in which the conjecture is true
will be treated in \cite{gmn6-to-appear}.

As we noted above, if $\wnet$ and $\wnet'$ are equivalent
nondegenerate spectral networks then
$\Psi_{\wnet}$ and $\Psi_{\wnet'}$ are related by parallel transport, in the sense
explained under \eqref{eq:equalpt}.
Thus we can roughly say that $\Psi_\wnet$ is constant as $\wnet$ varies within one of
the cells of $X$.  If $\wnet$ and $\wnet'$ are in neighboring cells, then the story is
more interesting:  taking a path $\wnet(t)$ from $\wnet$ to $\wnet'$, there is a critical
$t = t_c$ where $\wnet(t)$ becomes a degenerate spectral network.
At this moment $\Psi_{\wnet(t)}$ jumps discontinuously.  This jump should be determined by the arguments of \S\ref{sec:f-jump}.
Indeed, by the same formulas we used there (and assuming the same conjectures we
assumed there), the degenerate network $\wnet(t_c)$ determines integers $\Omega(\gamma)$
and distinguished lifts $\tilde\gamma$,
for $\gamma$ lying in a semilattice $\Gamma_c \subset H_1(\Sigma(t_c); \IZ)$ with a single generator.
Using these data we define a (birational) automorphism $\CK$ of
$\CM(\Sigma(t_c), GL(1); \fm)$, by its action on the holonomies $\tilde\CY_{\tilde\gamma}$
of $\nabla^\ab$ around 1-cycles $\tilde\gamma$:
\begin{equation} \label{eq:holonomy-jump}
 \CK(\tilde\CY_{\tilde\gamma}) =\tilde \CY_{\tilde\gamma} \prod_{\gamma' \in \Gamma_c}
 (1 + \tilde \CY_{\tilde\gamma'})^{\inprod{\gamma,\gamma'} \Omega(\gamma')}.
\end{equation}
When $t$ crosses $t_c$, the map $\Psi_{\wnet(t)}$ jumps by composition with $\CK$, so
\begin{equation}
\Psi_{\wnet}(\nabla^\ab) = \Psi_{\wnet'}(\CK^{\pm 1}(\nabla^{\ab}))
\end{equation}
(where the sign $\pm 1$ is determined by which direction we cross the cell boundary, and
we must bear in mind the comment under \eqref{eq:equalpsi}.)

Finally, we conjecture that the coordinate systems we have defined are actually
\ti{cluster} coordinates.\footnote{It was noted in \cite{ks1,Gaiotto:2010be} that
the $\CK$-transformations which appeared there
could be viewed as cluster transformations in an appropriate sense; see also
\cite{Cecotti:2010fi,Cecotti:1993rm,Alim:2011ae,Alexandrov:2011ac,Cecotti:2011gu} for further
discussion of the relation of cluster varieties to four-dimensional
$\CN=2$ theory.}  The spaces $\CM(C)$ which we are considering are indeed cluster varieties,
as shown by Fock-Goncharov in \cite{MR2233852}.  However, relatively few of the cluster
coordinate systems on $\CM(C)$ have been described explicitly.  We believe that for each $\wnet$, the coordinate system
induced by $\Psi_\wnet$ should be an element of the cluster atlas.
So in particular,
we conjecture that there is an algorithm which determines from a
$\wnet$ a finite set of charges $\gamma^{n}_{\wnet} \in \Gamma$.
The $\gamma^{n}_{\wnet}$ should be the elements of a cluster
seed, with exchange matrix given by the intersection pairings:
$\eps^{nm} = \inprod{\gamma^{n}, \gamma^{m}}$.
Informally speaking, the $\gamma^{n}_\wnet$ should be
the charges of finite webs which can appear inside of $\wnet$ when $\wnet$ degenerates;
said otherwise, the $\gamma^{n}_\wnet$
label faces of the boundary of the cell in $X$ containing $\wnet$.
Finally, the $\cK$-transformation \eqref{eq:holonomy-jump}, in the simplest case where there is only a
single $\Omega(\gamma) = 1$ contributing,
should be identified with the action of a mutation on cluster variables.

This conjecture is true when $K=2$, and in \cite{gmn6-to-appear} we will
give some additional evidence for it in some special cases with $K>2$.  In general, though,
we do not know how to prove it, or even how to read off the $\gamma^{n}_\wnet$ from $\wnet$.

It is natural to wonder further whether spectral networks might give \ti{all} of the coordinate
systems in the cluster atlas on $\CM$.

\subsection{WKB asymptotics}\label{subsec:WKB-Asymptotics}

So far in this section we have been considering the map $\Psi_\wnet$ associated to an arbitrary
spectral network $\wnet$.  These maps have especially interesting asymptotic properties if we choose $\wnet$
to be the particular network $\wnet_\vartheta$ which we studied in the rest of the paper.
In this section we briefly explain this point.  (We will gloss over the role of the twisting in this section,
and use spin structures to pass freely between twisted and untwisted connections.)

Equip $C$ with a complex structure.
Suppose given a $GL(K)$ \ti{Higgs bundle} over $C$, i.e. a triple $(V, \bar\partial, \varphi)$
where $V$ is a complex vector bundle
of rank $K$, $\bar\partial$ a holomorphic structure on $V$,
and $\varphi$ a meromorphic $\End(V)$-valued $1$-form, with poles at the
punctures on $C$.
Let $\Sigma$ be the spectral cover
determined by $\varphi$,
\begin{equation}
\Sigma = \{ \det(\varphi - \lambda) = 0 \} \subset T^* C.
\end{equation}
Now assume that $\Sigma$ is smooth (this is the case for generic $\varphi$.)  Then
we have a corresponding spectral line bundle over $\Sigma$,
\begin{equation}
 \CL = \ker (\varphi - \lambda),
\end{equation}
and the Higgs bundle $(V,\bar\partial,\varphi)$ is the pushforward of $(\CL, \lambda)$.\footnote{The passage from $(V,\bar\partial,\varphi)$ to $\CL$ is
often called abelianization and has been exploited heavily in the study of ``nonabelian theta functions''; see e.g. \cite{MR92b:57008}
where it was introduced, and \cite{dongaits} for a very precise description of the abelianization map
in the more general setting of an arbitrary Lie group $G$.
Abelianization for Higgs bundles is simpler than for flat connections in one important respect:  for
Higgs bundles the monodromy around branch points causes no problem and no spectral network is needed.  There is thus
only \ti{one} abelianization map for Higgs bundles, in contrast to the various $\Psi_\wnet$ we found
in this paper for flat connections.}

Now, given a Higgs bundle there is a corresponding solution of Hitchin's equations \cite{MR89a:32021,MR965220,MR887285,hbnc,wnh}.
In \cite{Gaiotto:2011tf} we proposed a new method for constructing that solution.
The key ingredient is a set of integral equations written in \S 5.6 of \cite{Gaiotto:2011tf}.
The equations are determined by the data of:
\begin{itemize}
 \item the Higgs bundle $(V,\bar\partial,\varphi)$,
 \item the BPS degeneracies $\mu$ and $\omega$, which (as we have explained in \S\S \ref{sec:F-basics},\ref{sec:wnet},\ref{sec:varying-theta}) are computed from the spectral networks $\wnet_\vartheta$ determined by $\Sigma$,
 \item a real parameter $R>0$.
\end{itemize}

These equations are expected to have a unique solution for large enough $R$ (constructed e.g. by iteration).
The solution produces a family of flat connections $\nabla(\zeta)$ in $V$,
together with a family of flat connections $\nabla^\ab(\zeta)$ in $\CL$,
and a family of isomorphisms
\begin{equation}
g(\zeta): \Psi_{\wnet_{\vartheta = \arg \zeta}}(\CL, \nabla^\ab(\zeta)) \to (V, \nabla(\zeta)).
\end{equation}

The integral equations give some control over the analytic structure
and $\zeta \to 0,\infty$ asymptotic behavior of $g(\zeta)$ and $\nabla^{\ab}(\zeta)$.
From this we deduce that the family $\nabla(\zeta)$ is of the form
\begin{equation} \label{eq:hitchin-connection}
 \nabla(\zeta) = R \zeta^{-1} \varphi + D + R \zeta \bar\varphi,
\end{equation}
where $D$ is a connection in $V$, unitary with respect to some Hermitian metric $h$ in $V$ (``harmonic metric''),
and $\bar\varphi$ is the adjoint of $\varphi$ in the metric $h$.
The fact that $\nabla(\zeta)$ is flat for all $\zeta \in \IC^\times$ is equivalent to the statement
that $(V, \varphi, D, \bar\varphi)$
constitute a solution of \ti{Hitchin's equations} on $C$.  (The natural conjecture is that it is the unique
such solution corresponding to the given Higgs bundle $(V, \bar\partial, \varphi)$).
So the integral equations are a machine for producing solutions to Hitchin's equations.

In addition, the integral equations give some more interesting asymptotic information.
This information is most naturally formulated in terms of the objects
\begin{equation}
\cY_a(\zeta) = X_a(\nabla^\ab(\zeta)) \in \Hom(V_{\tz_1}, V_{\tz_2}).
\end{equation}
If we think of $\cY_a(\zeta)$ as a function of $\tz_2$ then it is flat with respect to the connection
$\nabla(\zeta)$, except when $z_2$ meets the network $\wnet_{\vartheta = \arg \zeta}$:  when $z_2$
crosses the $\CS$-walls, $\cY_a(\zeta)$ jumps discontinuously.  Analogous remarks
hold for the dependence of $\cY_a$ on $\tz_1$.
From the integral equations we learn that
the $\zeta \to 0$ asymptotics of $\cY_a(\zeta)$ are given by
\begin{equation}
 \cY_a(\zeta) \sim \exp \left[ \pi \frac{R}{\zeta} Z_{\bar a} \right].
\end{equation}
Indeed, we can say something a bit stronger:
suppose that we define a new section $\cY_a^\vartheta(\zeta)$ by
\ti{analytic continuation} of $\cY_a(\zeta)$ from the
locus $\arg \zeta = \vartheta$.  In this case, we have the asymptotics
\begin{equation} \label{eq:ya-halfplane}
\cY_a^\vartheta(\zeta) \sim \exp \left[ \pi \frac{R}{\zeta} Z_{\bar a} \right],
\end{equation}
so long as $\zeta \to 0$ while remaining in the half-plane ${\mathbb H}_\vartheta$ centered on the ray $\arg \zeta = \vartheta$.
This is rather sharp asymptotic information about the flat sections of the family of flat connections
$\nabla(\zeta)$.

Our way of describing this asymptotic information is perhaps a bit unfamiliar, so let us relate it to
something better known:  the WKB analysis of 1-parameter families of differential equations.
The flatness equation $\nabla(\zeta)s=0$ takes the general form
\begin{equation}
\left[ \zeta \frac{\de}{\de z} + (R \varphi + \zeta(\cdots)) \right] s = 0.
\end{equation}
The WKB analysis of this equation involves studying formal solutions of the form
\begin{equation} \label{eq:psi-formal}
s^{(i)}_\formal(\zeta) = \exp\left[ \frac{1}{\zeta} \sum_{n=0}^\infty S_n^{(i)} \zeta^n \right] \sum_{n=0}^\infty T^{(i)}_n \zeta^n
\end{equation}
where each $S^{(i)}_n$ is a function and $T^{(i)}_n$ is a section of $V$.
One important difficulty in the WKB method is that these solutions really are only formal:  the series \eqref{eq:psi-formal}
typically has zero radius of convergence.  So one should interpret \eqref{eq:psi-formal} as an asymptotic series.
It is then natural to ask,
are there actual $\nabla(\zeta)$-flat sections $s^{(i)}(\zeta)$ which have these asymptotics?

Our discussion in this section provides the answer:  such sections do indeed exist, in the
following sense.
Fix a phase $\vartheta$, a basepoint $\tz_1 \in \tilde C'_{\wnet_\vartheta}$,
and basis vectors $e^{(i)} \in \CL_{\tz_1}^{(i)}$.  Then
letting $\tilde\wp$ be a path in $\tilde C'_{\wnet_\vartheta}$ from $\tz_1$ to $\tz$, we can define
\begin{equation}
s^{(i)}(\zeta) = e^{(i)} \cY^\vartheta_{\tilde\wp^{(i)}}(\zeta).
\end{equation}
$s^{(i)}(\zeta)$ is a solution of $\nabla(\zeta) s^{(i)}(\zeta) = 0$, and
\eqref{eq:ya-halfplane} implies that $s^{(i)}$ indeed has an asymptotic expansion of the form \eqref{eq:psi-formal},
as $\zeta \to 0$ in the half-plane ${\mathbb H}_\vartheta$.
So, in each connected component of $\tilde C'_{\wnet_\vartheta}$, we have a basis of
$\nabla(\zeta)$-flat sections which have the asymptotics predicted by WKB,
as $\zeta \to 0$ in ${\mathbb H}_\vartheta$.  As we move from one component of $\tilde C'_{\wnet_\vartheta}$ to another,
these bases change.  Thus the walls of $\wnet_\vartheta$ have an interpretation as
Stokes lines.

Finally, we should say that some of the structures which appeared in this paper
have appeared before in the WKB literature (although we arrived at them independently.)  In the case $K=2$,
our constructions here and in \cite{Gaiotto:2009hg}
seem to be closely related to a line of development pursued by Voros and others
(e.g. \cite{Bender1969,Zinn-Justin1984,MR729194,MR1209700}).  For
$K>2$, the phenomenon that new Stokes lines can be born at intersections between old ones was apparently first noticed in
\cite{berk:988}, and has been followed up in a few works since then; we note in particular the reference
\cite{MR2132714}, which contains examples of Stokes diagrams
in the case $K=3$ which look identical to our $\wnet_\vartheta$.  What we have here called \ti{joints} are there called
\ti{virtual turning points}.  There are also results of Simpson on asymptotics
of monodromy which seem likely to be related to ours, e.g. \cite{simpson-agci,MR1166192}.  We have not understood
the precise relation.

\appendix

\section{Joint rules for two-way streets}\label{app:joint-rules}

In this appendix we describe the rules governing the behavior of the solitons near a
rather general kind of joint, where we have solitons oriented into the joint on up to six
distinct trajectories.  This kind of joint does not occur in a nondegenerate spectral network,
but it can occur for a degenerate one.  In the language of \S\ref{sec:wnet} this means that
this kind of joint does not occur in $\wnet_\vartheta$ at generic values of $\vartheta$, but
it can occur for non-generic $\vartheta$.
For generic values of $\vartheta$, the simpler rules of \S\ref{sec:wnet} suffice.

\subsection{Rules for soliton degeneracies}

\insfigscaled{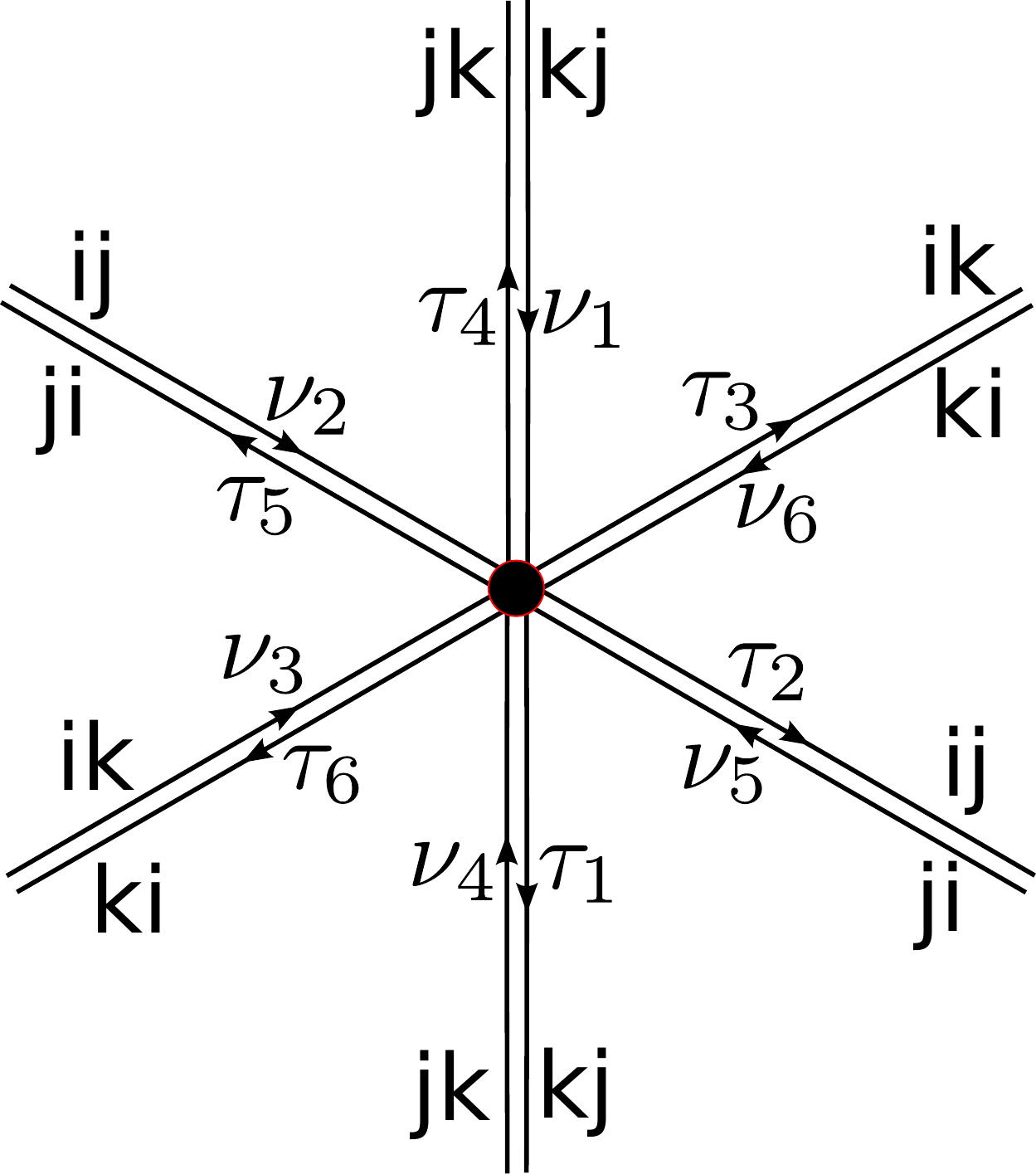}{0.34}{The British resolution of a joint with two-way streets.
Incoming soliton degeneracies are described by generating functions $\nu_n$, $n=1, \dots, 6$. Outgoing
soliton degeneracies are described by generating functions $\tau_n$, $n=1, \dots, 6$.}

We consider six two-way streets $p_n$ entering a joint
as in Figure \ref{fig:six-way-junction-2}.
It is convenient to summarize the soliton spectrum by defining
generating functions for the soliton degeneracies on these six streets:
\begin{equation}
 \nu_n = \sum_{\bar a} \mu(a, p_n) X_{a}, \qquad \tau_n = \sum_{\bar b} \mu(b, p_{n+3}) X_{b},
\end{equation}
where in $\nu_n$ the sum runs over all ingoing charges $\bar{a}$ supported on $p_n$ (i.e.
charges for which $\abs{Z_{\bar a}}$ increases as we go into the joint),
in $\tau_n$ the sum runs over all outgoing charges $\bar{b}$ supported on $p_{n+3}$,
and $n$ is always taken mod $6$.
Applying the constraint of homotopy invariance to artfully chosen paths one shows
\begin{align} \label{eq:hom-inv-jj}
 \tau_1 = \nu_1 + \nu_6 \tau_2, \qquad \tau_2 = \nu_2 + \tau_3 \nu_1, \qquad \tau_3 = \nu_3 + \nu_2 \tau_4, \\ \nonumber
 \tau_4 = \nu_4 + \tau_5 \nu_3, \qquad \tau_5 = \nu_5 + \nu_4 \tau_6, \qquad \tau_6 = \nu_6 + \tau_1 \nu_5.
\end{align}
(These equations have a symmetry under a cyclic shift of the index by $1$ combined with reversing order of
all products, corresponding to the symmetry of Figure \ref{fig:six-way-junction-2} under a rotation by
$\pi/3$ combined with reversing the order of the labels on all lines.)
Using \eqref{eq:hom-inv-jj} repeatedly we obtain the outgoing degeneracy $\tau_1$
in terms of the incoming degeneracies $\nu_i$,
\begin{equation}\label{eq:outgoing-2wayjoint}
\begin{split}
\tau_1
& = \nu_1 + \nu_6 \nu_2 + \nu_6 \nu_3 \nu_1 + \nu_6 \nu_2 \nu_4 \nu_1 + \nu_6\nu_2\nu_5\nu_3\nu_1 + \nu_6\nu_2\nu_4\nu_6\nu_3 \nu_1 + \\
& \qquad + \nu_6 \nu_2\nu_4 \nu_1\nu_5\nu_3\nu_1 + \nu_6\nu_2\nu_4\nu_6 \nu_2\nu_5\nu_3\nu_1 + \nu_6\nu_2\nu_4\nu_6\nu_3\nu_1\nu_5\nu_3\nu_1 + \cdots \\
& = \frac{1}{1 - \overrightarrow{\nu_6\nu_2\nu_4} \otimes \overleftarrow{\nu_5 \nu_3\nu_1}}\left(
 \nu_1 + \nu_6 \nu_2 + \nu_6 \nu_3 \nu_1 + \nu_6 \nu_2 \nu_4 \nu_1 + \nu_6\nu_2\nu_5\nu_3\nu_1 +  \nu_6\nu_2\nu_4\nu_6\nu_3 \nu_1\right).
\end{split}
\end{equation}
The arrows indicate that the the denominator is to be expanded as a
geometric series and then ordered so that factors $(\nu_6\nu_2\nu_4)$ are to be
multiplied successively on the left and the factors $(\nu_5 \nu_3\nu_1)$ are to
be multiplied successively on the right.

Note that if we put $\nu_3 = \nu_4 = \nu_5 = 0$,
so that we have only three incoming streets, then we find that $\tau_3 = \tau_4 = \tau_5 = 0$,
and
\begin{equation}
\tau_2 = \nu_2, \quad \tau_1 = \nu_1 + \nu_6\nu_2, \quad \tau_6 = \nu_6.
\end{equation}
This reproduces the rules for one-way streets which we obtained in \S\ref{sec:wnet-basics},
as expected.

For the record, the analog of \eqref{eq:hom-inv-jj} with the American resolution is
\begin{align} \label{eq:hom-inv-jj-american}
 \tau_1 = \nu_1 + \tau_6 \nu_2, \qquad \tau_2 = \nu_2 + \nu_3 \tau_1, \qquad \tau_3 = \nu_3 + \tau_2 \nu_4, \\ \nonumber
 \tau_4 = \nu_4 + \nu_5 \tau_3, \qquad \tau_5 = \nu_5 + \tau_4 \nu_6, \qquad \tau_6 = \nu_6 + \nu_1 \tau_5,
\end{align}
which leads to
\begin{equation}\label{eq:outgoing-2wayjoint-american}
\tau_1  =\frac{1}{1- \overrightarrow{\nu_1\nu_5\nu_3} \otimes \overleftarrow{\nu_4 \nu_6\nu_2}}\left(
\nu_1 + \nu_6 \nu_2 + \nu_1 \nu_5 \nu_2 + \nu_1 \nu_4 \nu_6 \nu_2 + \nu_1\nu_5\nu_3\nu_6\nu_2 +   \nu_1\nu_5\nu_2\nu_4\nu_6 \nu_2\right).
\end{equation}

\insfig{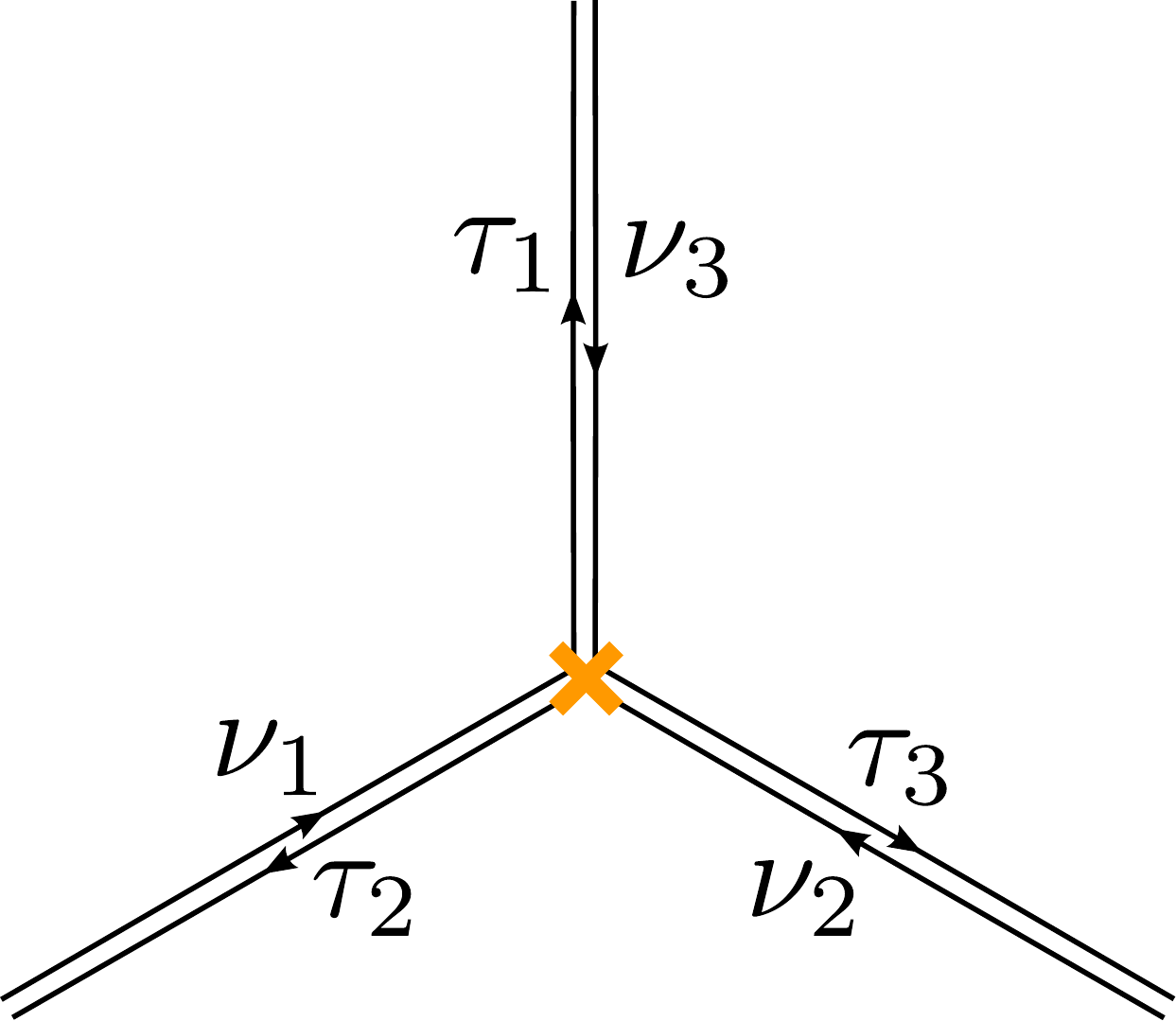}{Three two-way streets around a branch point.}

For a branch point with two-way streets we also have new rules.  As explained in \S\ref{sec:determining-wnet}
each of the three streets $p_i$ emerging from the branch point carries a charge $\bar a_i$
with a natural lift $a_i$, and $\mu(a_i, p_i) = 1$.  What changes in the two-way case is that
there may also be other charges supported on these streets.  Nevertheless, the outgoing degeneracies
$\tau_n$ are fully determined in terms of the incoming ones $\nu_n$:  the general statement
is
\begin{equation}
 \tau_n = X_{a_n} + \nu_n.
\end{equation}
(To make sense of this equation we have to say something about how we continue from
one street to another, since $\tau_n$ and $\nu_n$ are defined on different streets.  Unlike
the case of the joint above, we cannot simply continue from both streets to the branch point,
because there is monodromy there; we have to specify which way we go around the branch point.
We follow the short path around, traversing an angle $2 \pi/3$ rather than $4 \pi/3$.)
If we take all $\nu_n = 0$ then we simply recover
\begin{equation}
 \tau_n = X_{a_n}
\end{equation}
which we found in \S\ref{sec:determining-wnet}.

\subsection{Joint rules for soliton sets}

In the last subsection we discussed the rules for the soliton degeneracies $\mu$ at a joint.
These rules can be determined by the constraint of homotopy invariance.  However, the strange-looking series \eqref{eq:outgoing-2wayjoint} seems to be crying out for some more geometric interpretation.
Here we provide one. As a bonus, this interpretation also motivates a natural set of rules
for soliton sets (not only soliton degeneracies); these rules were included in \S\ref{subsec:Soliton-content}
as part of our definition of the soliton content of a general spectral network.

\insfig{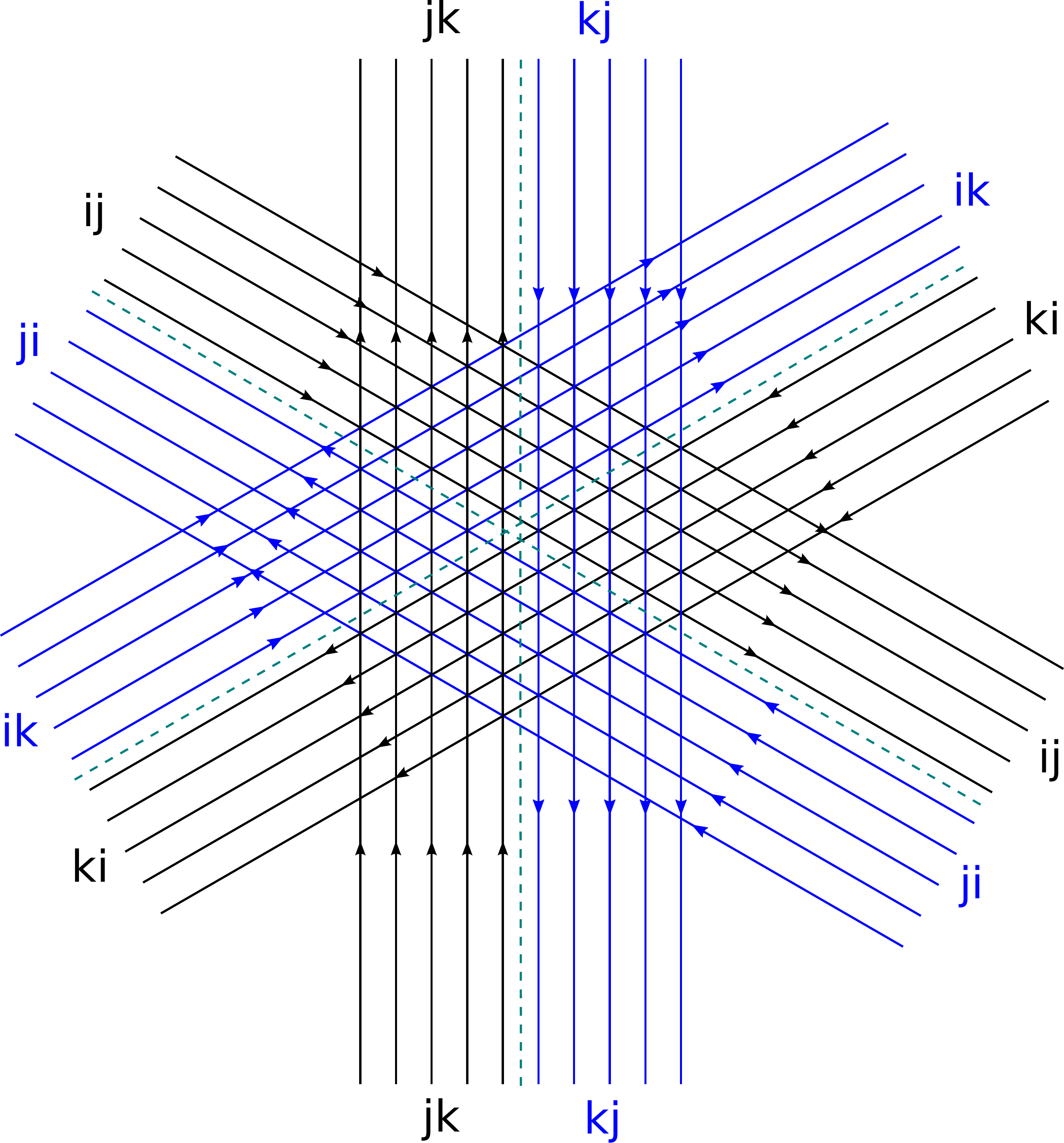}{The further resolution of the joint of Figure
\ref{fig:six-way-junction-2}, replacing each two-way street by infinitely many
parallel one-way ones.  (For clarity we only indicate five
one-way streets in each direction.)
Note that there is a vertical ``highway divider'' separating the $kj$ and $jk$ walls, and similarly for the other two pairs.  Accordingly, the plane is separated into $6$ regions, resembling the Weyl chambers of $su(3)$.}

We consider a joint of the type shown in Figure \ref{fig:six-way-junction-2}
which arises in some spectral network
$\wnet_{\vartheta_c}$.  In this picture, by assumption, each of the three $\cS$-walls supports
charges oriented in \ti{both} directions, e.g. the vertically oriented wall
supports charges $\bar a \in \Gamma_{jk}$ and $\bar a' \in \Gamma_{kj}$.
Concatenating these two
gives a closed cycle, $\cl(\bar a+\bar a') \in \Gamma$, with $e^{- \I \vartheta_c} Z_{\cl(\bar a+\bar a')} \in \R_-$.  Now, if the situation is generic enough, the semilattice of
$\gamma \in \Gamma$ with the property $e^{- \I \vartheta_c} Z_\gamma \in \R_-$ is
generated by a single element; let us assume we are in that situation, and let $\gamma$
denote the generator.  Then without loss of generality
the charges supported on the vertical wall can be
parameterized as $\{\bar a, \bar a + \gamma, \bar a + 2 \gamma, \dots\} \cup
\{\bar a', \bar a' + \gamma, \bar a' + 2 \gamma, \dots\}$, where $\cl(\bar a + \bar a') = \gamma$.
Similarly for the
other two walls, we replace $\bar a$ with $\bar b \in \Gamma_{ji}$ and $\bar c \in \Gamma_{ik}$,
and $\bar a'$ with $\bar b' \in \Gamma_{ij}$ and $\bar c' \in \Gamma_{ki}$.
Both $\cl(\bar a + \bar b + \bar c)$ and $\cl(\bar a' + \bar b' + \bar c')$
are positive multiples of $\gamma$, from which it follows that one of them
is $\gamma$ and the other is $2 \gamma$.  Again without loss of generality, let
us assume $\cl(\bar a + \bar b + \bar c) = \gamma$.

Now, suppose we perturb $\vartheta$ away from $\vartheta_c$ slightly, in the positive direction.
After the perturbation, these charges are no longer all supported on three walls:  rather,
each of the three breaks into an infinite set of walls, each supporting a single charge.
We thus obtain a further ``resolution'' of the spectral network in a
small neighborhood of the joint, pictured in Figure \ref{fig:resolved-six-way-scattering}.

Following the soliton lines in this picture, we arrive at a natural interpretation of
the solitonic traffic rule illustrated in
Figures \ref{fig:joint-rule}, \ref{fig:joint-rule-examples}.
Indeed, the ``highway divides'' pictured in Figure \ref{fig:resolved-six-way-scattering}
partition a neighborhood of the joint into 6 chambers, and in each chamber there is a
unique way in which two solitons can merge to produce a third.

Now we are ready to explain the series \eqref{eq:outgoing-2wayjoint}.
The first few terms correspond to the pictures in Figure \ref{fig:joint-rule-examples}.
For example, the first term $\nu_1$ corresponds to a single soliton which just travels through the
vicinity of the joint unmolested, while the second term $\nu_2 \nu_6$ corresponds to a
pair of incoming solitons which merge to produce an outgoing soliton.
To continue the series we build up longer and longer soliton trajectories by concatenating
successive mergings. At each step this amounts to replacing a factor $\nu_n$
by a pair of factors $\nu_{n+1}\nu_{n-1}$ or $\nu_{n-1}\nu_{n+1}$, alternating
between these possibilities. Each factor successively replaced is
(again alternately) one of the two factors inserted at the previous step. Thus, we begin
with $\nu_1$ and then replace $\nu_1 \to \nu_6 \nu_2$;
then in the product $\nu_6 \nu_2$ we replace $\nu_2 \to \nu_3 \nu_1$;
then in the product $\nu_3 \nu_1$ we replace $\nu_3 \to \nu_2\nu_4$; then
in the product $\nu_2 \nu_4$ we replace $\nu_4 \to \nu_5\nu_3$; and so on.
We recognize the seventh term as $(\nu_6\nu_2\nu_4)\nu_1 (\nu_5 \nu_3\nu_1)$,
and then the process repeats itself beginning with the factor $\nu_1$ in the
middle.

\section{A categorical approach}\label{App:MathSummary}

In this appendix we introduce some categorical constructions
which allow us to summarize some of the main results of the
paper in a language which, we hope, some of our mathematically
inclined readers will find congenial.

We begin in \S\S \ref{subsec:Ring-Category}  to \ref{subsec:WindingIdeal} with
a number of definitions.  Then in \S\ref{subsec:formal-pt-path} we state a
theorem regarding the ``formal parallel transport,'' or ``path-lifting,''
and its homotopy properties.
Finally, in \S\ref{subsec:2d4d-degen} we indicate how one can work
with paths on $C$ and $\Sigma$, rather than on $\tilde C$ and $\tilde \Sigma$,
at the price of introducing some tricky signs in the multiplication laws.

\subsection{The ring of a category}\label{subsec:Ring-Category}

To any category $\CC$ we may associate a ring $R(\CC)$.
As an abelian group, $R(\CC)$ is the free group on the
space of morphisms:
\begin{equation}
R(\CC) = \bigoplus_{f \in {\mathrm{Mor}}(\CC)} \IZ \cdot \ell_f .
\end{equation}
The ring structure in $R(\CC)$ is defined by
\begin{equation}
\ell_{f_1} \cdot \ell_{f_2} := \begin{cases} 0 & \text{if $f_1$ and $f_2$
are not composable}, \\
\ell_{f_1 f_2} & \text{if $f_1$ and $f_2$ are composable.}
\end{cases}
\end{equation}
Any functor $F:\CC \to \CD$
induces a canonical ring homomorphism $F: R(\CC) \to R(\CD)$.

More generally,
we could define a twisted version of $R(\CC)$ by writing
\begin{equation}
\ell_{f_1} \cdot \ell_{f_2} = \begin{cases} 0 & \text{if $f_1$ and $f_2$
are not composable}, \\
b(f_1, f_2) \ell_{f_1 f_2} & \text{if $f_1$ and $f_2$ are composable,}
\end{cases}
\end{equation}
where $b(f_1, f_2)$ is $\Z_2$-valued.
Associativity requires that $b$ is a cocycle.  A change
of basis changes it by a coboundary.  We denote the cohomology class of $b$ by $\sigma$ and the
corresponding twisted ring (up to isomorphism) by $R(\CC,\sigma)$.

\subsection{Categories of paths}\label{subsec:Category-Path}

Let $X$ be any topological space.  By a  \emph{path in $X$} we mean a continuous map
 $\wp: [T_1,T_2] \to X$ for some interval $[T_1, T_2] \subset \R$.
We define the \emph{path groupoid} ${\CP}(X)$ as follows:  the objects of $\CP(X)$ are
points of $X$, and the morphism space
${\CP}(X)(x_1,x_2)$ is the set of
all paths from $x_1$ to $x_2$, with the obvious composition.

By taking a quotient on the morphism spaces in $\CP(X)$ we can reduce to
the \emph{fundamental groupoid} $\pi_{\leq 1}(X)$. This is the
groupoid whose objects are points of $X$ and morphism spaces
$\pi_{\leq 1}(X)(x_1,x_2)$ are homotopy classes of paths from $x_1$ to $x_2$.

In \S \ref{subsec:formal-pt-path} and \S \ref{subsec:2d4d-degen}
 we will also make use of a further ``quotient'' of the
fundamental groupoid, which we call the \emph{first homology groupoid}
and denote $H_{\leq 1}(X)$.  It is a groupoid whose objects are again points of $X$.
The morphism space between two points
$x_1,x_2$ in $X$, denoted $H_{\leq 1}(x_1,x_2)$, is defined as follows.
First let $C_1(x_1, x_2)$ be the set of all 1-chains
$\mathfrak{c}$ in $X$ with $\p \mathfrak{c} = x_2 - x_1$. Note that if $x_1=x_2$ these
are simply 1-cycles.  In general $C_1(x_1,x_2)$ is a torsor for the
group of 1-cycles. We identify
\begin{equation}
H_{\leq 1}(x_1,x_2) := C_1(x_1,x_2)/\sim
\end{equation}
where $\mathfrak{c_1} \sim \mathfrak{c_2}$ if
$\mathfrak{c_1} - \mathfrak{c_2}$ is a 1-boundary.
The morphism space $H_{\leq 1}(X)(x_1,x_2)$ is an affine
subspace of the relative homology  $H_1(X, \{ x_1, x_2\};\IZ)$, and
is a torsor for $H_1(X;\IZ)$.
Composition of morphisms is induced by addition of chains.
The automorphism group of any object is canonically
$H_1(X;\IZ)$.
Note that there are natural functors $\CP(X) \to \pi_{\leq 1}(X) \to H_{\leq 1}(X)$.

\subsection{The winding ideal}\label{subsec:WindingIdeal}

Let $\pi: \tilde X \to X$ be a circle bundle.
There is an exact sequence
\begin{equation} \label{eq:homotopy-sequence}
 \pi_1(S^1) \stackrel{\iota}{\to} \pi_1(\tilde X) \stackrel{\pi_*}{\to} \pi_1(X) \to \pi_0(S^1).
\end{equation}
Choosing the standard generator $L$ of $\pi_1(S^1)$, let $W$ be the group generated by $\iota(L)$.
Then \eqref{eq:homotopy-sequence} induces an exact sequence
\begin{equation}
 1 \to W \to \pi_1(\tilde X) \stackrel{\pi_*}{\to} \pi_1(X) \to 1.
\end{equation}
So any class $[\wp] \in \pi_1(\tilde X)$ with $\pi_*([\wp]) = 1$ has a natural $W$-valued
``winding'' $w([\wp])$.

In the case where $X = S$ is a surface and $\tilde X = \tilde S$
is its bundle of tangent directions,
$W$ is either $\Z$ (if $X$ is punctured)
or $\Z / \chi(X) \Z$ (if $X$ is unpunctured).  In either case
there is a unique nontrivial map $W \to \Z / 2\Z$.  Applying this map to $w([\wp])$
we obtain a $\Z_2$-valued ``winding'' which by abuse of notation we also
call $w([\wp])$.  Given a pair of classes $[\wp_1],[\wp_2]\in
\pi_{\leq 1}(\tilde X)(x_1,x_2)$ with $\pi_*([\wp_1 \wp_2^{-1}]) = 1$,
we define the ``relative winding'' $w([\wp_1], [\wp_2]) = w([\wp_1 \wp_2^{-1}])$.

Now let the \emph{winding ideal} $I \subset R(\pi_{\leq 1}(\tilde X))$ be
the ideal generated by $[\wp_1] - (-1)^{w(\wp_1,\wp_2)} [\wp_2]$ for all pairs $[\wp_1],[\wp_2]\in
\pi_{\leq 1}(\tilde X)(x_1,x_2)$
such that  $\pi_*([\wp_1\circ \wp_2^{-1}])=1$.
This construction also descends to homology, giving an ideal $I \subset R(H_{\leq 1}(\tilde X))$.

\subsection{Formal parallel transport for the path groupoid}\label{subsec:formal-pt-path}

Now let $\Sigma \to C$ be a $K$-fold branched cover of an
oriented surface with at least one puncture, and let $\wnet$
be a spectral network subordinate to that cover, as defined in
\S\ref{subsec:GenSpecNet}.
The construction of \S\ref{subsec:Path-Lifting} defines a notion of \emph{formal parallel transport},
or \emph{path lifting}, along paths in $C''$, where $C''$ is $C$ with all joints of $\wnet$ and branch points of $\Sigma \to C$
removed.  The formal parallel transport is a ring homomorphism
\begin{equation}\label{eq:2d4d-deg-2}
\bF(\cdot, \wnet): R(\CP(\tilde C'')_{\off}) \to R(\CP(\tilde \Sigma)).
\end{equation}
Here $\tilde C''$ and $\tilde \Sigma$ denote the
circle bundles of tangent directions as usual, and $\CP(\tilde C'')_{\off}$ is the subcategory
whose objects do not lie on $\tilde\wnet$.

The formal parallel transport is closely related to flat connections,
as we saw in \S \ref{subsec:PushAbelian}.
The hallmark of a flat connection is that its parallel transport depends only on
the \emph{homotopy class} of the path. In order to see the homotopy
invariance   we must
project $R(\CP(\tilde \Sigma)) \to R(\pi_{\leq 1}(\tilde \Sigma)) \to R(\pi_{\leq 1}(\tilde \Sigma))/I$:
after doing so, and only after doing so, $\bF(\wp, \wnet)$   depends only on the
homotopy class of  $\wp$ in $\tilde C''$. Moreover, a check of
several special cases (equivalent to the computations of \S\ref{sec:homotopy-invariance})
shows that in fact $\bF(\wp, \wnet)$ only depends on the homotopy class
in $\tilde C$.  In establishing this it is crucial that we divide the codomain by
the winding ideal $I$, which enables various homotopically distinct paths in $\bF(\wp, \wnet)$ to
cancel one another.  Hence one can summarize the result of \S \ref{sec:homotopy-invariance} as

\medskip
\textbf{Theorem:} $\bF(\cdot, \wnet)$ descends to a homomorphism
\begin{equation}\label{eq:ROFF-HOM}
R(\pi_{\leq 1}(\tilde C)_{\off} ) \to R(\pi_{\leq 1}(\tilde \Sigma))/I.
\end{equation}

Moreover, the homomorphism \eqref{eq:ROFF-HOM} actually
factors through a homomorphism
\begin{equation}
R(\pi_{\leq 1}(\tilde C)_{\off} )/I \to R(\pi_{\leq 1}(\tilde \Sigma))/I.
\end{equation}
Passing from homotopy to homology in the codomain we obtain a map:
\begin{equation}\label{eq:R-MOD-I}
F(\cdot, \wnet): R(\pi_{\leq 1}(\tilde C)_{\off} )/I \to R(H_{\leq 1}(\tilde \Sigma))/I.
\end{equation}
Expanding  $F([\wp],\wnet)$ on a ${\mathbb Z}$-basis $X_{a}$ for  $R(H_{\leq 1}(\tilde \Sigma))/I$,
where $a$ runs over $\tilde\Gamma(\tilde z_1, \tilde z_2)$,
the coefficients are  the degeneracies $\fro'(L_{\wp},a)$.

\subsection{Cocycles}\label{subsec:2d4d-degen}

We would now like to relate the formalism of this paper to the formalism using the
$X_{\gamma_{ij'}}$ used in our previous
paper \cite{Gaiotto:2011tf}.  The essential problem here is one of signs.
In this section we explain how one can construct a twisted ring, which
allows us to work directly with paths on $C$ and $\Sigma$, at the cost of introducing
some subtle cocycles in the multiplication rules. The precise statement is
equation \eqref{eq:2d4d-deg-3} below. What we actually prove is the closely
related statement \eqref{eq:2d4d-deg-1}, but to pass to  \eqref{eq:2d4d-deg-3}
we rely on a conjecture stated below \eqref{eq:whole-disc}.

In order to define the framed 2d-4d degeneracies in terms of paths on $C$ rather
than on $\tilde C$, we will need a way of lifting from $C$ to $\tilde C$. To do this we introduce
$\CI(C)$, the \emph{immersion category} of $C$. The objects are points on $C$
together with tangent directions, i.e. points of $\tilde C$, and the morphisms are
immersions of intervals $[T_1,T_2]$ into $C$. Note that
there is no identity morphism so this ``category'' is a category without identity
morphisms, i.e., a \emph{semicategory}.
We can view it as a subsemicategory of $\CP(\tilde C)$.  If we
mod out by regular homotopy (i.e. homotopy through immersions) we obtain a semicategory $\pi^{Reg}_{\leq 1}(C)$.
We also have an obvious functor $\pi^{Reg}_{\leq 1}(C) \to \CP(\tilde C)$, hence a ring homomorphism
\begin{equation}\label{eq:Lift-map}
{\rm Lift}: R(\pi^{Reg}_{\leq 1}(C)) \to R(\pi_{\leq 1}(\tilde C)).
\end{equation}
Composing with the projection $R(\pi_{\leq 1}(\tilde C)) \to R(\pi_{\leq 1}(\tilde C)) / I$ we get
\begin{equation}\label{eq:map-to-quotient}
  R(\pi^{Reg}_{\leq 1}(C)) \to R(\pi_{\leq 1}(\tilde C))/I.
\end{equation}
Recall that the objects of $\pi^{Reg}_{\leq 1}(C)$ are points of
$\tilde C$. Thus, a point $z\in C$ with tangent vector corresponds
to a point $\tilde z\in \tilde C$. A little closed loop beginning at $\tilde z$
and wrapping around the fiber of $\tilde C$ over $z$
maps to  to $-1_{\tilde z}$.

The statement we are aiming for, \eqref{eq:2d4d-deg-3} below,
concerns homotopy classes of paths, rather than regular homotopy classes. Therefore,
the first step is to find a homomorphism $R(\pi_{\leq 1}(C)) \to R(\pi^{Reg}_{\leq 1}(C))$.
We could attempt to find such a homomorphism
by choosing, for each homotopy class, an immersion in the same homotopy
class. In general this will \emph{not} produce anything like a homomorphism
into $ R(\pi^{Reg}_{\leq 1}(C))$. First of all, composable paths in
$\pi_{\leq 1}(C)$ will, in general, not map to composable immersions in
$\pi^{Reg}_{\leq 1}(C)$. Thus, the first thing we should do is choose a
nowhere-zero vector field on $C$ and, for each homotopy class in $\pi_{\leq 1}(C)$,
choose an immersion whose initial and final tangent vectors match the
tangent vector at a point.  However, even once we have done this, if we
lift two composable homotopy classes $[\bar\wp_1]$ and $ [\bar \wp_2]$
to regular homotopy classes of immersions $[\iota_1]$ and $[\iota_2]$
there is no guarantee that the lift $[\iota_{12}]$ of
$[\bar\wp_1 \circ \bar \wp_2]$ is the same as $[\iota_1 \circ \iota_2]$.
That is,
 \begin{equation}
 [\iota_{12}] \not= [\iota_1\circ \iota_2] := [\iota_1]\cdot [\iota_2]
 \end{equation}
so we still do not get a homomorphism.
However, if we compose our
non-canonical non-homomorphism  with the map to the
quotient \eqref{eq:map-to-quotient} then we \emph{will} get a homomorphism
from a twisted ring:
\begin{equation}\label{eq:possible-iso}
R(\pi_{\leq 1}(C),\sigma) \to  {\rm Lift}\left( R(\pi^{Reg}_{\leq 1}(C)) \right)/I
\end{equation}
The cocycle in question is
\begin{equation}\label{eq:immersion-cocycle}
\sigma([\wp_1],[\wp_2]) = (-1)^{w[\iota_1\circ \iota_2\circ \iota_{12}^{-1,a}]}
\end{equation}
where $\iota_{12}^{-1,a}$ is the \emph{anti-lift} of the inverse immersion
$\iota_{12}^{-1}$. (Given an immersion,  we can lift it in the usual way, but then we can compose with the antipodal
map in the fiber. Call that the \emph{anti-lift}.
Given an immersion $\bar \wp$ in $C$ with canonical lift $\wp$ in $\tilde C$,
$\wp^{-1}$ is the antilift of $\bar\wp^{-1}$.) It would be nice to
have a more conceptual and invariant description of this cocycle.

Now let us reconsider the framed 2d-4d BPS degeneracies. We can
compose the map \eqref{eq:possible-iso} with the formal monodromy
and finally project to the homology groupoid to get a homomorphism of the form
\begin{equation}\label{eq:2d4d-deg-1}
\bar F: R(\pi_{\leq 1}(C),\sigma) \to R(H_{\leq 1}(\tilde \Sigma))/I
\end{equation}

The whole discussion above for the curve $C$ can be repeated
for $\Sigma$ to construct a map
\begin{equation}\label{eq:whole-disc}
R(H_{\leq 1}(\Sigma), \sigma^{\rm ir}) \to R(H_{\leq 1}(\tilde \Sigma))/I
\end{equation}
where the cocycle $\sigma^{\rm ir}$ is analogous to that
appearing in \eqref{eq:2d4d-deg-1}. It is natural to conjecture that
\eqref{eq:2d4d-deg-1} in fact factors through a homomorphism
\begin{equation}\label{eq:2d4d-deg-3}
\bar F: R(\pi_{\leq 1}(C),\sigma^{\rm uv}) \to R(H_{\leq 1}(\Sigma), \sigma^{\rm ir})
\end{equation}
where $\sigma^{\rm uv}$ is the cocycle \eqref{eq:immersion-cocycle}
appearing in \eqref{eq:2d4d-deg-1}. This would be a gauge-invariant
formulation of the 2d-4d degeneracies directly involving paths on
$C$ and $\Sigma$.

\section{Convention convention}\label{app:Conv-conv}

We summarize here some conventions used in this
and our preceding papers.

\begin{enumerate}

\item Homology classes on the IR (Seiberg-Witten)
curve $\Sigma$ are denoted in this paper by $\bar a \in \Gamma_{ij'}(z,z')$.
In our previous paper \cite{Gaiotto:2011tf} they were denoted by $\gamma_{ij'}$.
They are represented by paths oriented from $z^{(i)}$ to ${z'}^{(j')}$.
Thus $\p \gamma_{ij'} = {z'}^{(j')} - z^{(i)}$.
The sum $\gamma_{ij'} + \gamma_{j'k''} \in \Gamma_{ik''}$ is thus
oriented from $z^{(i)}$ to ${z''}^{(k'')}$. Note that the $+$
is not commutative:  the sum in the opposite order
$\gamma_{j'k''}+\gamma_{ij'}$ in general does not make sense.

\item Oriented open paths on the UV curve $C$ are generally
denoted by $\wp$, and often the
initial point is $z_1$ and the final point is $z_2$.
If $\wp' $ is another such path, with initial
and final points $z_1'$ and $z_2'$, respectively,
and $z_2=z_1'$ and moreover the canonical lifts of
$\wp $ and $\wp'$ to $\tilde C$ can be concatenated,
then we denote by $\wp\wp'$ the composed path, which goes from $z_1$ to $z_2'$.
Note the ``later'' path is written to the right.

\item $Z_{\gamma} = \frac{1}{\pi} \oint_{\gamma} \lambda$ is the central charge of a 4d state,
$Z_{\bar a} = \frac{1}{\pi} \int_{\bar a} \lambda$ is the central charge of a 2d soliton
for $\bar a \in \Gamma_{ij}(z,z)$, $i\not=j$, and the same expression is
also the central charge of a framed 2d-4d state with $\bar a \in \Gamma(z_1,z_2)$.
If $\bar a \in \Gamma_{ij}(z_1,z_2)$ we can regard $Z_{\bar a}$ as a function
 on $\CU_{i} \times \CU_{j}$, where
$\CU_i$ is a sufficiently small open region around $z_1^{(i)}$ and
$\CU_{j}$ is a sufficiently small open region around $z_2^{(j)}$.
In this case  $\de Z_{\bar a} = \frac{1}{\pi} ( \lambda^{(j)} - \lambda^{(i)})$.

\item For an ordered pair $(i,j)$ of sheets,
BPS walls of type $ij$ are defined by $e^{-\I \vartheta} Z_{\gamma_{ij}} <0$ with $\mu(\gamma_{ij})\not=0$.
Therefore they are oriented on $C$ so that $e^{-\I \vartheta} \langle \lambda^{(i)} - \lambda^{(j)}, \p_t \rangle > 0$,
where $\p_t$ is a tangent vector to the wall in the direction of the orientation.
Note that this implies that the integral  $e^{-\I \vartheta}\int^{z(t)}   \lambda^{(i)} - \lambda^{(j)}
= - e^{-\I \vartheta} \int_{\gamma_{ij}} \lambda$
\emph{increases} as $z(t)$ moves
along the direction of the orientation of the wall.

\item The mass of a soliton with charge $\bar a\in \Gamma_{ij}(z,z)$
is $M(z) = - \frac{e^{-\I \vartheta}}{\pi} \int_{\bar a} \lambda$.

\item Lifting to the circle bundle of directions in the tangent bundle is denoted by
a tilde.  Thus $\tilde \Sigma$ is the circle bundle of tangent directions over $\Sigma$, and so forth.

\item $F(\wp, \vartheta)$, $\textbf{F}(\wp,\wnet)$ and $\bF(\tilde \wp, \wnet, \nabla^\ab)$
all compose in the same way we compose paths in item 2 above:
$F(\wp, \cdot) F(\wp', \cdot) = F(\wp\wp', \cdot)$.

\item The parallel transport $\bF(\tilde \wp, \wnet, \nabla^\ab) \in {\rm Hom}(E_{\tilde z_1}, E_{\tilde{z_2}})$
if $\wp$ goes from $z_1$ to $z_2$.
In \cite{Gaiotto:2011tf}
$\hat E$ was denoted by $V$, $\CL_i$ was denoted
by $V_i$, and  $F(\wp,\wnet_\vartheta,\nabla^\ab)$ was denoted $\langle L_{\wp} \rangle$
and expanded in $\CY_{\gamma_{ij'}} \in {\rm Hom}(\CL_i, \CL_{j'})$.
(The conventions of \S 5 of \cite{Gaiotto:2011tf} are opposite to those of \S\S 7-9 of
\cite{Gaiotto:2011tf}; we use the conventions of \S\S 7-9.)

\item We compose linear transformations on the \emph{right} so that the composition
of linear operators $T_1\in {\rm Hom}(V_1,V_2)$
and $T_2 \in {\rm Hom}(V_2,V_3)$ is written $T_1 T_2\in {\rm Hom}(V_1,V_3)$.
The parallel transport of a section $s(z)$ from
$z_1$ to $z_2$ by the connection $\nabla = \Psi_\wnet(\nabla^\ab)$ is thus
given by $s(z_2) = s(z_1) \bF(\tilde \wp, \wnet, \nabla^\ab)$.
This differs from the more usual convention and in particular differs from
\cite{Gaiotto:2011tf}. In a local trivialization we can write this as
\be\label{eq:pexp-sign}
s(z_2)= s(z_1) {\rm Pexp} \left( - \int_{z_1}^{z_2} \CA \right)
\ee
where $\CA$ is the $\fg\mathfrak{l}(K,\IC)$-valued 1-form representing $\nabla$.

\item In \cite{Gaiotto:2011tf}, equations $(5.31)$ and $(5.32)$, we
introduced flat sections  $\CY_{\gamma_{ij'}}(z,z')$ of the vector
bundle $\hat E^* \otimes \hat E$ on $C \times C$ obtained by
projection and parallel transport.
For a homology class $a \in \Gamma(\tilde z,\tilde z')$ connecting $\tilde z^{(i)}$ and $\tilde z'^{(j')}$
we define $\CY_{a}$ to be parallel transport with respect to
$\nabla^\ab$ along $a$ from $\tilde z^{(i)}$ to  $\tilde z'^{(j')}$.
The two notions coincide for $z,z'$ away from $\CS$-walls if we
make use of the isomorphism \eqref{eq:e-simp-decomp}.
From either point of view we have, in local coordinates
\be
\begin{split}
\partial_z \CY_{\gamma_{ij'}} - \CA_z \CY_{\gamma_{ij'}} & = 0, \\
\partial_{z'} \CY_{\gamma_{ij'}} +   \CY_{\gamma_{ij'}} \CA_{z'} & = 0. \\
\end{split}
\ee

\item The map between solutions $(\varphi, A)$ of Hitchin equations
and flat connections $\CA$ (used in
 \cite{Gaiotto:2008cd,Gaiotto:2009hg,Gaiotto:2010be,Gaiotto:2011tf} although not in this paper)
is $\CA = \frac{R}{\zeta}\varphi + A + R \zeta \overline{\varphi}$.

\item The WKB asymptotics of flat sections are formally given by
\begin{equation}
\CY_{\gamma_{ij}} \sim \exp \frac{ \pi R}{\zeta} Z_{\gamma_{ij}} + \cdots
\end{equation}
This implies that if $\zeta$ lies in the half-plane $\IH_{\vartheta}$,
then $\CY_{\gamma_{ij}}(z)$ is exponentially \emph{small} as $\zeta \to 0$
when $z$ lies on a wall of type $ij$ supporting the charge $\gamma_{ij}.$

\end{enumerate}

\section*{Acknowledgements}

We would like to thank P. Boalch, T. Bridgeland, A. Goncharov, N. Hitchin, S. Keel, I. Smith and E. Witten for discussions, and T. Mainiero for pointing out several typos in a previous preprint version.

GM and AN thank the KITP for hospitality, where
some of this work was done. GM would also like to
thank the Institute for Advanced Study and the Ambrose Monell
Foundation for partial support during part of this research.
This research was supported in part by DARPA under grant
HR0011-09-1-0015 and by the NSF under grant PHY05-51164.
The work of DG is supported in part by NSF grant PHY-0969448 and in part
by the Roger Dashen membership in the Institute for Advanced Study.
The work of GM is supported by the DOE under grant
DE-FG02-96ER40959. The work of AN is supported by
the NSF under grant DMS-1006046.

\bibliographystyle{utphys}

\bibliography{swn-paper}

\end{document}